%% file: compile_baseline_v24.tex
\documentclass[12pt,english]{article}
\usepackage{amsmath}
\usepackage{amssymb}
\usepackage{amsthm}
\usepackage{mathptmx}
\usepackage{newtxmath}
\usepackage[T1]{fontenc}
\usepackage[utf8]{inputenc}
\usepackage{geometry}
\usepackage{comment}
\usepackage{color}
\usepackage{babel}
\usepackage{float}
\usepackage{placeins}
\usepackage{url}
\usepackage{graphicx}
\usepackage{booktabs}
\usepackage{longtable}
\usepackage{setspace}
\usepackage[authoryear,round]{natbib}
\let\citepbase\citep
\let\citetbase\citet
\let\citebase\cite
\let\citealtbase\citealt
\let\citealpbase\citealp
\renewcommand{\citep}{\citepbase*}
\renewcommand{\citet}{\citetbase*}
\renewcommand{\cite}{\citebase*}
\renewcommand{\citealt}{\citealtbase*}
\renewcommand{\citealp}{\citealpbase*}
\usepackage{lineno}

\usepackage[unicode=true,pdfusetitle,
 bookmarks=true,bookmarksnumbered=false,bookmarksopen=false,
 breaklinks=false,pdfborder={0 0 0},pdfborderstyle={},backref=false,colorlinks=true]
 {hyperref}
\hypersetup{
 linkcolor=black,urlcolor=black,citecolor=black}

\usepackage{eurosym}

\theoremstyle{plain}

\makeatletter

\providecommand{\tabularnewline}{\\}

\usepackage[bottom]{footmisc}
\usepackage{indentfirst}
\usepackage{pdflscape}
\usepackage{xcolor}

\usepackage{titlesec}
\usepackage{caption}
\usepackage{subcaption}
\makeatother

\graphicspath{{figures/}{./}}

\title{\textbf{\Large The Cross-Section of Stock Returns and AI Exposure}}
\author{Nicola Borri, Yukun Liu, and Aleh Tsyvinski\thanks{Nicola Borri is with LUISS University. Yukun Liu is with the University of Rochester, Simon Business School. Aleh Tsyvinski is with Yale University and NBER. Data provided by OpenRouter, Inc. Used under license. The analysis is based only on aggregated and anonymized derivative data constructed by the authors. The original OpenRouter data are not redistributed or disclosed. We are grateful to OpenRouter for providing access to the data, especially to Justin Summerville. We thank Daron Acemoglu, Andrew Atkeson, Alejandro Lopez-Lira, and Pascual Restrepo for detailed comments, Ron Borzekowski and the Yale Data-Intensive Social Science Center for assistance in onboarding the data. This paper was previously circulated under the title ``AI Premium''.}}
\date{September 24, 2026}

\begin{document}
\pagecolor{white}

\hypersetup{pageanchor=false}
\maketitle

\begin{abstract}
\begin{singlespace}

\noindent 
We study 380 trillion tokens of realized AI consumption across more than four hundred LLMs. We build a high-frequency AI factor and show that a long-short strategy based on firms’ AI exposure earns significantly positive returns. The average strategy return is larger based on intensive, frontier-oriented AI consumption but smaller based on casual or open-weight usage. Internationally, the return spread is significant in developed countries but insignificant in emerging markets. Examining occupational AI exposure, we find more positive exposure in occupations intensive in nonroutine interactive tasks and more negative exposure in those intensive in nonroutine analytical tasks.
\end{singlespace}
\end{abstract}

\newpage
\hypersetup{pageanchor=true}

\setcounter{page}{1}

\section{Introduction}

Using 380 trillion tokens of realized AI consumption across more than four hundred large language models from the OpenRouter dataset covering approximately 2 percent of current global monthly AI token consumption from January 2024 through April 2026,\footnote{OpenRouter's May 2026 run rate is about 100 trillion tokens per month \citep{dasMurphy2026openrouter}. For comparison, Google alone disclosed processing 3.2 quadrillion tokens per month at I/O 2026 \citep{pichai2026io}, so accounting for the other major providers puts total global LLM token consumption at roughly 5 to 7 quadrillion per month and OpenRouter at about 2 percent, also consistent with the projections by Goldman Sachs Research \citep{tan2026goldmanAiBoom}.} we analyze how AI affects firms, markets, and workers. Leveraging the unprecedented size, scope and granularity of OpenRouter AI data, we construct and analyze a forward-looking measure of firms' exposure to AI consumption growth. We use this measure to study subsequent stock returns and the distribution of AI exposure across firms, occupations, tasks, and skills. Specifically, we start by constructing the \emph{AI Factor}---a high-frequency measure of growth in realized worldwide AI consumption, of which we further analyze its various important components. We then measure how strongly each public firm's equity price co-moves with this factor---firms' \emph{AI Betas} (or AI exposure)---and characterize the associated spread in subsequent average stock returns, the \emph{AI Premium}.

The paper has four main parts. First, we build a high-frequency, time-varying AI factor from the full universe of OpenRouter AI token consumption data and decompose it into its important components: frontier closed-source versus open-weight models, core versus new users, seasoned versus casual users, long versus short prompts, prompt-content categories (e.g., programming, science, roleplay), and agentic versus ordinary requests. The AI factor and its different components show that the diffusion is uneven, token consumption is concentrated in programming, technology, and sciences, and the share of agentic tokens increased dramatically from negligible to more than half of total AI consumption. Second, we show that AI exposure is significantly associated with the cross-section of future stock returns, and the AI premium is both large and heterogeneous. A value-weighted long-short strategy earns 60.4 basis points per week. We then show that the premium extends beyond frontier model-release weeks: excluding those weeks, the spread remains 34.3 basis points per week. Moreover, the AI premium is associated with intensive, frontier-oriented AI consumption---closed-source models, paid/core and seasoned users, and long prompts, but is not present in open-weight or casual use. Third, the AI exposure reaches beyond technology firms into consumer-facing and capital-heavy parts of the economy: industry exposure is most positive in retail and turns most negative in health and non-durable goods. We also show that the AI premium is larger in regions where AI development and adoption are more advanced---about 10 basis points per week across developed markets, such as Europe, but small and insignificant in emerging markets, including China. Fourth, we build a market-implied AI exposure of the tasks and skills across the US economy. We show that the nonroutine, interactive work---communication, persuasion, instruction, and the installation and repair of new systems---loads more positively on the AI factor, and analytical, scientific, and operations-control skills load less. Specifically, the most positive loadings are on the interaction-and-communication skills. An occupation which is one standard deviation higher in interaction-and-communication content has a 0.26 standard deviation higher market-implied AI exposure. Our measures are constructed using market information, building on the literature of using stock-market prices to value innovation \citep{KoganPapanikolaouSeruStoffman2017}, and differ from task-based approaches \citep{AcemogluRestrepo2019}, usage-based approaches such as the Anthropic Economic Index \citep{anthropic2026learningCurves}, and survey-based approaches (\citealp{bick2026mind,baslandze2026artificial,lindenlaub2026beyond}).

We now provide a detailed description of the underlying data. OpenRouter, Inc.\ is a global AI platform that routes requests between users and more than four hundred large language models through a single interface---from frontier closed-source models such as OpenAI's GPT, Anthropic's Claude, and Google's Gemini to open-weight models such as Meta's Llama, DeepSeek, and Qwen. Our licensed data is built from the universe of OpenRouter data, anonymized and aggregated to the user--model--day level, from January 2024 through April 2026. For each anonymized user, model, and day, we observe the number of requests, the completion and prompt tokens, and the dollar cost, as well as many other characteristics. The panel covers millions of anonymized accounts, allowing us to track individual token consumption of each anonymized user over more than two years. Two additional features of the data are important to note. First, token consumption in the dataset is realized consumption rather than reported consumption, since every observation is a paid request carrying its own token count and dollar cost. Second, the dataset spans every major provider rather than just a single model: the average user account draws on more than five distinct AI models, and 70\% of users use at least two providers. The unprecedented granularity of the data allows us to separate frontier closed-source from open-weight AI consumption, intensive from casual users, to identify the content category of each request, to determine agentic requests that call external tools, and track a large variety of other salient characteristics of anonymized user token consumption over time and across models. We start by building three weekly series---total tokens, dollars spent, and distinct active users---that measure the quantity, expenditure, and breadth of global AI consumption.\footnote{For comparison, the Anthropic Economic Index, currently the most prominent and comprehensive public measure of AI demand, classifies only about one million Claude conversations from a single provider over a short sample period \citep{anthropic2026learningCurves}.}

First, using the OpenRouter data, we build an AI factor using realized AI consumption series as the first principal component of the weekly growth rates of tokens, dollars, and active users. The weekly AI factor loads positively and evenly on all three measures. Moreover, the user--model--day granularity also allows us to construct separate sub-factors for different components of token consumption by model tier, user intensity and tenure, prompt complexity, content category, and agentic use, to study the subcomponents of AI consumption.

Second, we show that a company's AI exposure positively and significantly predicts its future stock returns, which we denote as the AI premium. Firms whose returns covary more positively with the AI factor---high AI beta firms---earn higher subsequent returns: in value-weighted quintile portfolios, the most exposed firms outperform the least exposed by 60.4 basis points per week, which is significant at the 1 percent level, and by about 53 basis points after Fama-French five-factor and momentum adjustment, which is significant at the 5 percent level. Under the Fama-MacBeth regressions, the coefficients on the AI exposures are also positive and significant after controlling for size, value, profitability, investment, momentum, reversal, leverage, and accruals. Two caveats qualify this result. First, the sample covers an early, fast-moving phase of AI diffusion, so the estimates describe realized return differences during this period. Second, data from OpenRouter, Inc., though exceptionally rich, likely overrepresents sophisticated and developer users, so the premium is tied to the market pricing of realized observed AI consumption toward these users.

We then test whether the AI factor repackages the recent rally or potential hype in technology stocks and AI-themed equities. Specifically, we construct first-stage AI exposures controlling for returns to high-technology stocks, a value-weighted semiconductor portfolio, or an AI/robotics ETF basket. The value-weighted high-minus-low spread based on these firms' AI exposure remains significant at 48.7, 35.0, and 53.1 basis points per week, respectively. The spread is also 54.1 basis points, significant at the 1 percent level, after removing the Fama-French 30 industry component, and 69.1 basis points, significant at the 1 percent level, when the first-stage AI exposure is estimated controlling for Google Trends attention to AI.

Third, we analyze the heterogeneity of the AI premium. We start by splitting AI token consumption into its important components. We find that the AI premium is primarily related to intensive, frontier-oriented AI consumption---frontier models, experienced users, and complex tasks---rather than to casual or open-weight model usage. The AI premium is larger when exposure is estimated from frontier closed-source model consumption: the high-minus-low spread is significant at 51.6 basis points per week using closed-source AI consumption compared with 23.0 basis points for AI consumption of open-weight models. Similar results also hold across the other measures of intensive and sophisticated AI consumption: 44.5 basis points for paid/core users versus 18.5 basis points for new users, 54.0 basis points for seasoned users versus 31.6 basis points for non-seasoned users, and 47.7 basis points for long prompts versus 29.6 basis points for short prompts. Next, we examine the AI premium by country. Repeating the portfolio sorting exercise gives a significant 10.0-basis-point weekly spread in developed markets but only an insignificant 5.0-basis-point spread in emerging markets, including China. The international results are consistent with the view that the premium is more pronounced in regions where listed firms and investors are closer to the development and adoption of frontier AI. 

We next examine how the AI premium varies around major model releases and whether it extends beyond release weeks. In our event study of AI model releases, high-AI-beta firms outperform low-AI-beta firms around frontier-provider releases by 1.7 percent over the five-day window and by 0.9 percent after five-factor and momentum adjustment, both of which are statistically significant, at the 1 and 10 percent levels, respectively, while the responses to non-frontier releases are more muted. When we instead remove the frontier model-release weeks from the sample, the value-weighted high-minus-low spread remains positive and significant at the 5 percent level at 0.343 percent per week. That is, the premium is stronger around releases, but a significant portion accrues gradually throughout the sample rather than only on AI release dates, suggesting that an important fraction is not directly tied to AI news surprises.

An important paper by \citet{koganPapanikolaou2014growth} establishes that a premium for technology risk generically arises in models in which firm exposures to technological change are heterogeneous. \citet{papanikolaou2011investment} further shows that the sign of the premium depends both on the investment-specific technology (IST) shock they study and, under recursive preferences, on whether households prefer early or late resolution of uncertainty. The premium of the IST shock depends on two effects: reallocation toward investment today and news about future growth. Under late resolution, both channels lead to a negative premium of the IST shock. Under early resolution, the reallocation channel carries a negative premium while the second channel carries a positive premium. That is, the sign of the premium of the IST shock depends on its composition. \citet{koganPapanikolaou2014growth} infer the negative price of the IST shock from the cross-section, and \citet{kogan2020left} obtain a negative sign through limited risk sharing between innovators and other households. Our cross-sectional results point to a positive price of AI exposure, and we show that AI shocks are empirically different from the IST shock.\footnote{Our results are further consistent with the transition-risk channel of \citet{pastorVeronesi2009}, in which new technologies reallocate rents across firms, sectors, and workers, and a technology's risk becomes systematic, and priced, once adoption is large-scale.} The large realized AI premium we document during a period when the AI phenomenon is rapidly unfolding is also consistent with a non-risk-based mechanism: revisions during the sample to expectations about which firms' cash flows AI will raise or displace. The AI release date results described above bear on both risk and non-risk interpretations. The large responses of high-exposure firms around frontier releases suggest revisions of expectations during these release periods, while the persistence of the spread outside release weeks indicates that it is not only confined to AI release news.

Fourth, we use the same baseline firm-level exposures to study the market-implied AI exposure of different occupations and skills in the current organization of work. We first map firm and industry exposures into occupations using BLS employment weights and then map occupations into skills using O*NET ratings. Then, we compare AI exposures of occupations with the task taxonomies in \citet{AutorLevyMurnane2003} and \citet{AcemogluAutor2011}, and the skill measures in \citet{deming2017growing}. We construct a market-based AI exposure for the occupations, which is different from the technical automatability score or labor-displacement forecast that are the focus of task- and capability-based exposure measures such as \citet{eisfeldt2023generative}, \citet{kogan2023technology}, and \citet{liu2025technology}. We find that a one-standard-deviation increase in nonroutine interactive content is associated with a significant 0.16-standard-deviation higher market-implied AI exposure, suggesting that the current market implies these tasks may benefit from AI development. On the other hand, a one-standard-deviation increase in nonroutine analytic content is associated with a significant 0.17-standard-deviation lower exposure. Regressing AI exposure on the skill content of occupations, we find that the coefficient estimates of social skills tend to be positive and significant. For example, the coefficient estimate of interaction-and-communication is large at 0.26, significant at the 1 percent level. The coefficient estimate of information use, on the other hand, is negative and significant at the 5 percent level. The market-based occupation-level AI-exposures are significantly different from existing technical automatability measures---existing occupation-level AI-exposure measures explain less than 2 percent of this variation.

Finally, we provide early evidence on the rise and market-implied effect of agentic AI consumption, defined as models executing multi-step workflows that plan, call external tools, take actions, and coordinate with other agents (see, e.g., \citealp{hassabisKavukcuoglu2024gemini2,ng2024agenticpatterns,ng2024tooluse,yang2025perplexityAgents}). We measure this component in the OpenRouter data as requests in which the model invokes an external tool. We document the rise in the agentic token share from near zero in 2024 to roughly half of all tokens by the end of the sample, and the agentic dollar share rises more slowly as the realized price per agentic token falls. The granularity of the OpenRouter data allows us to track three salient token-level components of this activity: tool calls, the model's internal reasoning tokens, and cache reads of recurring prompt prefixes. We show that tool-call and cache-read shares rise to roughly two-fifths to one-half of total token consumption, while reasoning tokens rise more slowly. The decline in unit cost is consistent with more usage of cache tokens and more efficient routing of agentic requests to cost-effective backing models as firms struggle with skyrocketing AI token budgets. Moreover, in the cross-section of equity returns, exposure to these agentic-consumption components tends to have positive, although insignificant, point estimates for the individual agentic factors. Because these estimates are small and imprecise since agentic token consumption is a very recent phenomenon, we interpret this result as early and suggestive evidence of a positive agentic premium.

We next briefly relate our paper to the relevant literature. \citet{AcemogluRestrepo2019aiWork,AcemogluRestrepo2019,acemogluRestrepo2022tasks} provide the task-based mechanism through which automation displaces labor from some tasks, creates new tasks, and reallocates rents across workers and firms. In asset pricing, \citet{koganPapanikolaou2014growth} link technology shocks and growth opportunities to expected returns, and \citet{garleanuKoganPanageas2012} and \citet{kogan2020left} show that displacement risk is priced. Recent work also studies AI and market beliefs in asset prices \citep{andrewsFarboodi2025transformative}, and \citet{blumenfeld2026software} use firms' stock-return exposure to an AI equity index to measure the market-implied software engineering productivity gains from AI. \citet{lopezLiraTang2026chatgpt} show that large language models' assessments of news headlines predict initial stock-price reactions and subsequent return drift. \citet{babinaFedykHeHodson2024systematicRisk} show that firms investing more in AI experience higher market betas, consistent with AI creating growth options. \citet{babinaFedykHeHodson2024firmGrowth,babina2026understanding} measure firms' AI efforts and study their effects on growth, innovation, and broader economic outcomes. An important paper by \citet{eisfeldt2023generative} uses AI to classify task exposure and then maps those task scores to occupations and firms. They show that the returns of the high-exposure firms rose by about 5 percent relative to low-exposure firms after OpenAI introduced ChatGPT, initially based on GPT-3.5, that the effect is distinct from product-market AI exposure and stronger for firms with data assets. Our event-study evidence shows similar results. The OpenRouter--a16z report by \citet{openrouter2025state} documents stylized facts about the structure of the token consumption market from a 100-trillion-token OpenRouter dataset: substantial open-weight adoption, heavy use for roleplay and coding, the rise of agentic inference, global variation, cost-use patterns, and persistent early-user cohorts. \citet{demirer2025emerging} use early scraped, aggregate data from OpenRouter's public website to study LLM pricing and demand. Using micro data from the AI coding platform Cursor, \citet{sarkar2026agents} and \citet{sarkar2026returns} show that AI agents shift developer effort from implementation toward supervision and that agent usage rises sharply after model releases.

\section{Data}\label{sec:data}

This section describes the data used in the empirical analysis. Section~\ref{sec:openrouter_data} introduces the OpenRouter data. Section~\ref{sec:other_data} describes the remaining data used in the empirical analysis.

\subsection{OpenRouter Data}\label{sec:openrouter_data}

We measure AI consumption (which we refer to interchangeably as \emph{AI use}) using a panel of realized inference requests from OpenRouter, Inc., an AI inference provider that connects users and developers to hundreds of large language models through a single API endpoint. Users submit a prompt and select which model runs it---an OpenAI, Anthropic, Google, Meta, DeepSeek, or open-weight model among more than four hundred available models---and OpenRouter routes the request to a backing provider that hosts that model. One benefit of the data is the large number of models available on OpenRouter, making the data particularly useful for measuring aggregate AI consumption: observed consumption is unlikely to reflect the product cycle of any single provider, the design of any single consumer application, or the marketing strategy of any single model lab \citep{openrouter2025state}. The data are also realized rather than reported AI consumption. That is, every observation is measured on actual paid requests.

We work with user-level OpenRouter data structured as a panel at the user--model--day level, where model identifies a specific model version (e.g., Anthropic's Opus 4.7 is a separate unit from the earlier Opus 4.6). Each observation summarizes the activity of a single anonymized user with a particular model on a particular day: for example, the total requests, prompt and completion tokens, and dollar cost generated by a particular user when interacting with Opus 4.7 on a given day. Importantly, the panel structure allows us to track each anonymized user's consumption over time, across models, and across providers. Starting from this data, we construct three weekly aggregate series that capture important components of AI consumption: total tokens (the quantity of language-model work demanded), dollar usage (the expenditure on that work, combining quantity and the actual price paid), and distinct users (the extensive margin of adoption). Our sample covers the weekly period from January 2024 through April 2026. Over this time period, OpenRouter intermediates 380 trillion cumulative tokens, which, to our knowledge, is the largest cross-provider record of realized AI consumption available outside the model labs themselves.

The user--model--day granularity is one feature that distinguishes our panel from alternative existing datasets, which are typically tied to a single model, provider, or product. The resulting panel is both large and detailed, with the dataset covering millions of anonymized accounts. Most accounts also use multiple models and providers: the average account draws on more than five distinct models, and seven in ten accounts use at least two providers (see Internet Appendix Table~\ref{tab:user_model_granularity}). The granularity of the data allows us to decompose AI consumption along multiple margins we exploit in the asset-pricing tests: frontier closed-source models versus open-weight models, paid/core users versus the rest, seasoned users versus new arrivals, and long, complex prompts versus short ones. The same granularity also supports classification of activity by prompt-content category and identification of requests that invoke external tools (e.g., agentic use).

Our baseline AI factor aggregates the three main AI consumption series into a single index, defined as the first principal component of their standardized weekly log growth. We use growth rather than levels because asset returns tend to respond to the unexpected component of AI consumption---changes in adoption and intensity---rather than to the mechanical upward trend of a diffusing technology. The first principal component places weights of 0.665, 0.559, and 0.496 on the three series and accounts for 56.5 percent of their joint variance, so a single common factor captures the bulk of the comovement across these three margins of AI consumption (see Internet Appendix Table~\ref{tab:ia_pca3_loadings}). Section \ref{sec:app_portfolios} in the Internet Appendix also reports asset pricing results based on the simple cross-sectional average, and each component series separately as robustness checks, delivering similar conclusions.

\begin{figure}[!htbp]
\centering
\makebox[\linewidth][c]{\includegraphics[width=0.95\linewidth]{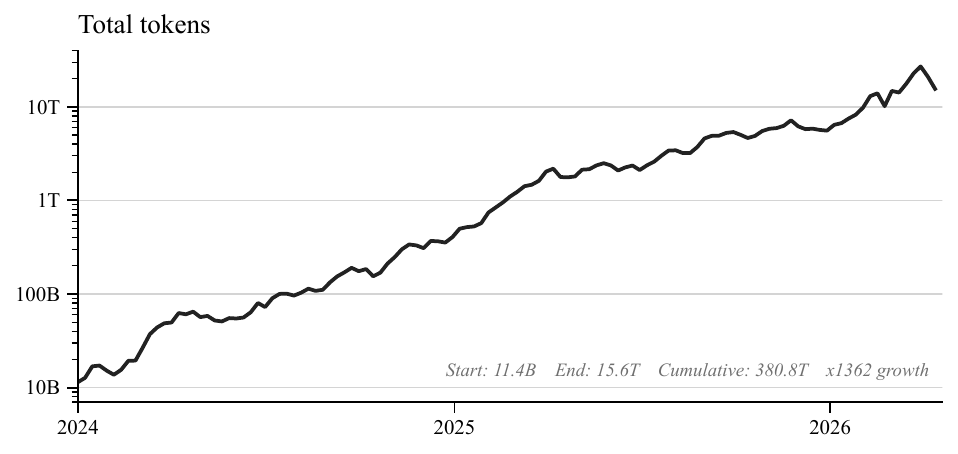}}
\caption{Weekly Total Tokens}
\label{fig:pc3_levels}

\begin{minipage}{\textwidth}
\footnotesize
\textit{Note:} The figure plots weekly total token consumption on a log scale. Total tokens are prompt tokens plus completion tokens. The in-figure annotation reports the first-week level, final observed-week level, cumulative total token consumption of 380.8 trillion, and growth multiple. The corresponding standardized weekly log growth series is plotted in Internet Appendix Figure~\ref{fig:pc3_growth}.
\end{minipage}
\end{figure}

Figure~\ref{fig:pc3_levels} plots weekly total token consumption on a log scale. Total tokens rise from 11.4 billion in the first week to 15.6 trillion in the final observed week, with cumulative total token consumption of 380.8 trillion over the sample period. The growth persists during the sample and is not concentrated around a single model release, which is consistent with steady diffusion of the technology across users and applications.

Table~\ref{tab:summary_statistics} reports the full sample summary statistics for the AI variables (Panel~A) and the firm characteristics used in the asset pricing tests (Panel~B). Panel~A reports the baseline factor and its three standardized input growth series, along with the AI sub-factors computed on different subsets of the data: $AI^{Closed}$ and $AI^{Open}$ from closed-source and open-weight model consumption, $AI^{Core}$ and $AI^{New}$ from paid/core and new accounts, $AI^{Seasoned}$ and $AI^{Nonseasoned}$ from seasoned and non-seasoned users, and $\Delta\ln Tok^{Long}$ and $\Delta\ln Tok^{Short}$ from long- and short-prompt total token growth. We use these additional factors in the analysis of Section~\ref{sec:AI_stock_returns}.

\input{tables_v22/table_summary_statistics.tex}

Section \ref{sec:app_add_facts} in the Internet Appendix reports additional descriptive information on other aspects of the data. Figure~\ref{fig:agentic_total_token_usage} decomposes total token consumption into agentic and non-agentic consumption. Figure~\ref{fig:category} reports the share of weekly tokens by prompt-content category for all models and separately for closed- and open-weight models. Figure~\ref{fig:decomposition} plots the token share evolution of the four model- and user-side decompositions used to compute the additional factors reported in Panel~A of Table~\ref{tab:summary_statistics}.

\subsection{Additional Data}\label{sec:other_data}

For the U.S. equity analysis, we use CRSP common stocks listed on the NYSE, AMEX, and Nasdaq matched to Compustat accounting data. The baseline sample keeps firms with formation-week market capitalization of at least \$1 million, and we compute weekly returns and the standard firm characteristics: log market equity, log book-to-market, gross profitability, investment, and momentum. For factor adjustments, we use the Fama-French five factors \citep{famaFrench1993,famaFrench2015} and the momentum factor \citep{jegadeeshTitman1993,carhart1997}. For industry adjustments, we use the Fama-French 30 industry classification. For technology-sector returns, we use the Fama-French 10 high-technology industry return from Kenneth French's data library. We also construct a semiconductor portfolio return as the value-weighted return on CRSP common stocks in three semiconductor-related NAICS industries, as semiconductors are crucial for the development of AI infrastructure.\footnote{NAICS 334413 (Semiconductor and Related Device Manufacturing), 333242 (Semiconductor Machinery Manufacturing), and 334418 (Printed Circuit Assembly Manufacturing).}

For the international analysis, we use Compustat Global common stocks split into MSCI developed and emerging markets, applying the same market-capitalization filter in U.S. dollars and measuring returns in local currency. For the China test, we use the Datastream China A-share dataset together with a China-specific AI factor built from mainland-China OpenRouter consumption. To capture broader AI attention and traded AI exposure, we use Google Trends search interest for the topic ``Artificial intelligence'' worldwide and an equal-weighted basket of seven AI- and robotics-themed exchange-traded funds, respectively.\footnote{The ETF basket includes the following ETFs: BOTZ, AIQ, IRBO, ROBO, ARTY, WTAI, and CHAT.} Finally, for the event-study analysis, we hand-collect a calendar of major AI model releases from official model cards (the labs' technical release documentation), company blog posts, and developer announcements, separating releases by the five frontier providers---Anthropic, DeepSeek, Google, Meta, and OpenAI---from those of other providers. The full calendar with sources is reported in Internet Appendix Table~\ref{tab:ia_release_calendar}.

\section{The AI Exposure and the Cross-Section of Equity Returns}\label{sec:AI_stock_returns}

This section examines whether firms' AI exposure is significantly related to future cross-sectional equity returns. Section~\ref{sec:ai_portfolios} forms portfolios sorted on firm-level AI exposure and documents the AI premium. Section~\ref{sec:price_ai_risk} presents the results of Fama-MacBeth (\citealp{famaMacBeth1973}) cross-sectional regressions. Section~\ref{sec:additional_results} reports results on robustness tests, international evidence, and heterogeneity across subcomponents of AI consumption. Section~\ref{sec:stock_exposure} presents firm-level exposures within the S\&P 500.

\subsection{AI Portfolios}\label{sec:ai_portfolios}

We define firm $i$'s market-implied AI exposure at week $t$, $\beta^{AI}_{i,t}$, as the coefficient estimate from a rolling weekly regression of its log excess return on the baseline AI factor, controlling for the log excess market return:

\begin{equation}
ret_{i,\tau} = \alpha_{i,t} + \beta^{AI}_{i,t} \times AI_\tau + \beta^{rm}_{i,t} \times r_{m_\tau} + \epsilon_{i,\tau}, \quad \tau \in \mathcal{W}_t,
\end{equation}
where $\mathcal{W}_t$ is the 13-week estimation window ending at formation week $t$. The $t$ subscript on $\beta^{AI}_{i,t}$ indicates that exposure is re-estimated every week and can change as investors revise forward-looking valuations in response to AI consumption shocks. Our exposure measure is time-varying and is inferred from realized stock price comovement. %

Portfolios are formed by sorting all firms at the beginning of each week into quintiles by their current AI exposure $\beta^{AI}_{i,t}$, with breakpoints based on the full all-stock universe, and are rebalanced weekly. Our baseline portfolios are value-weighted. Each quintile contains about 680 stocks on average. We also report equal-weighted portfolios and NYSE-breakpoint robustness checks. NYSE breakpoints follow the standard Fama-French sorting convention and reduce the influence of smaller, less liquid non-NYSE stocks on the portfolio cutoffs \citep{famaFrench1992}. 

Table~\ref{tab:us_portfolio_chars} reports the properties of the quintile portfolios. The first row shows a large and statistically significant spread in AI exposure across the portfolios. The high-low spread in $\beta^{AI}\sigma_w(AI)$ is 0.0624, significant at the 1 percent level. The five quintiles are otherwise similar in size, book-to-market, profitability, and investment: average $\ln ME$ is 14.02 in the lowest-$\beta^{AI}$ quintile and 14.29 in the highest, $\ln BM$ is $-0.99$ versus $-0.94$, profitability is 0.231 versus 0.254, and investment is 0.133 versus 0.128. The small cross-quintile differences in $\ln ME$ and profitability are statistically significant although small in magnitude. The one characteristic that differs systematically across the quintiles is momentum: $Mom$ rises from 0.034 in the lowest-$\beta^{AI}$ quintile to 0.138 in the highest, a 10 percent spread that is highly significant. This pattern motivates the inclusion of momentum among the factor-model controls in the pricing tests below.

\input{tables_v22/table_us_portfolio_chars.tex}

Table~\ref{tab:us_portfolio_p5_p1} reports the portfolio sorting results. Panels~A and B use value-weighted portfolios with all-stock breakpoints and NYSE breakpoints, respectively. The results are similar across the two breakpoint choices. With all-stock breakpoints in Panel~A, the high-low excess-return spread is 60.4 basis points per week, significant at the 1 percent level. With NYSE breakpoints in Panel~B, the spread is 46.1 basis points per week, significant at the 5 percent level. Panels~C and D repeat the exercise with equal-weighted portfolios. The spreads are smaller but remain positive and statistically significant at the 10 percent level, at 29.3 and 25.2 basis points per week. Hence, the result is stronger for the larger companies but is not driven only by a few mega-cap firms. Figure~\ref{fig:ai_premium_overview} plots the cumulative return of the high-minus-low portfolio over the sample period and the mean weekly excess return of each exposure quintile, and previews the robustness of the spread to the technology-sector, industry, and attention controls examined in Section~\ref{sec:additional_results}. The spread is also robust to standard risk adjustment. In Panel~A, the alpha from the Fama--French five-factor model \citep{famaFrench2015} is 54.5 basis points per week, and the alpha from the five-factor model augmented with momentum \citep{carhart1997} is 52.6 basis points per week, both of which are significant at the 5 percent level.\footnote{Internet Appendix Table~\ref{tab:us_hml_ff_controls} reports the factor regressions for the value-weighted, all-stock-breakpoint long-short strategy, including factor coefficient estimates and $R^2$ values. Internet Appendix Table~\ref{tab:ia_small_sample_robustness} shows that the result remains positive under block-bootstrap inference and when sorting on empirical-Bayes-shrunk AI betas.}

\input{tables_v22/table_us_portfolio_p5_p1.tex}

\input{figure_us_premium_v22.tex}

Additional Internet Appendix results further support this baseline portfolio evidence. Table~\ref{tab:us_components} repeats the portfolio sorting exercises using each of the three input series of the AI factor ($\Delta\ln Tok$, $\Delta\ln Dol$, $\Delta\ln User$) separately and finds high-low spreads of 49.4, 35.5, and 39.5 basis points per week, respectively. Table~\ref{tab:us_indadj} sorts on industry-neutral AI exposure (the difference between $\beta^{AI}$ and the within-period Fama-French 30 industry median) and finds a spread of 54.1 basis points per week, close to the 60.4-basis-point baseline. Finally, to control for broad public attention and sentiment toward AI, Table~\ref{tab:ia_gtrend_controlled} re-estimates $\beta^{AI}$ controlling for the log growth of Google Trends search interest in ``Artificial intelligence'' and finds a spread of 69.1 basis points per week with comparable factor-model alphas.

\subsection{Regression Method}\label{sec:price_ai_risk}

The portfolio sorting results above document a return spread between high- and low-exposure firms. Table~\ref{tab:us_fama_macbeth} further reports estimates of the AI premium using the Fama-MacBeth procedure \citep{famaMacBeth1973} at the individual-stock level. Each week, we regress individual stock returns on their pre-formation AI exposure, $\beta^{AI}$, and standard firm-level characteristics, and we average the weekly coefficient estimates over time. The coefficient estimate on $\beta^{AI}$ is our variable of interest. Panel~A weights stocks equally, so the weekly cross-sectional regressions are estimated by OLS regressions. Panel~B weights each stock by its market capitalization, so they are estimated by weighted least squares regressions.

The point estimates of $\beta^{AI}$ are positive and statistically significant across specifications. When $\beta^{AI}$ is the only regressor, the coefficient estimate is 6.26, which is significant at the 10 percent level in Panel~A and 17.04, which is significant at the 1 percent level in Panel~B. After controlling for size, book-to-market, profitability, and investment rate, the coefficient estimate for AI exposure remains positive and significant. The result is also robust to a more demanding specification, which additionally controls for momentum, short-term reversal, leverage, and accruals. Under this specification, the equal-weighted result is 6.69, which is significant at the 5 percent level, and the value-weighted result is 14.11, which is significant at the 1 percent level. Thus, Table~\ref{tab:us_fama_macbeth} provides regression-based evidence that the AI premium is positive.

\input{tables_v22/table_us_fama_macbeth.tex}

\subsection{Additional Results}\label{sec:additional_results}

This subsection reports three sets of additional results. First, we show that the AI premium is robust to controlling for technology-sector and AI-themed equity returns. Second, we examine which subcomponents of AI consumption are most informative for asset prices. Third, we extend the test to international markets as a cross-market check. 

\subsubsection{Technology-Sector and AI-Themed Controls}

As a robustness test, we test whether the AI premium reflects returns to technology-sector or AI-themed equity portfolios rather than information about AI consumption from OpenRouter. This is especially relevant in our short sample, a period in which technology stocks and AI-themed valuations rose sharply. We address this by re-estimating each firm's AI exposure with one of three traded benchmarks added to the market control: the high-technology industry return from the Fama-French 10 industry classification, the value-weighted semiconductor portfolio return, and an equal-weighted AI/robotics ETF basket return. We summarize these results in Table~\ref{tab:us_industry_controlled}. Panels A, B, and C report the results for high-low spreads for these three robustness tests. The spreads remain positive and statistically significant at the 10\% level or better across the three specifications, at 48.7, 35.0, and 53.1 basis points per week, respectively.\footnote{Controlling directly for these benchmark returns, rather than adding them to the first-pass regression, leads to the same conclusion. Internet Appendix Table~\ref{tab:ia_aietf_spanning} regresses the baseline value-weighted high-minus-low return on each benchmark, alone and together with the market, the Fama-French five factors, or the five factors plus momentum. The intercepts range from 52.6 to 60.3 basis points per week with all-stock breakpoints, all significant at the 5 percent level, and from 38.3 to 45.7 basis points per week with NYSE breakpoints, all significant at the 10 percent level or better.} Panel~C of Figure~\ref{fig:ai_premium_overview} plots these spreads together with the industry-demeaned and attention-controlled spreads.

The AI factor is also empirically distinct from the conventional investment-specific technology shock \citep[e.g.,][]{koganPapanikolaou2014growth}. The results are presented in Internet Appendix Table~\ref{tab:ia_imc_ist}. Net of the market, its correlation with the investment-minus-consumption (IMC) portfolio ranges from $-0.02$ to $-0.14$ across different industry classification vintages, and firms' AI betas are weakly and negatively related to their IMC betas. We also show that the AI premium remains 49 to 52 basis points per week after controlling for IMC, the Fama-French five factors, and momentum.

\input{tables_v22/table_us_industry_controlled.tex}

\subsubsection{Important Components of AI Consumption}\label{sec:ai_subcomponents}

A unique feature of the OpenRouter AI dataset is its granularity. For each anonymized user and model on each day, we observe realized requests, tokens, and dollar usage. This user--model--day structure lets us ask which components of AI consumption best capture the AI premium. We decompose AI consumption along four margins: closed-source versus open-weight models, paid/core versus new accounts, seasoned versus non-seasoned users, and long versus short prompts. For each of the resulting eight specific AI factors, we re-estimate firm-level exposures and sort firms on that exposure separately. Table~\ref{tab:subsamples_combined} reports the resulting high-minus-low spreads.

The four subcomponents are about different margins of AI consumption. First, following OpenRouter's model taxonomy, closed-source models are proprietary or restricted-API models whose weights are not publicly released, while open-weight models have publicly available weights. Second, paid/core users are defined as the accounts with at least 13 weeks of tenure, at least eight prior active weeks, at least \$10 of prior OpenRouter usage, and activity in at least two of the previous eight weeks. New users are accounts with less than four weeks of tenure, and middle/developing accounts are excluded from this split. Third, seasoned users are a broader experience measure: before week $t$, they have been observed for at least 13 weeks, were active in at least eight prior weeks, and made at least 20 prior requests. The remaining active users are non-seasoned. Finally, long prompts are requests above the prior-13-week rolling 75th percentile of prompt length, and short prompts are the remaining requests. We make sure that all user and prompt-length classifications are predetermined using information available before the portfolio-formation week.

The pattern is consistent across the four splits: the estimated AI premium is larger for exposure to more intensive, frontier-oriented AI consumption. The high-low spread is 51.6 basis points per week based on exposure to closed-source model consumption versus 23.0 basis points based on open-weight consumption (Panel~A), 44.5 basis points versus 18.5 basis points based on paid/core accounts versus new accounts (Panel~B), 54.0 basis points versus 31.6 basis points based on seasoned versus non-seasoned users (Panel~C), and 47.7 basis points versus 29.6 basis points based on long- versus short-prompt token growth (Panel~D). The AI premium is better captured by AI consumption on frontier models, among experienced users, and for complex tasks, rather than on casual consumption. 

\input{tables_v22/table_subsamples_combined.tex}

\subsubsection{International Evidence}

We next extend the test to international markets. We ask whether the AI premium is only a US phenomenon and, if not, whether it is concentrated in the markets closer to leading-edge AI development, infrastructure, and commercial deployment. This builds on the ``distance to frontier'' framework of \citet{AcemogluAghionZilibotti2006distance}, in which an economy's position relative to the global technology frontier shapes its innovation and growth dynamics. In our setting, the local equity markets whose listed firms and investors are closer to AI development and deployment are what we consider the AI frontier markets. Based on the transition-risk view (\citealp{pastorVeronesi2009}), if AI exposure matters more for firms and investors nearer the frontier, the spread may be larger there. We define a local market as a country-level equity market. Using Compustat Global common stocks, each week and separately within each local market, we sort stocks into quintiles by AI exposure. We then compute the high-low return spread within each local market and average these local-market spreads across markets classified by MSCI as developed or emerging.

Panel~A of Table~\ref{tab:international_combined} combines both local markets classified by MSCI as developed and emerging. The overall international high-low spread is 7.3 basis points per week, with a significant local-market alpha of 7.8 basis points. Panels~B and~C show that this average is mainly driven by developed local markets. In developed markets, the average high-low spread is 10.0 basis points per week with a 10.8-basis-point local-market alpha, both statistically significant at the 5\% level. In emerging local markets, the average spread is 5.0 basis points per week and is statistically insignificant. The results suggest that the AI premium is more pronounced in developed equity markets, which may be closer to the global AI frontier.

China is a particularly interesting test because Chinese AI consumption has become an important share of the global total AI consumption and AI is highly salient to local investors in China. Internet Appendix Table~\ref{tab:ia_china_hml} examines China A-shares, sorting either on the baseline AI factor or on a China-specific AI factor built from mainland-China OpenRouter consumption. In both specifications, the high-minus-low spreads are negative and statistically insignificant.

\input{tables_v22/table_international_combined.tex}

\subsubsection{Event-Study Evidence}\label{sec:event_study}

AI is a recent phenomenon, and a series of unanticipated positive surprises could, in principle, explain part of the observed average excess return. If high-AI-beta firms are those who benefit most as AI advances, then AI advancing faster than the market expected would also be consistent with high realized returns over our sample. We provide two complementary results on the timing of the return spread: an event study around AI model releases, and an analysis that removes release weeks from the sample.

First, we use major AI model releases as the news events, separating releases by the five frontier providers---Anthropic, DeepSeek, Google, Meta, and OpenAI---from those of other providers, and compare the cumulative abnormal returns (CARs) of high- and low-AI-beta firms in narrow windows around release dates. Our constructed release calendar, containing 29 frontier and 36 non-frontier events, is reported in the Internet Appendix (Table~\ref{tab:ia_release_calendar}). Figure~\ref{fig:event_study_caar} plots the high-minus-low CAR difference from five days before to ten days after each release, with 95 percent confidence intervals. The effect is concentrated in frontier-provider releases (Panel~A): the CAR difference reaches about 3.5 percent, with evidence of pre-drift before the release and a plateau about five days after, and is statistically significant within the window. For non-frontier releases (Panel~B), it reaches only about 2 percent, with little pre-drift. These results support the interpretation of AI beta as a measure of firms' sensitivity to economically important AI developments. We report the full set of risk-adjusted CAR differences across event windows in the Internet Appendix (Table~\ref{tab:event_study_all}). Using the [0,+5] window,  high-AI-beta firms outperform low-AI-beta firms around frontier-provider releases by 1.7 percent over the five-day window and by 0.9 percent after five-factor and momentum adjustment, both of which are statistically significant, at the 1 and 10 percent levels, respectively, while the responses to non-frontier releases are more muted.

\input{figure_oa_event_study_v22.tex}

Second, we remove the model-release weeks from the sample and re-examine the AI premium result. If the premium were primarily about the ex-post realization of AI news, it should be concentrated in those weeks, and excluding them should eliminate the AI premium. Table~\ref{tab:ia_news_window_quintiles} repeats the value-weighted quintile sort with these weeks excluded, identified from the model-release calendar in Table~\ref{tab:ia_release_calendar}. Dropping weeks with a frontier-provider release, the high-minus-low spread remains positive and significant at the 5 percent level at 0.343 percent per week, and additionally dropping non-frontier-release weeks leaves 0.331 percent per week that is significant at the 10 percent level. In both cases, the high AI-beta quintile has a higher average return than the low quintile. The premium is stronger around releases, as the event study shows, but an important fraction of it accrues gradually throughout the sample rather than only on model release news dates.

\input{tables_v22/table_ia_news_window_quintiles.tex}

Overall, we find that firms with higher AI exposure earn higher future returns than firms with lower exposure. These findings can be interpreted through the framework of \citet{koganPapanikolaou2014growth}, in which technological change creates heterogeneous firm exposures and gives rise to a technology-risk premium. As \citet{papanikolaou2011investment} shows, the sign of the technology-risk premium depends on the nature of the shock and, under recursive preferences, on whether households prefer early or late resolution of uncertainty. For the investment-specific technology (IST) shock they study, reallocation toward investment today and news about future growth are both priced negatively under late resolution. Under early resolution, reallocation is priced negatively and growth news positively, making the sign depend on the shock's composition. The negative price of IST risk is inferred from the cross-section in \citet{koganPapanikolaou2014growth} and arises through limited risk sharing between innovators and other households in \citet{kogan2020left}. Our results point to a positive price of AI exposure and show that AI shocks are empirically distinct from the IST shock. The findings are also consistent with the transition-risk channel of \citet{pastorVeronesi2009}, in which new technologies reallocate rents across firms, sectors, and workers, and technological risk becomes systematic and priced once adoption is large-scale. Consistent with this interpretation, the AI premium is concentrated in frontier-oriented AI consumption and developed markets, where AI risk may be most systematic.

The large realized AI premium during this period of rapid AI development may also reflect revisions to expectations about which firms' cash flows AI will raise or displace. The release-date evidence bears on both interpretations. The large responses of high-exposure firms around frontier releases suggest revisions to expectations during these periods. The spread nevertheless persists outside release weeks, indicating that it is not confined to AI release news.

\subsection{Firm-Level AI Exposure}\label{sec:stock_exposure}

The rolling regressions used in the portfolio tests also give us a firm-level measure of AI exposure. For each firm, we estimate $\beta^{AI}$ from weekly stock returns and then average the estimated exposure across valid formation weeks. This allows us to move beyond portfolios and ask which individual firms' returns comove most strongly with innovations to AI consumption.

Figure~\ref{fig:exposure_industries_firms} reports these exposures. Panel~A shows the value-weighted average exposure for the Fama-French 10 industries, expressed as the weekly industry return in percentage points associated with a one-standard-deviation AI-factor shock. Most industries have positive exposure, with the largest values in retail. Non-durable goods and health have negative exposure, and the high-technology industry average is close to zero.\footnote{Internet Appendix Figure~\ref{fig:industry_full_sample} shows that the coefficient estimates on the AI factor based on industry-level regressions. Similarly, the health and non-durable goods are at the bottom and retail is near the top.} Panel~B shows the fifteen S\&P 500 firms with the most positive and the fifteen with the most negative average exposure, with firms in the high-technology industry highlighted.\footnote{Internet Appendix Figure~\ref{fig:sp500_top_bottom_company_ai_exposure} reports the full ranking of the top and bottom 50 S\&P 500 firms on a common radial scale, and Internet Appendix Figure~\ref{fig:ff10_beta_heatmap} reports industry exposures for the AI factor and its subcomponents.}

\input{figure_us_exposure_industry_firm_v22.tex}

The highest-ranked firm, AppLovin, a software company whose platform helps mobile-app developers advertise and monetize apps, has the highest estimated positive AI exposure among current S\&P 500 firms. At the other extreme, Moderna, a biotechnology firm focused on mRNA medicines and vaccines, has the most negative estimated AI exposure. The positive tail also includes natural AI-infrastructure names such as Western Digital and Lumentum, digital-platform and travel firms such as Expedia, and power firms such as NRG Energy. The negative tail includes Estée Lauder, as well as technology and semiconductor firms such as AMD and ON Semiconductor. The main point of this result is that firm-level AI exposure is not the same as a simple technology-sector classification, our AI exposure measure identifies firms whose returns covary with AI consumption across many parts of the economy.

\section{AI Exposure across Occupations, Tasks and Skills}\label{sec:skill_exposure}

This section provides the occupation-level,  task-level, and skill-level counterparts of the firm-level results. We use the firm-level AI betas to link market-implied AI exposure to occupations, tasks and skills in the current organization of work. We proceed in two steps. First, we map the firm-level AI betas into occupations using BLS occupation-by-industry employment weights (the occupation ranking is in Internet Appendix Figure~\ref{fig:occupation_ai_exposure}). Second, we map these occupation exposures into skills using O*NET, the U.S. Department of Labor database that describes the tasks, abilities, and skills used in detailed occupations, or the skill-importance ratings for each occupation.

The firm-level AI exposure comes directly from stock returns and AI-consumption innovations as described above, the occupation mapping adds employment weights, and the skill mapping adds O*NET skill-importance ratings. In the end, this procedure produces a market-based measure of AI exposure for occupations and skills. This measure is thus different from worker surveys, expert labels, or AI labels which ask AI models to judge which skills are exposed. To the best of our knowledge, this is the first exercise that uses realized AI consumption and equity market valuations to build a market-implied occupation and skill-level measure of AI exposure.

Table~\ref{tab:occupation_task_alm_deming} compares the market-implied occupation exposures with task and skill measures from the literature on technological change and the demand for labor. Tasks are the activities a job involves, and skills are the worker capabilities those activities require. Columns (1) and (2) of the table use task-content measures. Column (1) follows the work of \citet{AutorLevyMurnane2003} and assigns each occupation four scores: routine, nonroutine analytic, nonroutine interactive, and manual/physical.  Column (2) keeps the two nonroutine cognitive measures and splits the rest into routine cognitive, routine manual, and nonroutine manual, following the five-group taxonomy of \citet{AcemogluAutor2011}. Columns (3) through (5) assign the skill-based measures of \citet{deming2017growing} to each occupation, including the three headline measures in column (3) (social skills, nonroutine analytical and mathematical content, and routine tasks), the interaction between social skills and analytical content that captures his social-cognitive complementarity in column (4), and the full set of ten O*NET task and skill measures in column (5). All variables are standardized, so coefficients are in standard-deviation units.

We find a similar pattern across these different task and skill taxonomies, from the routine versus nonroutine split of \citet{AutorLevyMurnane2003} and \citet{AcemogluAutor2011} to the skill-based measures of \citet{deming2017growing}. Market-implied AI exposure is positive on nonroutine interactive tasks and social skills, suggesting that the current market implies these skills may benefit from AI development. On the other hand, AI exposure is negative for nonroutine analytical and math tasks. Specifically, in column (1), the coefficient estimate to nonroutine interactive content is positive and significant at 0.16 and the coefficient estimate to nonroutine analytic content is negative and significant at $-0.17$. The Deming measures in column (3) show the same pattern, the coefficient estimate is $0.11$ for social skills and is $-0.14$ for nonroutine analytical and math, both statistically significant. The coefficient estimates to composite routine and manual measures in column (1) are small and insignificant, but splitting them in more detail in column (2) shows two offsetting effects: routine manual content has a strongly negative coefficient ($-0.36$) and nonroutine manual content a positive one ($0.30$), in line with the installation and repair skills near the top of Figure~\ref{fig:occupation_ai_exposure} in the Internet Appendix, and the two nonroutine cognitive coefficients keep their signs and significance. Column (4) shows that the interaction between social skills and analytical and math content is small and insignificant. In other words, AI exposure is associated with social skills directly, not with the social-cognitive combination that \citet{deming2017growing} emphasizes. Column (5) splits the composite measures into ten task and skill measures and shows where the pattern comes from in detail. The positive coefficient comes from interaction/communication tasks, at 0.26 (significant at the 1 percent level), and the negative coefficients from deductive and inductive reasoning and from information use. That is, with this finer measure, the composite analytical and math measure of an occupation is no longer significantly associated with the AI exposures of the occupation. Service tasks also have a negative coefficient, in line with the negative AI exposure of healthcare occupations in Internet Appendix Figure~\ref{fig:occupation_ai_exposure}.

The positive coefficients on nonroutine interactive tasks and social skills indicate that market-implied AI exposure is concentrated in occupations built around communication, persuasion, instruction, and coordination. The negative coefficients on nonroutine analytical and math content, deductive and inductive reasoning, and information use indicate that this AI exposure is not only a measure of abstract analytical or scientific work. If anything, the occupations most intensive in those tasks have the most negative AI exposure. Overall, the explanatory power using these existing task and skill measures of occupations is modest, where task and skill contents explain between 2 and 11 percent of the cross-sectional variation in market-implied AI exposure. \citet{jonesTonetti2026weakLinks} build on the task framework and emphasize that output is constrained by tasks where labor improves only slowly relative to capital---the ``weak links'' that bind even when most other tasks are automated. The positively-exposed skills in our map---interaction, persuasion, instruction, installation and repair---may be the weak links the market expects to bind production as AI automates other tasks.

Table~\ref{tab:occupation_ai_exposure_measures} reports regressions of the market-implied ranking on existing occupation-level measures of AI exposure. Column (1) uses three technical measures: the ability-based AI occupational exposure of \citet{felten2021occupational}, the GPT-4 task-exposure measure of \citet{eloundou2023gpts}, and the patent-based measure of \citet{webb2020impact}. Column (2) uses the generative-AI task exposure of \citet{eisfeldt2023generative}, split into exposure through core tasks, which they tie to labor substitution, and exposure through supplemental tasks, which they tie to complementarity. Column (3) combines the five measures. We find that market-implied AI exposure is largely unrelated to these technical automatability measures. Its correlation with each is below 0.08 in absolute value, and no specification explains more than 2 percent of the cross-sectional variation. The coefficient estimate to the GPT-4 exposure measure is positive and significant at 0.13, and the coefficient estimates to the \citet{eisfeldt2023generative} measures are small and statistically insignificant. The coefficient estimates to the \citet{felten2021occupational} and \citet{webb2020impact} measures are significantly negative. In the combined column, the coefficient estimates to the GPT-4 and \citet{eisfeldt2023generative} measures become individually insignificant, while the coefficient estimates to the \citet{felten2021occupational} and \citet{webb2020impact} measures stay significantly negative.

\input{tables_v22/table_occupation_task_alm_deming.tex}

\section{The Rise of Agentic AI Consumption}\label{sec:agentic_economy}

This last section presents early evidence on the rise of agentic AI consumption---models executing multi-step workflows by planning, calling external tools, taking actions, and coordinating with other agents (see, e.g., \citealp{ng2024agenticpatterns,ng2024tooluse}). Tool use is especially central for this measurement---a model becomes agentic when it is given functions it can request to call in order to gather information, manipulate data, or take action. We measure the agentic component of AI consumption in the OpenRouter data as requests whose normalized finish reason is \texttt{tool\_calls}, in which the model invoked an external tool.

These agentic requests have grown rapidly. Figure~\ref{fig:agentic_total_token_usage} plots weekly agentic and non-agentic token volumes from January 2024 through April 2026. Both series scale up by several orders of magnitude over the sample, but agentic tokens grow faster. The agentic share is small in 2024, rises sharply through 2025 and into 2026, and reaches 52.2 percent of total tokens by the latest full week (see also \citealp{yang2025perplexityAgents}).

\input{figure_oa_agentic_v16.tex}

Figure~\ref{fig:agentic_paradox} plots agentic token volume against the realized dollars per agentic token over the same period. Volume rises by orders of magnitude, while the realized dollars per token decline at the same time. Two forces compress the per-token price: prompt caching serves prefix tokens without recomputing them through the model, and providers route agentic requests toward cheaper backing models. Both forces reduce the revenue earned on each agentic token even as agentic usage expands.

\input{figure_oa_agentic_paradox_v16.tex}

Internet Appendix Figure~\ref{fig:agentic_combined} brings these two patterns together. Panel~A plots the agentic share of total tokens against its share of total usage dollars. The token share rises much faster than the dollar share, because volume growth (Figure~\ref{fig:agentic_total_token_usage}) outpaces the per-token price decline (Figure~\ref{fig:agentic_paradox}). Panel~B plots three token-level salient components of agentic activity, each as a share of total tokens. The tool-call share counts tokens on requests that call an external tool. The cache-read share counts prompt prefixes the platform serves from cache rather than recomputing through the model. The reasoning share counts the model's internal work plan tokens. A single agentic exchange can require all three: asked ``What is the temperature in Rome?'', the model produces reasoning tokens to lay out the steps, issues a tool call to an external weather API, and reads the recurring system prompt from cache rather than re-processing it. The tool-call and cache-read shares are near zero in 2024 and rise to roughly two-fifths to one-half of all tokens by the end of the sample, while the reasoning share rises far more slowly.

We then test whether exposure to these agentic-consumption components is priced in the cross-section of equity returns---whether firms whose returns covary more with these components earn higher average returns. For each factor we constructed above---agentic, cache-read, and reasoning tokens, and agentic dollars---we estimate firm-level betas from rolling weekly return regressions, sort firms into quintiles, and compute the value-weighted high-minus-low spread. Internet Appendix Table~\ref{tab:agentic_pricing} reports these spreads, raw and as alphas from the Fama--French three-factor model \citep{famaFrench1993}, the Fama--French five-factor model \citep{famaFrench2015}, and the five-factor model augmented with momentum \citep{carhart1997}. The point estimates are positive for the individual agentic factors and of the same sign as the overall AI premium. They are, however, estimated imprecisely because of the short sample. We thus take this as early and suggestive evidence of a positive agentic premium.

\clearpage
\begin{singlespace}
\bibliographystyle{jae}
\bibliography{references_v16}
\end{singlespace}

\clearpage
\appendix
\hypersetup{pageanchor=false}
\setcounter{page}{1}
\setcounter{section}{0}
\setcounter{subsection}{0}
\setcounter{equation}{0}
\setcounter{footnote}{0}
\setcounter{table}{0}
\setcounter{figure}{0}
\renewcommand{\thepage}{\arabic{page}}
\renewcommand{\thesection}{IA.\arabic{section}}
\renewcommand{\thesubsection}{IA.\arabic{section}.\arabic{subsection}}
\renewcommand{\theequation}{IA.\arabic{equation}}
\renewcommand{\thetable}{IA.\arabic{table}}
\renewcommand{\thefigure}{IA.\arabic{figure}}

\section*{Internet Appendix for ``The Cross-Section of Stock Returns and AI Exposure''}
\addcontentsline{toc}{section}{Internet Appendix for The Cross-Section of Stock Returns and AI Exposure}

\noindent This is the Internet Appendix to Borri, Liu, and Tsyvinski, ``The Cross-Section of Stock Returns and AI Exposure.''

\section{AI Data}

This section documents the scale and broad composition of the OpenRouter, Inc. data used to construct the AI token consumption measures. Figure~\ref{fig:pc3_levels} shows the rapid growth in total-token consumption on the platform and reports cumulative total-token consumption over the sample. Cumulative tokens rise to 380.8 trillion by the end of the sample, with most of the growth concentrated in the second half of the period and no obvious single-event jump. Figure~\ref{fig:category} decomposes AI consumption across twelve prompt-content categories, separately for all models, closed-source models, and open-weight models. The prompt labels are assigned by OpenRouter as discussed in \citet{openrouter2025state}. Across all models, programming and technology together account for the majority of categorized tokens at the end of the sample (37 percent and 26 percent), followed by roleplay and science. Open-weight models tilt more toward programming and roleplay. Closed-source models tilt more toward technology. Category coverage begins in May 2025, when OpenRouter started populating the field.

\input{figure_oa_ai_data_v18.tex}

Table~\ref{tab:user_model_granularity} reports summary statistics for OpenRouter consumption at the account-day, account, and account-model levels, corresponding to the highest level of granularity of the data we use. Panel A summarizes active account-days, Panel B summarizes account-level breadth and concentration, and Panel C summarizes account-model daily consumption and volatility.

\input{tables_v22/table_user_model_granularity.tex}

\FloatBarrier
\section{AI Factor}

This appendix documents the construction of the AI factor introduced in Section~\ref{sec:openrouter_data} and of the split-specific factors used in Section~\ref{sec:ai_subcomponents}. Figure~\ref{fig:pc3_levels} of the main text plots total tokens in levels; we begin in growth space, which is what enters the construction of the factor.

Figure~\ref{fig:pc3_growth} plots the weekly log growth of total tokens, usage dollars, and distinct users after full-sample standardization. The three series move closely together but are not identical, motivating the use of their common principal component.

\input{figure_oa_growth_v16.tex}

Table~\ref{tab:ia_pca3_loadings} then reports the principal-component loadings used to combine those three growth series into the baseline AI factor of Section~\ref{sec:openrouter_data}. The first principal component places weights of 0.665, 0.559, and 0.496 on total tokens, dollars, and users, and explains 56.5 percent of the joint variance. The roughly equal loadings indicate that a single common factor captures the bulk of co-movement across the three margins.

\input{tables_v22/table_ia_pca3_loadings.tex}

We next turn to the four user- and model-side splits that underlie the split-specific factors used in Table~\ref{tab:subsamples_combined} of Section~\ref{sec:ai_subcomponents}. Figure~\ref{fig:decomposition} plots each as a weekly token-share time series: the closed-source share, the paid/core account share, the seasoned-user share, and the long-prompt share. The four series are distinct and not collinear---the closed-source share fluctuates with frontier-model release cycles, the paid/core and seasoned-user shares evolve more smoothly, and the long-prompt share is the most volatile---so each generates a different split-specific AI factor in the main analysis.

\input{figure_oa_decomposition_v18.tex}

\FloatBarrier
\section{AI Portfolios\label{sec:app_portfolios}}

This section collects portfolio evidence and robustness tests that complement the main-text sorts. We organize the discussion around four questions about the headline AI-premium result: whether it depends on how the AI factor is built, whether it survives standard risk adjustment and small-sample inference, whether it merely relabels broad technology, attention, or disclosure exposures, and how it varies across geography and the rise of agentic AI consumption.

We start with the factor construction. Table~\ref{tab:us_components} repeats the value-weighted, all-stock-breakpoint quintile sort using each of the three input series ($\Delta\ln Tok$, $\Delta\ln Dol$, $\Delta\ln User$) in place of the principal-component AI factor. The high-minus-low spreads are 49.4, 35.5, and 39.5 basis points per week, so the headline result is not driven by a single input.

\input{tables_v22/table_us_components.tex}

Turning from factor construction to risk adjustment, Table~\ref{tab:us_hml_ff_controls} reports the time-series regression of the value-weighted, all-stock-breakpoint AI-beta long-short return on the FF3, FF5, and FF5-plus-momentum models. The intercept is 54.0, 54.5, and 52.6 basis points per week respectively, all significant at the 5 percent level, and none of the factor loadings is large. The H$-$L spread is not absorbed by size, value, profitability, investment, or momentum.

\input{tables_v22/table_us_hml_ff_controls.tex}

Beyond the headline portfolio sorts, Table~\ref{tab:saas_recent_exposure} looks directly at incumbent application-layer firms. It reports value-weighted SaaS-portfolio regressions on the baseline AI factor and the market. Over the full sample the SaaS-portfolio loading on $AI_t$ is 0.183, which is statistically insignificant, and restricted to the last calendar quarter the loading is 0.956 per one-standard-deviation AI shock, significant at the 1 percent level.

\input{tables_v22/table_saas_exposure_recent_v9.tex}

We next address two inference-related concerns about the headline test. Table~\ref{tab:ia_small_sample_robustness}, Panel~A, reports a circular moving-block bootstrap that resamples consecutive holding-week observations: the $[2.5, 97.5]$-percentile interval for the H$-$L excess return is $[24.1, 96.6]$ basis points per week, with similar coverage for the FF5 and FF5+Mom alphas, so the interval excludes zero in all three cases. Panel~B replaces the raw rolling AI beta with an empirical-Bayes-shrunk beta toward the cross-sectional mean. The value-weighted, all-stock-breakpoint H$-$L spread is 46.2 basis points per week, significant at the 5 percent level, so the result is not an artifact of estimation noise in the firm-level betas.

\input{tables_v22/table_small_sample_robustness_v3.tex}

We then ask whether the AI premium is simply a relabeling of broader cross-sectional patterns. Industry tilts are the first candidate. Table~\ref{tab:us_indadj} replaces the AI beta with its deviation from the within-period Fama-French 30 industry median and re-runs the quintile sort. The value-weighted, all-stock-breakpoint H$-$L spread is 54.1 basis points per week, significant at the 1 percent level, close to the 60.4-basis-point baseline, so the AI premium is a within-industry phenomenon and not a relabeling of broad industry tilts.

\input{tables_v22/table_us_indadj.tex}

Broad public attention to AI is the second candidate. Table~\ref{tab:ia_gtrend_controlled} re-estimates each firm's AI beta after adding the weekly log growth of Google Trends search interest in ``Artificial intelligence'' to the first-stage market regression, and sorts on the residual AI beta. The value-weighted, all-stock-breakpoint H$-$L spread is 69.1 basis points per week, significant at the 1 percent level, with comparable factor-model alphas, so the premium is not a relabeling of broad public attention to AI.

\input{tables_v22/table_ia_gtrend_controlled.tex}

A third candidate is firms' own AI narratives. Table~\ref{tab:ia_sec_ai_mention_spanning} regresses the baseline AI-beta H$-$L return on a control H$-$L portfolio built from firms' AI mentions in 10-K, 10-Q, and 8-K filings, using unscaled AI-mention counts in Panel~A and word-scaled counts in Panel~B. The spanning intercept ranges from 52.5 to 62.1 basis points per week, comparable to the unconditional H$-$L return, and the SEC-mention loading is small and statistically insignificant in all rows. The realized-consumption signal is distinct from firms' own AI narratives in their disclosures.

\input{tables_v22/table_sec_attention_spanning_v9.tex}

Stepping back to factor construction, Table~\ref{tab:ia_avgstd_baseline} replaces the principal-component AI factor with an alternative aggregator: the simple equal-weighted average of the three full-sample-standardized input series. The value-weighted, all-stock-breakpoint H$-$L spread is 58.5 basis points per week, significant at the 1 percent level, and the FF5 and FF5+Mom alphas are 52.3 and 50.6 basis points respectively. The headline result is not specific to the principal-component construction.

\input{tables_v22/table_ia_avgstd_baseline.tex}

We also split AI consumption into agentic and non-agentic components and sort on exposure to each separately. Table~\ref{tab:ia_agentic_nonagentic_ttok} builds two alternative factors from token growth split this way: agentic tokens come from requests with normalized finish reason \texttt{tool\_calls}, in which the model invoked an external tool, and non-agentic tokens are the remainder. Sorting on exposure to non-agentic token growth yields an H$-$L spread of 46.1 basis points per week, significant at the 5 percent level. Sorting on exposure to agentic token growth yields a smaller spread of 2.4 basis points, which is statistically insignificant. Agentic use is a fast-growing but still small share of total consumption in this sample, and the priced AI signal loads on broader, non-agentic activity.

\input{tables_v22/table_ia_agentic_nonagentic_ttok.tex}

\FloatBarrier

Table~\ref{tab:agentic_pricing} extends the agentic portfolio evidence to the agentic-token, cache-read-token, reasoning-token, and agentic-dollar factors described in Table~\ref{tab:agentic_defs}.

\input{tables_v22/table_ia_agentic_pricing_v15.tex}

Returning to the headline sort, Table~\ref{tab:ia_baseline_decile} replaces the quintile portfolios with deciles. The value-weighted, all-stock-breakpoint H$-$L spread widens to 69.4 basis points per week, significant at the 5 percent level, and the FF5+Mom alpha is 53.8 basis points, significant at the 10 percent level. The relationship between AI exposure and future returns is approximately monotonic and strengthens at the extreme tails.

\input{tables_v22/table_ia_baseline_decile.tex}

A final candidate relabeling is exposure to traded technology baskets. Table~\ref{tab:ia_aietf_spanning} reports spanning regressions of the AI-beta H$-$L portfolio return on three benchmarks: the Fama-French 10 HiTec industry return (Panel~A), the value-weighted NAICS semiconductor portfolio (Panel~B), and an equal-weighted AI/robotics ETF basket (Panel~C). Each panel runs the benchmark alone, on top of the market, on top of FF5, and on top of FF5 plus momentum. The spanning intercepts range from 38.3 to 60.3 basis points per week and remain individually significant, and the loadings on the additional benchmarks are small and insignificant in most rows. The AI premium is not absorbed by exposure to technology-sector, semiconductor, or AI-themed equity baskets.

\input{tables_v22/table_ia_aietf_spanning.tex}

We then turn to geography. Table~\ref{tab:ia_china_hml} repeats the quintile sort on the Datastream China A-share panel, using either the baseline AI factor (Panel~A) or a China-specific AI factor built from OpenRouter consumption originating in mainland China (Panel~B). The H$-$L spreads are uniformly negative and statistically insignificant in both panels and under both weighting schemes. The AI premium is concentrated in developed equity markets near the global AI frontier and is absent from a market where AI is salient to investors but listed firms sit further from that frontier.

\input{tables_v22/table_ia_china_hml.tex}

Finally, Figure~\ref{fig:ai_vix_hml_realized_volatility} plots the 13-week rolling realized volatility of the value-weighted, all-stock-breakpoint AI-beta long-short return, annualized by $\sqrt{52}$. The series varies meaningfully over the sample, with peaks around major AI-news episodes, summarizing how the conditional risk in AI exposure has evolved through the sample.

\input{figure_oa_ai_vix_v22.tex}

\FloatBarrier
\section{Event Study of AI Model Releases}

This appendix reports the details behind the event study in Section~\ref{sec:event_study}, where high-$\beta^{AI}$ firms are shown to outperform low-$\beta^{AI}$ firms in narrow windows around major AI model releases. We classify releases into two groups by provider. Frontier providers are Anthropic, DeepSeek, Google, Meta, and OpenAI, and yield 29 events in the sample; releases from other providers add 36 more. Table~\ref{tab:ia_release_calendar} documents each event with its sources.

Table~\ref{tab:event_study_all} reports the CAR comparison in statistical form. Panel~A uses frontier-provider releases and Panel~B uses releases from other providers, each over three event windows ($(-1,+1)$, $(0,+1)$, and $(0,+5)$) and adjusted for the market model, the Fama-French three- and five-factor models, and FF5 plus momentum. The frontier-provider effect dominates throughout. In Panel~A, the value-weighted CAR difference over the $(0,+5)$ window is 1.731 percent (0.932 percent after FF5+Mom adjustment), against 0.982 percent for non-frontier releases (and weaker after the same controls).

\input{tables_v22/table_event_study_all.tex}
\input{tables_v22/table_ia_release_calendar.tex}

\FloatBarrier
\section{AI Exposure}

This section reports the industry- and firm-level exposure maps behind the portfolio evidence. Figure~\ref{fig:ff10_beta_heatmap} shows how the baseline AI factor and the split-specific AI factors revalue the Fama-French 10 industries. Each cell reports the value-weighted average firm-level AI beta within an industry, scaled by the weekly standard deviation of the relevant AI factor, so the numbers can be read as the percentage-point response of an industry-average weekly return to a one-standard-deviation shock to that AI factor. Most industries have positive baseline AI exposure, with the largest values in retail. Non-durable goods and health have negative exposure on average. The high-technology sector is only modestly exposed on the baseline factor, masking strongly positive exposure of semiconductor and AI-infrastructure firms and negative exposure elsewhere in technology. The cross-factor columns show that exposures are larger and more concentrated when the factor is built from closed-source models, paid/core users, seasoned users, and long prompts. The figure reads AI exposure as a broad diffusion and deployment shock rather than as a narrow technology-sector label.

\input{figure_oa_industry_v22.tex}

Figure~\ref{fig:industry_full_sample} compares the industry exposures in Panel~A of Figure~\ref{fig:exposure_industries_firms}, which average the firms' 13-week rolling exposures, with full-sample industry exposures. The full-sample exposure of an industry is the coefficient on the AI factor in a single regression of the industry's value-weighted weekly log-excess return on the AI factor, the Fama-French five factors, and momentum over the full sample. The two measures order the industries similarly, with a Spearman rank correlation of 0.73. Health and non-durable goods are the two most negative industries under both measures, and retail/wholesale ranks first under the average measure and second under the full-sample measure. These three industries are also the only ones whose full-sample exposure is statistically significant, at the 1 percent level for health and at the 5 percent level for non-durable goods and for retail and wholesale.

\input{figure_oa_industry_full_sample.tex}

Figure~\ref{fig:sp500_top_bottom_company_ai_exposure} reports the firm-level exposure ranking discussed in Section~\ref{sec:stock_exposure}. Panel~A shows the 50 current S\&P 500 firms with the largest positive average rolling AI exposures, and Panel~B shows the 50 firms with the most negative exposures.

\input{figure_us_baseline_v22.tex}

\FloatBarrier
\section{Additional AI Stylized Facts\label{sec:app_add_facts}}

This section collects additional stylized facts on AI consumption based on the OpenRouter data.

Table~\ref{tab:agentic_defs} lists the agentic-economy concepts we extract from the OpenRouter data and the weekly factor that proxies each. These salient components of AI token consumption associated with agentic calls are used in the analysis of firms' exposure to agentic consumption discussed in Section~\ref{sec:agentic_economy} of the paper and specifically in Table~\ref{tab:agentic_pricing}.

\input{tables_v22/table_ia_agentic_definitions_v15.tex}

\begin{figure}[!htbp]
\centering
\textit{Panel A. Agentic Token Share versus Dollar Share}\\[2pt]
\makebox[\linewidth][c]{\includegraphics[width=0.80\linewidth]{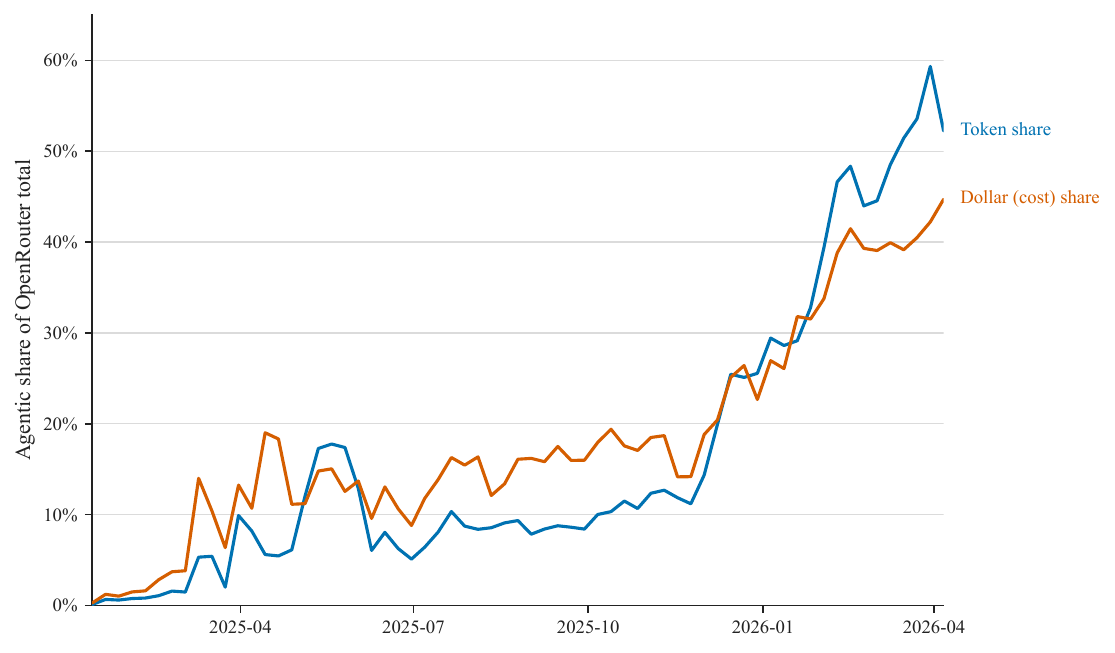}}

\vspace{0.3em}

\textit{Panel B. Tool-Call, Cache-Read, and Reasoning Token Shares}\\[2pt]
\makebox[\linewidth][c]{\includegraphics[width=0.80\linewidth]{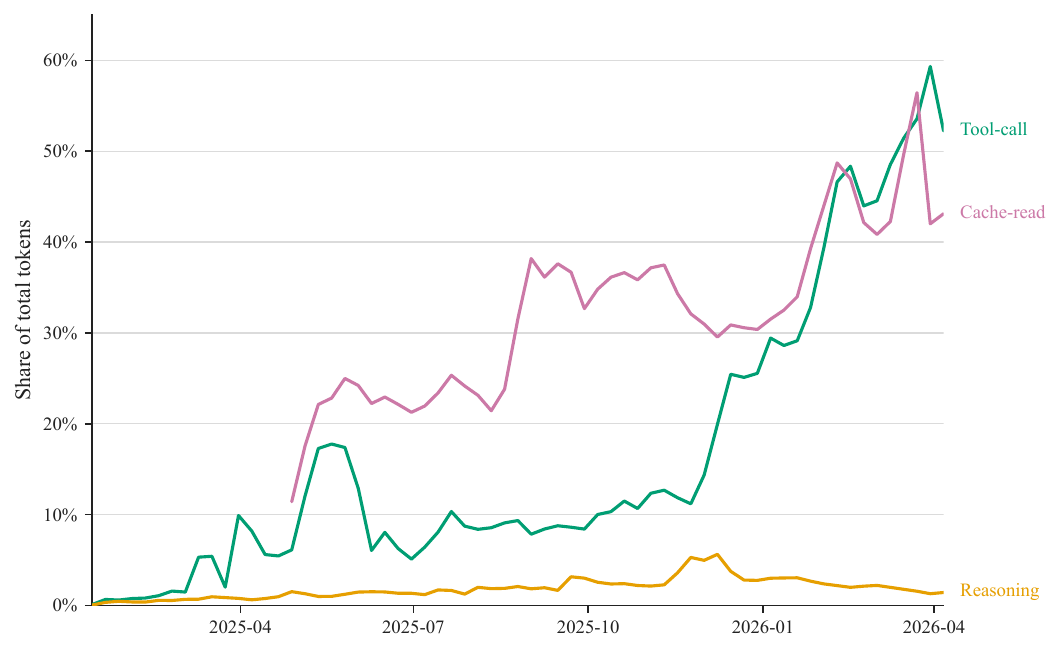}}
\caption{Token, Dollar, and Intensity Shares of Agentic AI Consumption}
\label{fig:agentic_combined}
\label{fig:agentic_shares}
\label{fig:agentic_shares_cost}

\begin{minipage}{\textwidth}
\footnotesize
\textit{Note:} The two panels plot the weekly evolution of agentic AI consumption in the OpenRouter data, January 2024 through April 2026. Panel~A compares the agentic share of total tokens with its share of total usage dollars. The agentic margin is defined as all tokens on requests whose normalized finish reason is \texttt{tool\_calls}. Panel~B plots three token-level margins, each as a share of total tokens: the agentic share, the cache-read share (cached tokens divided by total tokens), and the reasoning share (reasoning tokens divided by total tokens).
\end{minipage}
\end{figure}

Figure~\ref{fig:agentic_combined} reports the corresponding token, dollar, cache-read, and reasoning shares.

Table~\ref{tab:nber_fact_rewrites} compares the granular OpenRouter user--model--day panel with the scraped, aggregate data from the OpenRouter public website used by \citet{demirer2025emerging}.

\input{tables_v22/table_nber_fact_rewrites_v9.tex}

Their public scrape records posted prices and aggregate token counts but cannot observe realized spending, individual users, or within-model variation, and four of their headline findings change once these margins are visible in our licensed user--model--day panel. First, \citet{demirer2025emerging} report that the average price per token is roughly constant. The series they recover is a token-weighted list price, which the scrape cannot distinguish from the price actually paid. We find that the price paid fell by about 30 percent within their 2025 window and by about half over our full sample. Second, \citet{demirer2025emerging} put the open-weight share at about 48 percent of tokens. The same models earn only about 15 percent of realized dollars, overstating open-weight's economic weight by roughly threefold. Third, \citet{demirer2025emerging} document rapid turnover among leading models and creators. Weighted by realized dollars rather than tokens, the market instead consolidates: creator-level concentration rises by 72 percent. Fourth, \citet{demirer2025emerging} estimate short-run prompt-price elasticities just above one. With model-date and provider fixed effects in our panel, these elasticities become small and statistically insignificant. The table also reports application-level multihoming: only 35.8 percent of applications use at least two models in the latest month, with the median top-model token share at 100 percent.

Figures~\ref{fig:retention_providers} and~\ref{fig:retention_margins} show how long AI consumption persists at the account level. Retention at week $k$ is the share of the accounts that start using a model, or any model of a provider, in a given week and are active on it in week $k$ or in a later week within twelve weeks. Among the five frontier providers in Figure~\ref{fig:retention_providers}, 41 to 58 percent of new accounts use the provider again within twelve weeks, and 6 to 13 percent are still active in the twelfth week. DeepSeek retains the most accounts over the first five weeks, and Anthropic retains the most from the sixth week on. In the twelfth week, 12.8 percent of Anthropic's new accounts are still active, against 6.4 percent of Meta's.

\input{figure_oa_retention_providers.tex}

Figure~\ref{fig:retention_margins} splits model adopters along two of the margins in Section~\ref{sec:ai_subcomponents}. Closed-source and open-weight models retain their adopters at almost the same rate, with 5.4 and 4.8 percent still active in the twelfth week, even though exposure to closed-source consumption earns the larger premium in Table~\ref{tab:subsamples_combined}. Seasoned users stay longer than non-seasoned users at every horizon. Within twelve weeks, 47.2 percent of seasoned adopters use the model again, against 39.8 percent of non-seasoned adopters, and 6.9 against 4.4 percent are still active in the twelfth week. Exposure to seasoned-user consumption also earns the larger premium in Table~\ref{tab:subsamples_combined}.

\input{figure_oa_retention_margins.tex}

\FloatBarrier
\section{Skill Exposure to AI}

This section reports the occupation-level exposure map that underlies the skill-exposure exercise in the main text. Figure~\ref{fig:occupation_ai_exposure} maps firm and industry AI exposures into SOC minor occupation groups using BLS occupation-by-industry employment weights, and plots the resulting employment-weighted occupation exposure to the baseline AI factor. Panels~A and~B together rank all 109 SOC minor occupation groups, with the colored rail flagging coherent economic themes. Positive exposure clusters in installation, maintenance, persuasion, instruction, and systems-integration occupations. Negative exposure clusters in science, healthcare, and operations-control occupations. The pattern is the occupation-level counterpart of the firm-level finding that the AI premium is broad rather than confined to the technology sector.

\input{figure_oa_skill_exposure_v22.tex}

\FloatBarrier
\section{Determinants of Firm-Level AI Betas}

This section asks which firm characteristics are associated with the firm-level AI betas that underlie the portfolio sorts. Table~\ref{tab:beta_determinants} reports pooled panel regressions of the weekly rolling AI beta on firm characteristics, with week fixed effects (Panel~A) or week and Fama-French 30 industry fixed effects (Panel~B) and standard errors clustered by week and firm. High-AI-beta firms are past winners and are more profitable and more levered: in the full specification of Panel~A, the momentum, gross-profitability, and leverage coefficients are 0.464, 0.402, and 0.358, all significant at the 1 percent level. AI exposure declines with idiosyncratic volatility and with R\&D intensity, with coefficients of $-10.814$ and $-1.568$ that are both significant at the 1 percent level, and the unconditional size gradient in column~(1) largely disappears once the full set of controls is included. The within-industry estimates in Panel~B show a similar pattern, so the characteristic profile of high-AI-beta firms is not a by-product of industry composition.

\input{tables_v22/table_beta_determinants.tex}

\FloatBarrier
\section{Attention-Orthogonalized AI Factor}

Table~\ref{tab:ia_gtrend_controlled} shows that the AI premium survives controlling for Google Trends attention in the first-pass beta regression. This section reports a complementary two-step exercise that instead strips attention out of the factor itself. Panel~A of Table~\ref{tab:gtrend_residual_sort} regresses the weekly AI factor on the weekly log growth of Google Trends search interest in ``Artificial intelligence.'' Attention comoves with the factor, with a coefficient of 2.419 that is significant at the 10 percent level, but explains only 2.7 percent of its weekly variance, so the AI factor is largely orthogonal to broad public attention to AI. Panel~B re-estimates the first-pass rolling betas with the regression residual replacing the raw AI factor and repeats the value-weighted, all-stock-breakpoint quintile sort. The H$-$L spread is 58.2 basis points per week, significant at the 1 percent level, and the FF5 and FF5+Mom alphas are 53.3 and 51.1 basis points, close to the 60.4-basis-point baseline. The premium reflects exposure to the component of AI consumption growth that is orthogonal to public attention.

\input{tables_v22/table_gtrend_residual_sort.tex}

\FloatBarrier
\section{The AI Factor and Investment-Specific Technology Risk}

This section asks whether the AI factor is the investment-specific technology (IST) shock priced in \citet{papanikolaou2011investment} and \citet{koganPapanikolaou2014growth}, whose estimated price of risk is negative. Because the IST shock is not directly observable, that literature prices its tradable proxy, the investment-minus-consumption (IMC) portfolio, which is long investment-good producers and short consumption-good producers. We build IMC on our sample in three ways. The first uses the published industry classification of \citet{gomesKoganYogo2009}, which assigns each industry to the final-demand category with the largest share of its output in the 1987 benchmark input-output accounts. The other two rebuild the same classification on the 2012 and 2017 benchmark accounts. The vintage matters for the industries at the center of our sample. The 1987 accounts predate the treatment of software as fixed investment, and in that year the largest final-demand destination of semiconductors, electronic components, and computer and data processing services was government, reflecting defense electronics and federal computing, so these industries belong to neither leg and much of the technology sector is absent from the 1987 portfolio, although computer manufacturing was already an investment good. On the modern accounts software publishers and computer systems design enter the investment leg, and data processing and hosting enters the consumption leg on the 2012 accounts and the investment leg on the 2017 accounts, so the modern portfolios contain the technology firms the AI factor is about. Within each leg we value-weight the following week's return by formation-week market equity, matching the convention of the portfolio tests.

Table~\ref{tab:ia_imc_ist} reports three sets of results. Net of the market, the correlation between the AI factor and IMC is small and negative in every classification. The cross-sectional correlation between firms' AI betas and their IMC betas is close to zero on the 1987 classification and small and negative on the modern ones. The AI premium is essentially unchanged by controlling for IMC: the high-minus-low alpha on the five factors, momentum, and IMC ranges from 0.49 to 0.52 percent per week, against 0.53 with no IMC control. On the 2017 classification the high-minus-low portfolio does load positively on IMC, so high-AI-beta firms carry some modern investment-specific exposure, but that exposure does not explain the premium. The AI factor is empirically distinct from these IMC proxies.

\input{tables_v22/table_ia_imc_ist.tex}

\FloatBarrier
\section{Variable Definitions}

Table~\ref{tab:definitions_appendix} reports the definitions of the variables used in the analysis and their correspondence in the OpenRouter dataset and the additional datasets and sources used in the paper.

\input{tables_v22/variable_definitions_v9.tex}

\end{document}

%% file: tables_v22/table_summary_statistics.tex
\begin{table}[!htbp]
\centering
\small
\caption{Summary Statistics}
\label{tab:summary_statistics}
\begin{tabular}{lrrrrrrr}
\toprule
 & Mean & Std. & p10 & p25 & p50 & p75 & p90 \\
\midrule
\multicolumn{8}{l}{\textit{Panel A. AI variables}} \\
$AI$ & 0.000 & 1.302 & -1.561 & -0.872 & -0.107 & 0.726 & 1.596 \\
$\Delta\ln Tok$ & 0.000 & 1.000 & -1.274 & -0.555 & 0.050 & 0.608 & 1.249 \\
$\Delta\ln Dol$ & 0.000 & 1.000 & -1.299 & -0.609 & -0.052 & 0.581 & 1.233 \\
$\Delta\ln User$ & 0.000 & 1.000 & -1.168 & -0.607 & 0.004 & 0.525 & 1.150 \\
$AI^{Closed}$ & 0.000 & 1.341 & -1.555 & -0.911 & -0.177 & 0.816 & 1.381 \\
$AI^{Open}$ & 0.000 & 1.297 & -1.422 & -0.784 & -0.103 & 0.720 & 1.642 \\
$AI^{Core}$ & 0.000 & 1.272 & -1.343 & -0.849 & -0.028 & 0.773 & 1.362 \\
$AI^{New}$ & 0.000 & 1.271 & -1.516 & -0.863 & 0.153 & 0.898 & 1.537 \\
$AI^{Seasoned}$ & 0.000 & 1.267 & -1.279 & -0.884 & -0.070 & 0.759 & 1.336 \\
$AI^{Nonseasoned}$ & 0.000 & 1.249 & -1.724 & -0.709 & -0.050 & 0.786 & 1.719 \\
$\Delta\ln Tok^{Long}$ & 0.069 & 0.183 & -0.132 & -0.053 & 0.064 & 0.184 & 0.295 \\
$\Delta\ln Tok^{Short}$ & 0.053 & 0.142 & -0.079 & -0.030 & 0.039 & 0.137 & 0.225 \\
\addlinespace[3pt]
\multicolumn{8}{l}{\textit{Panel B. Firm characteristics}} \\
$Ret_{t+1}$ & 0.212 & 9.762 & -6.667 & -2.744 & 0.098 & 2.924 & 6.736 \\
$\ln ME$ & 14.42 & 2.06 & 11.76 & 12.93 & 14.40 & 15.80 & 17.21 \\
$\ln BM$ & -0.90 & 1.00 & -2.25 & -1.51 & -0.76 & -0.14 & 0.21 \\
$Profit$ & 0.240 & 0.253 & 0.018 & 0.055 & 0.206 & 0.364 & 0.553 \\
$Inv$ & 0.100 & 0.352 & -0.108 & -0.024 & 0.037 & 0.121 & 0.298 \\
$Mom$ & 0.086 & 0.421 & -0.387 & -0.122 & 0.082 & 0.295 & 0.549 \\
\bottomrule
\end{tabular}

\begin{minipage}{\textwidth}
\footnotesize\textit{Note:}
{\footnotesize Panel A reports full-sample summary statistics for the weekly AI variables, and Panel B reports full-sample summary statistics for the firm characteristics. $AI$ is computed as the first principal component of standardized weekly log growth in total tokens ($\Delta\ln Tok$), usage dollars ($\Delta\ln Dol$), and distinct users ($\Delta\ln User$). The table further reports summary statistics for AI variables based on closed- and open-weight models, paid/core and new users, seasoned and non-seasoned users, and long- and short-prompt total-token growth.}
\end{minipage}
\end{table}

%% file: tables_v22/table_us_portfolio_chars.tex
\begin{table}[!htbp]
\centering
\small
\caption{Properties of U.S.\ AI-Beta Quintile Portfolios}
\label{tab:us_portfolio_chars}
\begin{tabular}{lcccccc}
\toprule
 & Low & 2 & 3 & 4 & High & H$-$L \\
\midrule
$\beta^{AI}\sigma_w(AI)$ & -0.0338 & -0.0090 & -0.0001 & 0.0083 & 0.0286 & 0.0624 \\
\textit{t(H$-$L)} &  &  &  &  &  & (35.63) \\
\addlinespace[1pt]
$\ln ME$ & 14.02 & 14.65 & 14.59 & 14.73 & 14.29 & 0.27 \\
\textit{t(H$-$L)} &  &  &  &  &  & (5.89) \\
\addlinespace[1pt]
$\ln BM$ & -0.99 & -0.92 & -0.88 & -0.86 & -0.94 & 0.05 \\
\textit{t(H$-$L)} &  &  &  &  &  & (1.40) \\
\addlinespace[1pt]
$Profit$ & 0.231 & 0.245 & 0.237 & 0.235 & 0.254 & 0.023 \\
\textit{t(H$-$L)} &  &  &  &  &  & (3.73) \\
\addlinespace[1pt]
$Inv$ & 0.133 & 0.090 & 0.080 & 0.084 & 0.128 & -0.006 \\
\textit{t(H$-$L)} &  &  &  &  &  & (-0.90) \\
\addlinespace[1pt]
$Mom$ & 0.034 & 0.078 & 0.091 & 0.111 & 0.138 & 0.104 \\
\textit{t(H$-$L)} &  &  &  &  &  & (6.22) \\
\addlinespace[1pt]
$\sigma(R)_{13w}$ & 0.087 & 0.049 & 0.042 & 0.048 & 0.078 & -0.009 \\
\textit{t(H$-$L)} &  &  &  &  &  & (-6.21) \\
\addlinespace[1pt]
Analyst disp. & 0.364 & 0.208 & 0.164 & 0.175 & 0.298 & -0.067 \\
\textit{t(H$-$L)} &  &  &  &  &  & (-4.61) \\
\addlinespace[1pt]
\# stocks & 680 & 680 & 680 & 680 & 680 &  \\
\bottomrule
\end{tabular}

\begin{minipage}{\textwidth}
\footnotesize\textit{Note:}
\footnotesize The table reports the properties of the AI portfolios. For each formation week, firms are sorted into quintiles by their pre-formation rolling AI beta, $\beta^{AI}$. Firm betas are estimated from a 13-week rolling regression of weekly log-excess returns on the log-excess market return and the AI factor. The $\beta^{AI}\sigma_w(AI)$ row reports the implied portfolio return using the weekly AI-factor standard-deviation scale. Within each quintile, we take the cross-sectional average of each firm characteristic, then average across formation weeks. $\ln ME$ is log market equity. $\ln BM$ is log book-to-market. $Profit$ is gross profitability (revenues minus cost of goods scaled by assets). $Inv$ is the year-over-year change in total assets scaled by lagged assets. $Mom$ is the cumulative return from week $t-52$ to $t-5$. $\sigma(R)_{13w}$ is the trailing 13-week standard deviation of weekly log returns. Analyst disp. is the standard deviation of analysts' FY1 annual EPS forecasts divided by the absolute value of the mean forecast, using the latest IBES consensus available before the formation week and requiring at least two analysts. The H$-$L column reports the time-series average of the weekly difference between the High and Low quintile means. $t(H$-$L)$ is the corresponding time-series $t$-statistic.
\end{minipage}
\end{table}

%% file: tables_v22/table_us_portfolio_p5_p1.tex
\begin{table}[!htbp]
\centering
\small
\caption{Excess Returns of AI-Beta Quintile Portfolios}
\label{tab:us_portfolio_p5_p1}
\begin{tabular}{lcccccc}
\toprule
 & Low & 2 & 3 & 4 & High & H$-$L \\
\midrule
\multicolumn{7}{l}{\textit{Panel A. Value-weighted, all-stock breakpoints}} \\
Excess return & -0.163 & 0.147 & 0.197 & 0.279 & 0.441 & 0.604 \\
\textit{$t$} & (-0.67) & (0.72) & (1.00) & (1.31) & (1.63) & (2.85) \\
\addlinespace[2pt]
FF5 $\alpha$ & -0.320 & -0.007 & 0.001 & 0.073 & 0.224 & 0.545 \\
\textit{$t(\alpha)$} & (-2.43) & (-0.08) & (0.02) & (0.90) & (1.74) & (2.53) \\
\addlinespace[2pt]
FF5+Mom $\alpha$ & -0.306 & 0.004 & 0.003 & 0.064 & 0.220 & 0.526 \\
\textit{$t(\alpha)$} & (-2.32) & (0.05) & (0.05) & (0.78) & (1.69) & (2.44) \\
\addlinespace[4pt]
\multicolumn{7}{l}{\textit{Panel B. Value-weighted, NYSE breakpoints}} \\
Excess return & -0.037 & 0.110 & 0.249 & 0.275 & 0.424 & 0.461 \\
\textit{$t$} & (-0.16) & (0.52) & (1.25) & (1.29) & (1.62) & (2.28) \\
\addlinespace[2pt]
FF5 $\alpha$ & -0.196 & -0.043 & 0.044 & 0.062 & 0.210 & 0.406 \\
\textit{$t(\alpha)$} & (-1.60) & (-0.50) & (0.55) & (0.77) & (1.67) & (1.97) \\
\addlinespace[2pt]
FF5+Mom $\alpha$ & -0.174 & -0.038 & 0.044 & 0.048 & 0.209 & 0.383 \\
\textit{$t(\alpha)$} & (-1.44) & (-0.43) & (0.55) & (0.60) & (1.65) & (1.86) \\
\addlinespace[4pt]
\multicolumn{7}{l}{\textit{Panel C. Equal-weighted, all-stock breakpoints}} \\
Excess return & -0.113 & 0.097 & 0.173 & 0.163 & 0.180 & 0.293 \\
\textit{$t$} & (-0.39) & (0.45) & (0.89) & (0.73) & (0.64) & (1.89) \\
\addlinespace[2pt]
FF5 $\alpha$ & -0.231 & -0.050 & 0.032 & -0.021 & -0.010 & 0.221 \\
\textit{$t(\alpha)$} & (-2.28) & (-0.98) & (0.82) & (-0.40) & (-0.11) & (1.43) \\
\addlinespace[2pt]
FF5+Mom $\alpha$ & -0.208 & -0.032 & 0.037 & -0.023 & -0.019 & 0.189 \\
\textit{$t(\alpha)$} & (-2.10) & (-0.68) & (0.97) & (-0.43) & (-0.20) & (1.25) \\
\addlinespace[4pt]
\multicolumn{7}{l}{\textit{Panel D. Equal-weighted, NYSE breakpoints}} \\
Excess return & -0.075 & 0.123 & 0.174 & 0.169 & 0.177 & 0.252 \\
\textit{$t$} & (-0.27) & (0.59) & (0.88) & (0.77) & (0.65) & (1.78) \\
\addlinespace[2pt]
FF5 $\alpha$ & -0.198 & -0.024 & 0.029 & -0.006 & -0.013 & 0.185 \\
\textit{$t(\alpha)$} & (-2.16) & (-0.51) & (0.72) & (-0.11) & (-0.15) & (1.30) \\
\addlinespace[2pt]
FF5+Mom $\alpha$ & -0.173 & -0.007 & 0.031 & -0.005 & -0.022 & 0.151 \\
\textit{$t(\alpha)$} & (-1.97) & (-0.16) & (0.76) & (-0.10) & (-0.25) & (1.09) \\
\bottomrule
\end{tabular}

\begin{minipage}{\textwidth}
\footnotesize\textit{Note:}
\footnotesize The table reports time-series properties of weekly returns on AI-beta quintile portfolios. Portfolios are rebalanced weekly. \textit{Excess return} is the average post-formation-week portfolio return in excess of the weekly risk-free rate. \textit{H$-$L} is the spread between the High and Low AI-beta quintiles. Returns and alphas are in percent per week. FF5 alphas are intercepts from regressing the portfolio's post-formation-week excess return on the \citet{famaFrench2015} five-factor model. FF5+Mom additionally includes the momentum factor of \citet{carhart1997}. Under \textit{all-stock breakpoints}, the quintile cutoffs are computed from the full eligible universe. Under \textit{NYSE breakpoints}, the quintile cutoffs are computed from NYSE-listed stocks only and applied to the full universe. AI consumption data is weekly from January 2024 through April 2026. The equity sample covers 108 formation weeks from March 2024 through March 2026.
\end{minipage}
\end{table}

%% file: figure_us_premium_v22.tex
\begin{figure}[!htbp]
\centering
\caption{The AI Premium: Performance and Robustness}
\label{fig:ai_premium_overview}
\makebox[\linewidth][c]{\includegraphics[width=\linewidth]{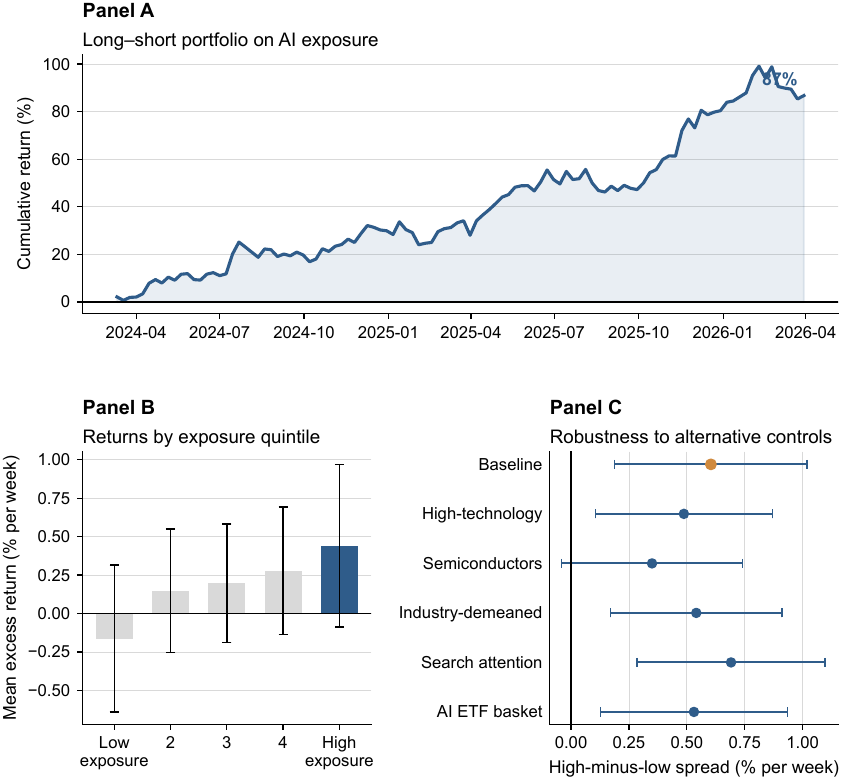}}

\begin{minipage}{\textwidth}
\footnotesize
\textit{Note:} The figure summarizes the portfolio evidence on the AI premium. Firms are sorted each week into quintiles by their current AI exposure $\beta^{AI}_{i,t}$, with breakpoints from the full eligible universe, and portfolios are value-weighted and held for one week. Panel~A plots the cumulative compounded return of the long-short portfolio that buys the highest-exposure quintile and sells the lowest. Panel~B reports the mean weekly excess return of each exposure quintile. Panel~C reports the high-minus-low weekly spread when the first-stage exposure regression adds one control at a time: the high-technology industry return and the value-weighted semiconductor portfolio return (Table~\ref{tab:us_industry_controlled}), exposures demeaned within Fama-French 30 industries (Internet Appendix Table~\ref{tab:us_indadj}), Google Trends search attention to AI (Internet Appendix Table~\ref{tab:ia_gtrend_controlled}), and the AI/robotics ETF basket return (Table~\ref{tab:us_industry_controlled}). Error bars in Panels~B and C are 95 percent confidence intervals. AI consumption data is weekly from January 2024 through April 2026. The equity sample covers 108 formation weeks from March 2024 through March 2026.
\end{minipage}
\end{figure}

%% file: tables_v22/table_us_fama_macbeth.tex
\begin{table}[!htbp]
\centering
\small
\caption{Fama-MacBeth Cross-Sectional Regressions}
\label{tab:us_fama_macbeth}
\begingroup
\setlength{\tabcolsep}{3pt}
\footnotesize
\begin{tabular}{lcccccccccc}
\toprule
 & $\beta^{AI}$ & $\ln ME$ & $\ln BM$ & Profit & Inv & Mom & Rev & Lev & Accr & $\bar{R}^2$ \\
\midrule
\multicolumn{11}{l}{\textit{Panel A. Equal-weighted}} \\
(1) & 6.26 &  &  &  &  &  &  &  &  & 0.006 \\
\textit{t} & (1.84) &  &  &  &  &  &  &  &  &  \\
\addlinespace[2pt]
(2) &  & 0.02 & 0.04 & -0.05 & -0.04 &  &  &  &  & 0.025 \\
\textit{t} &  & (0.27) & (0.49) & (-0.81) & (-0.84) &  &  &  &  &  \\
\addlinespace[2pt]
(3) &  & 0.01 & 0.06 & -0.04 & -0.05 & 0.12 &  &  &  & 0.032 \\
\textit{t} &  & (0.08) & (0.80) & (-0.63) & (-1.18) & (1.85) &  &  &  &  \\
\addlinespace[2pt]
(4) & 5.61 & 0.00 & 0.03 & -0.05 & -0.03 &  &  &  &  & 0.031 \\
\textit{t} & (1.77) & (0.06) & (0.44) & (-0.92) & (-0.83) &  &  &  &  &  \\
\addlinespace[2pt]
(5) & 5.46 & -0.01 & 0.05 & -0.04 & -0.05 & 0.11 &  &  &  & 0.038 \\
\textit{t} & (1.76) & (-0.13) & (0.76) & (-0.74) & (-1.16) & (1.78) &  &  &  &  \\
\addlinespace[2pt]
(6) &  & -0.01 & 0.03 & -0.03 & -0.05 & 0.11 & -0.03 & -0.02 & 0.09 & 0.044 \\
\textit{t} &  & (-0.12) & (0.53) & (-0.58) & (-1.25) & (1.83) & (-0.51) & (-0.58) & (1.82) &  \\
\addlinespace[2pt]
(7) & 6.69 & -0.02 & 0.03 & -0.04 & -0.05 & 0.11 & -0.03 & -0.01 & 0.08 & 0.049 \\
\textit{t} & (2.17) & (-0.38) & (0.42) & (-0.65) & (-1.23) & (1.74) & (-0.56) & (-0.44) & (1.77) &  \\
\addlinespace[6pt]
\multicolumn{11}{l}{\textit{Panel B. Value-weighted}} \\
(1) & 17.04 &  &  &  &  &  &  &  &  & 0.024 \\
\textit{t} & (2.91) &  &  &  &  &  &  &  &  &  \\
\addlinespace[2pt]
(2) &  & 0.06 & 0.00 & 0.00 & 0.06 &  &  &  &  & 0.122 \\
\textit{t} &  & (0.70) & (0.02) & (0.06) & (0.48) &  &  &  &  &  \\
\addlinespace[2pt]
(3) &  & 0.04 & 0.02 & 0.01 & 0.03 & 0.12 &  &  &  & 0.161 \\
\textit{t} &  & (0.54) & (0.30) & (0.16) & (0.32) & (0.88) &  &  &  &  \\
\addlinespace[2pt]
(4) & 11.50 & 0.05 & -0.02 & -0.01 & 0.05 &  &  &  &  & 0.141 \\
\textit{t} & (2.21) & (0.60) & (-0.36) & (-0.12) & (0.46) &  &  &  &  &  \\
\addlinespace[2pt]
(5) & 10.30 & 0.03 & 0.00 & 0.00 & 0.02 & 0.14 &  &  &  & 0.177 \\
\textit{t} & (2.03) & (0.44) & (-0.05) & (-0.02) & (0.21) & (1.01) &  &  &  &  \\
\addlinespace[2pt]
(6) &  & 0.05 & 0.00 & 0.00 & 0.02 & 0.09 & 0.07 & -0.04 & 0.07 & 0.209 \\
\textit{t} &  & (0.64) & (-0.03) & (0.05) & (0.23) & (0.69) & (0.67) & (-0.70) & (0.72) &  \\
\addlinespace[2pt]
(7) & 14.11 & 0.05 & -0.02 & -0.02 & 0.01 & 0.09 & 0.04 & -0.05 & 0.04 & 0.222 \\
\textit{t} & (2.86) & (0.59) & (-0.38) & (-0.28) & (0.11) & (0.74) & (0.40) & (-0.79) & (0.46) &  \\
\bottomrule
\end{tabular}
\endgroup

\begin{minipage}{\textwidth}
\footnotesize\textit{Note:}
\footnotesize This table reports the second step of the Fama-MacBeth procedure, estimated at the individual-stock level. Each week, individual stock returns are regressed cross-sectionally on the variables shown. Panel A weights stocks equally, so the weekly regressions are estimated by OLS. Panel B weights each stock by its market capitalization, so they are estimated by weighted least squares. $\beta^{AI}$ is the pre-formation rolling AI beta estimated in the first-step return regression. The remaining variables are firm-level characteristics included as controls. $\ln ME$ is log formation-week market equity, $\ln BM$ is log book-to-market, $Profit$ is gross profitability, $Inv$ is asset growth, $Mom$ is prior $52$-to-$5$-week return, $Rev$ is prior one-month return, $Lev$ is long-term debt to assets, and $Accr$ is accruals. Characteristics are weekly cross-sectionally winsorized at the $1$st/$99$th percentiles and standardized to unit variance. Reported coefficients are time-series averages of the weekly cross-sectional estimates. $t$-statistics are reported in parentheses and are time-series $t$-tests of those estimates (\citealp{famaMacBeth1973}). The $\bar{R}^2$ column is the time-series average of weekly cross-sectional $R^2$. AI consumption data is weekly from January 2024 through April 2026. The equity sample covers 108 formation weeks from March 2024 through March 2026.
\end{minipage}
\end{table}

%% file: tables_v22/table_us_industry_controlled.tex
\begin{table}[!htbp]
\centering
\small
\caption{High-Tech-, Semiconductor-, and AI-ETF-Controlled AI-Beta Quintile Portfolios}
\label{tab:us_industry_controlled}
\begin{tabular}{lcccccc}
\toprule
 & Low & 2 & 3 & 4 & High & H$-$L \\
\midrule
\multicolumn{7}{l}{\textit{Panel A. FF10 HiTec industry as control}} \\
\addlinespace[2pt]
Excess return & 0.020 & 0.145 & 0.125 & 0.198 & 0.508 & 0.487 \\
\textit{$t$} & (0.09) & (0.72) & (0.62) & (0.94) & (1.82) & (2.50) \\
\addlinespace[2pt]
FF5 $\alpha$ & -0.121 & -0.037 & -0.050 & -0.029 & 0.298 & 0.419 \\
\textit{$t(\alpha)$} & (-0.96) & (-0.48) & (-0.74) & (-0.36) & (2.45) & (2.20) \\
\addlinespace[2pt]
FF5+Mom $\alpha$ & -0.098 & -0.025 & -0.060 & -0.029 & 0.278 & 0.376 \\
\textit{$t(\alpha)$} & (-0.79) & (-0.33) & (-0.89) & (-0.36) & (2.30) & (2.02) \\
\addlinespace[6pt]
\multicolumn{7}{l}{\textit{Panel B. NAICS semiconductor industry as control}} \\
\addlinespace[2pt]
Excess return & 0.030 & 0.234 & 0.092 & 0.233 & 0.380 & 0.350 \\
\textit{$t$} & (0.12) & (1.13) & (0.45) & (1.07) & (1.47) & (1.75) \\
\addlinespace[2pt]
FF5 $\alpha$ & -0.140 & 0.084 & -0.114 & 0.003 & 0.158 & 0.298 \\
\textit{$t(\alpha)$} & (-1.06) & (1.05) & (-1.50) & (0.03) & (1.31) & (1.46) \\
\addlinespace[2pt]
FF5+Mom $\alpha$ & -0.122 & 0.096 & -0.108 & -0.004 & 0.138 & 0.260 \\
\textit{$t(\alpha)$} & (-0.93) & (1.20) & (-1.41) & (-0.05) & (1.15) & (1.29) \\
\addlinespace[6pt]
\multicolumn{7}{l}{\textit{Panel C. AI ETF basket as control}} \\
\addlinespace[2pt]
Excess return & -0.014 & 0.049 & 0.225 & 0.219 & 0.517 & 0.531 \\
\textit{$t$} & (-0.06) & (0.24) & (1.17) & (1.00) & (1.91) & (2.58) \\
\addlinespace[2pt]
FF5 $\alpha$ & -0.169 & -0.115 & 0.028 & 0.018 & 0.300 & 0.469 \\
\textit{$t(\alpha)$} & (-1.36) & (-1.36) & (0.37) & (0.21) & (2.29) & (2.23) \\
\addlinespace[2pt]
FF5+Mom $\alpha$ & -0.139 & -0.108 & 0.033 & 0.006 & 0.290 & 0.429 \\
\textit{$t(\alpha)$} & (-1.15) & (-1.27) & (0.43) & (0.08) & (2.20) & (2.07) \\
\bottomrule
\end{tabular}

\begin{minipage}{\textwidth}
\footnotesize\textit{Note:}
\footnotesize All panels report value-weighted quintile sorts using all-stock breakpoints. Panel~A expands the first-pass regression to include the Fama-French 10-industry HiTec return: $r^{e}_{i,t} = a + b_{\mathrm{mkt}}\,r^{e}_{m,t} + b_{\mathrm{HiTec}}\,r^{e}_{\mathrm{HiTec},t} + b_{\mathrm{AI}}\,AI_t$. Panel~B instead controls for the value-weighted weekly log return on the U.S.\ semiconductor industry: $r^{e}_{i,t} = a + b_{\mathrm{mkt}}\,r^{e}_{m,t} + b_{\mathrm{semi}}\,r_{\mathrm{semi},t} + b_{\mathrm{AI}}\,AI_t$. Panel~C controls for the equal-weighted AI ETF basket return: $r^{e}_{i,t} = a + b_{\mathrm{mkt}}\,r^{e}_{m,t} + b_{\mathrm{ETF}}\,r^{e}_{\mathrm{ETF},t} + b_{\mathrm{AI}}\,AI_t$. Returns and alphas are in percent per week. AI consumption data is weekly from January 2024 through April 2026. The equity sample covers 108 formation weeks from March 2024 through March 2026.
\end{minipage}
\end{table}

%% file: tables_v22/table_subsamples_combined.tex
\begingroup
\setlength{\tabcolsep}{4pt}
\renewcommand{\arraystretch}{0.90}
\begin{table}[p]
\centering
\footnotesize
\caption{AI-Beta Quintile Portfolios by Salient Consumption Components}
\label{tab:subsamples_combined}
\begin{tabular}{lcccc@{\hspace{10pt}}cccc}
\toprule
\multicolumn{9}{l}{\textit{Panel A. AI model class}} \\
 & \multicolumn{4}{c}{$\beta^{AI,Closed}$} & \multicolumn{4}{c}{$\beta^{AI,Open}$} \\
\cmidrule(lr){2-5}\cmidrule(lr){6-9}
 & Low & 3 & High & H$-$L & Low & 3 & High & H$-$L \\
\midrule
Excess return & -0.037 & 0.229 & 0.479 & 0.516 & 0.091 & 0.278 & 0.322 & 0.230 \\
\textit{$t$} & (-0.15) & (1.09) & (1.77) & (2.52) & (0.36) & (1.55) & (1.18) & (1.30) \\
\addlinespace[1pt]
FF5 $\alpha$ & -0.195 & 0.047 & 0.238 & 0.432 & -0.119 & 0.090 & 0.107 & 0.225 \\
\textit{$t(\alpha)$} & (-1.38) & (0.59) & (2.21) & (2.09) & (-1.07) & (1.20) & (0.93) & (1.25) \\
\addlinespace[1pt]
FF5+Mom $\alpha$ & -0.180 & 0.056 & 0.231 & 0.411 & -0.112 & 0.092 & 0.113 & 0.225 \\
\textit{$t(\alpha)$} & (-1.28) & (0.70) & (2.14) & (1.98) & (-1.00) & (1.22) & (0.97) & (1.23) \\
\addlinespace[5pt]

\multicolumn{9}{l}{\textit{Panel B. Paid/core versus new users}} \\
 & \multicolumn{4}{c}{$\beta^{AI,Core}$} & \multicolumn{4}{c}{$\beta^{AI,New}$} \\
\cmidrule(lr){2-5}\cmidrule(lr){6-9}
 & Low & 3 & High & H$-$L & Low & 3 & High & H$-$L \\
\midrule
Excess return & -0.046 & 0.164 & 0.398 & 0.445 & 0.110 & 0.210 & 0.294 & 0.185 \\
\textit{$t$} & (-0.19) & (0.80) & (1.17) & (1.72) & (0.36) & (0.97) & (1.03) & (0.78) \\
\addlinespace[1pt]
FF5 $\alpha$ & -0.226 & -0.016 & 0.144 & 0.370 & -0.138 & 0.016 & 0.098 & 0.236 \\
\textit{$t(\alpha)$} & (-1.55) & (-0.19) & (0.96) & (1.54) & (-0.98) & (0.20) & (0.75) & (1.09) \\
\addlinespace[1pt]
FF5+Mom $\alpha$ & -0.203 & -0.016 & 0.150 & 0.354 & -0.125 & 0.019 & 0.092 & 0.218 \\
\textit{$t(\alpha)$} & (-1.40) & (-0.19) & (1.00) & (1.46) & (-0.89) & (0.24) & (0.69) & (1.00) \\
\addlinespace[5pt]

\multicolumn{9}{l}{\textit{Panel C. Seasoned versus non-seasoned users}} \\
 & \multicolumn{4}{c}{$\beta^{AI,Seasoned}$} & \multicolumn{4}{c}{$\beta^{AI,Nonseasoned}$} \\
\cmidrule(lr){2-5}\cmidrule(lr){6-9}
 & Low & 3 & High & H$-$L & Low & 3 & High & H$-$L \\
\midrule
Excess return & -0.079 & 0.255 & 0.460 & 0.540 & 0.024 & 0.101 & 0.340 & 0.316 \\
\textit{$t$} & (-0.31) & (1.22) & (1.32) & (2.11) & (0.07) & (0.44) & (1.18) & (1.48) \\
\addlinespace[1pt]
FF5 $\alpha$ & -0.243 & 0.101 & 0.196 & 0.439 & -0.202 & -0.093 & 0.117 & 0.320 \\
\textit{$t(\alpha)$} & (-1.66) & (1.17) & (1.21) & (1.79) & (-1.35) & (-1.07) & (0.94) & (1.49) \\
\addlinespace[1pt]
FF5+Mom $\alpha$ & -0.238 & 0.105 & 0.209 & 0.447 & -0.191 & -0.099 & 0.090 & 0.281 \\
\textit{$t(\alpha)$} & (-1.61) & (1.19) & (1.28) & (1.81) & (-1.26) & (-1.13) & (0.73) & (1.32) \\
\addlinespace[5pt]

\multicolumn{9}{l}{\textit{Panel D. Prompt length}} \\
 & \multicolumn{4}{c}{$\beta^{Tok,Long}$} & \multicolumn{4}{c}{$\beta^{Tok,Short}$} \\
\cmidrule(lr){2-5}\cmidrule(lr){6-9}
 & Low & 3 & High & H$-$L & Low & 3 & High & H$-$L \\
\midrule
Excess return & -0.040 & 0.316 & 0.437 & 0.477 & 0.150 & 0.203 & 0.446 & 0.296 \\
\textit{$t$} & (-0.17) & (1.53) & (1.60) & (2.16) & (0.57) & (1.03) & (1.62) & (1.65) \\
\addlinespace[1pt]
FF5 $\alpha$ & -0.228 & 0.122 & 0.206 & 0.433 & -0.021 & -0.031 & 0.228 & 0.249 \\
\textit{$t(\alpha)$} & (-1.63) & (1.34) & (1.60) & (1.93) & (-0.18) & (-0.44) & (1.78) & (1.35) \\
\addlinespace[1pt]
FF5+Mom $\alpha$ & -0.191 & 0.132 & 0.172 & 0.363 & -0.022 & -0.031 & 0.237 & 0.259 \\
\textit{$t(\alpha)$} & (-1.42) & (1.44) & (1.39) & (1.72) & (-0.19) & (-0.43) & (1.84) & (1.40) \\
\bottomrule
\end{tabular}

\begin{minipage}{\textwidth}
\footnotesize\textit{Note:}
\footnotesize This table reports value-weighted, all-stock-breakpoint quintile sorts on split-specific AI exposures. Panels~A--D compare closed- and open-weight model consumption, paid/core and new accounts, seasoned and non-seasoned users, and long- and short-prompt token growth, respectively (see Section~\ref{sec:ai_subcomponents} for the definitions of the groups). For compactness, the table reports the Low, middle (3), High, and High-minus-Low portfolios. Quintiles 2 and 4 are omitted. Returns and alphas are in percent per week. AI consumption data is weekly from January 2024 through April 2026. The equity sample covers 108 formation weeks from March 2024 through March 2026.
\end{minipage}
\end{table}
\endgroup

%% file: tables_v22/table_international_combined.tex
\begin{table}[h!]
\centering
\small
\caption{International AI-Beta Quintile Portfolios}
\label{tab:international_combined}
\label{tab:international_full}
\label{tab:international_vw}
\begin{tabular}{lcccccc}
\toprule
 & Low & 2 & 3 & 4 & High & H$-$L \\
\midrule
\multicolumn{7}{l}{\textit{Panel A. Full international sample}} \\
Excess return & 0.141 & 0.157 & 0.138 & 0.199 & 0.214 & 0.073 \\
\textit{$t$} & (1.52) & (1.95) & (2.17) & (2.47) & (2.36) & (2.79) \\
\addlinespace[2pt]
Local-mkt $\alpha$ & 0.022 & 0.051 & 0.054 & 0.093 & 0.101 & 0.078 \\
\textit{$t(\alpha)$} & (0.46) & (1.34) & (1.79) & (2.41) & (2.00) & (2.98) \\
\addlinespace[6pt]
\multicolumn{7}{l}{\textit{Panel B. MSCI Developed}} \\
Excess return & 0.129 & 0.166 & 0.126 & 0.194 & 0.228 & 0.100 \\
\textit{$t$} & (1.10) & (1.68) & (1.70) & (2.02) & (2.07) & (2.52) \\
\addlinespace[2pt]
Local-mkt $\alpha$ & 0.013 & 0.061 & 0.049 & 0.093 & 0.121 & 0.108 \\
\textit{$t(\alpha)$} & (0.21) & (1.42) & (1.44) & (2.15) & (1.94) & (2.75) \\
\addlinespace[6pt]
\multicolumn{7}{l}{\textit{Panel C. MSCI Emerging}} \\
Excess return & 0.151 & 0.149 & 0.147 & 0.201 & 0.201 & 0.050 \\
\textit{$t$} & (1.71) & (1.95) & (2.29) & (2.61) & (2.28) & (1.32) \\
\addlinespace[2pt]
Local-mkt $\alpha$ & 0.041 & 0.050 & 0.066 & 0.103 & 0.093 & 0.053 \\
\textit{$t(\alpha)$} & (0.70) & (1.04) & (1.57) & (2.08) & (1.55) & (1.36) \\
\bottomrule
\end{tabular}

\begin{minipage}{\textwidth}
\footnotesize\textit{Note:}
\footnotesize The table reports returns and local-market alphas for international pre-formation AI-beta-sorted quintile portfolios. A local market is a country-level equity market. Panel~A pools local markets classified by MSCI as developed or emerging (the full international sample), and excludes frontier, standalone, and not-rated markets. Panel~B restricts the sample to developed local markets and Panel~C to emerging local markets. For each local market and week, stocks are sorted into quintiles using local-market breakpoints, and quintile returns are equal-weighted across firms. H$-$L is formed within each local market. Portfolio returns and H$-$L spreads are then equal-weighted across local markets. Firm betas are 13-week rolling log-excess regressions of weekly local-currency returns on the local-market log-excess return and the AI factor. Daily returns are JKP-style (\citealp{jensen2023there}) winsorized using same-month CRSP cutoffs before weekly compounding. \textit{Local-mkt $\alpha$} is the intercept from regressing the portfolio's next-week return on the value-weighted market return of the included local markets. AI consumption data is weekly from January 2024 through April 2026. Equity betas use 13-week rolling windows, so portfolio formation weeks begin in March 2024.
\end{minipage}
\end{table}

%% file: figure_oa_event_study_v22.tex
\begin{figure}[!htbp]
\centering
\begin{subfigure}{0.78\textwidth}
\centering
\includegraphics[width=\textwidth]{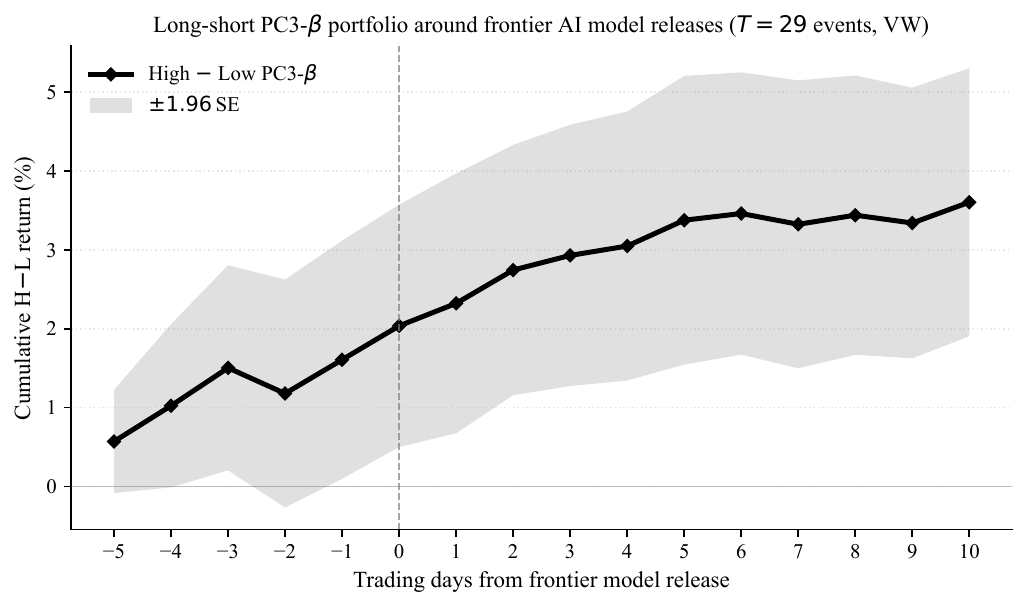}
\caption{Frontier Model Releases (29 Events)}
\label{fig:event_study_caar_frontier}
\end{subfigure}
\\[0.6em]
\begin{subfigure}{0.78\textwidth}
\centering
\includegraphics[width=\textwidth]{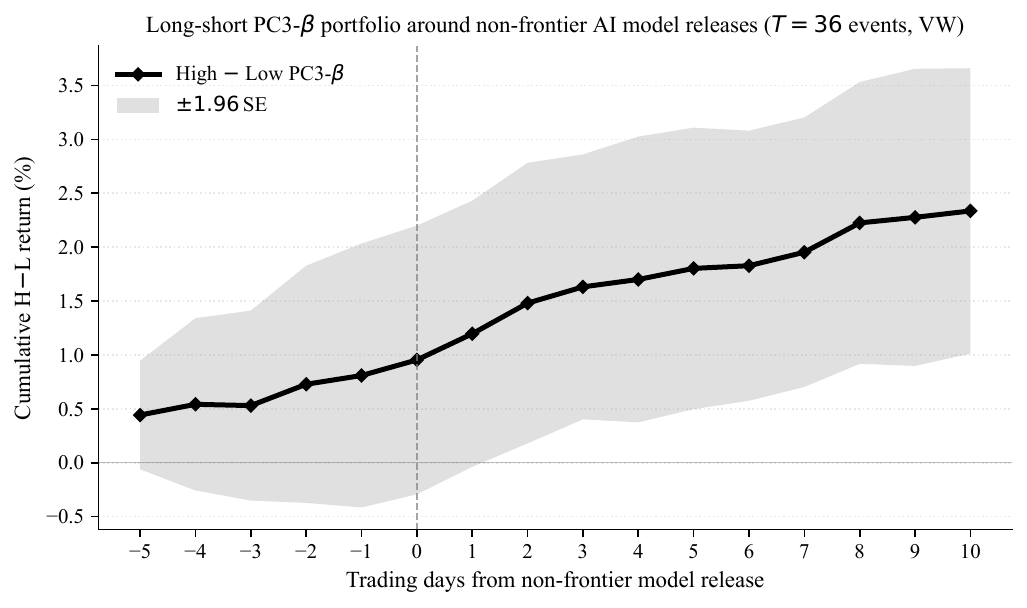}
\caption{Non-Frontier Model Releases (36 Events)}
\label{fig:event_study_caar_nonfrontier}
\end{subfigure}
\caption{Cumulative Average Returns Around AI Model Releases}
\label{fig:event_study_caar}

\begin{minipage}{\textwidth}
\footnotesize
\textit{Note:} Daily event-time CAARs to the value-weighted, all-stock-breakpoint long-short AI-$\beta$ quintile portfolio. In each panel, firms are sorted on their pre-event 13-week rolling log-excess AI beta and the H$-$L spread is the High-quintile minus Low-quintile cumulative return. Time is measured in trading days from the event anchor (same-or-next trading day on or after the announced release date). The shaded band is the $\pm 1.96$ cross-event standard error of the cumulated H$-$L return at each event-time day. Panel~A uses the 29 frontier-model events of Table~\ref{tab:ia_release_calendar} Panel~A; Panel~B uses the 36 non-frontier-model events of Table~\ref{tab:ia_release_calendar} Panel~B. Both panels are value-weighted. AI consumption data is weekly from January 2024 through April 2026. The equity sample covers 108 formation weeks from March 2024 through March 2026.
\end{minipage}
\end{figure}

%% file: tables_v22/table_ia_news_window_quintiles.tex
\begin{table}[!htbp]
\centering
\small
\caption{AI-Beta Quintile Portfolios Excluding Model-Release Weeks}
\label{tab:ia_news_window_quintiles}
\begin{tabular}{lcccccc}
\toprule
 & Low & 2 & 3 & 4 & High & H$-$L \\
\midrule
\multicolumn{7}{l}{\textit{Panel A. Excluding frontier-release weeks}} \\
Excess return & 0.022 & 0.093 & 0.158 & 0.228 & 0.365 & 0.343 \\
\textit{$t$} & (0.10) & (0.48) & (0.77) & (1.13) & (1.48) & (2.15) \\
\addlinespace[4pt]
\multicolumn{7}{l}{\textit{Panel B. Excluding frontier and non-frontier release weeks}} \\
Excess return & -0.096 & 0.035 & 0.030 & 0.146 & 0.235 & 0.331 \\
\textit{$t$} & (-0.35) & (0.13) & (0.12) & (0.65) & (0.68) & (1.68) \\
\bottomrule
\end{tabular}

\begin{minipage}{\textwidth}
\footnotesize\textit{Note:}
\footnotesize The table reports weekly value-weighted excess returns on AI-beta quintile portfolios, all-stock breakpoints, after excluding weeks that contain major AI model releases. Panel~A excludes holding weeks with a frontier-provider release (Anthropic, DeepSeek, Google, Meta, OpenAI); Panel~B additionally excludes weeks with releases from other providers. The sorting signal, universe, and portfolio construction match Table~\ref{tab:us_portfolio_p5_p1}. \textit{H$-$L} buys the High and shorts the Low AI-beta quintile. Returns are in percent per week; Newey--West $t$-statistics with five lags are in parentheses.
\end{minipage}
\end{table}

%% file: figure_us_exposure_industry_firm_v22.tex
\begin{figure}[!htbp]
\centering
\caption{AI Exposure across Industries and Firms}
\label{fig:exposure_industries_firms}
\makebox[\linewidth][c]{\includegraphics[width=\linewidth]{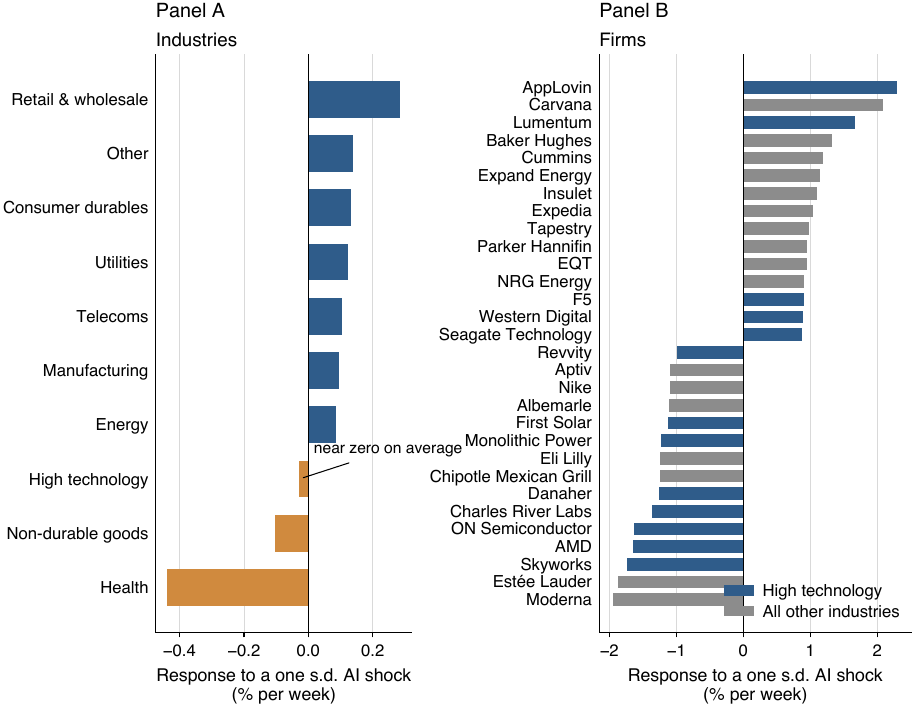}}

\begin{minipage}{\textwidth}
\footnotesize
\textit{Note:} The figure reports market-implied AI exposure, the coefficient on the AI factor in 13-week rolling regressions of a firm's weekly log-excess return on the AI factor and the log-excess market return. Values are the return response, in percentage points per week, to a one-standard-deviation AI-factor shock. Panel~A reports the value-weighted average exposure for the Fama-French 10 industries, using formation-week market equity weights. Panel~B reports the fifteen current S\&P 500 firms with the most positive and the fifteen with the most negative average exposure, with firm-level exposures averaged across all weeks in which a firm has a valid estimate. Bar colors indicate whether the firm belongs to the Fama-French high-technology industry. Internet Appendix Figure~\ref{fig:sp500_top_bottom_company_ai_exposure} reports the full ranking of the top and bottom 50 firms. AI consumption data is weekly from January 2024 through April 2026. The equity sample covers 108 formation weeks from March 2024 through March 2026.
\end{minipage}
\end{figure}

%% file: tables_v22/table_occupation_task_alm_deming.tex
\begin{table}[!htbp]
\singlespacing
\centering
\scriptsize
\setlength{\tabcolsep}{4pt}
\renewcommand{\arraystretch}{0.85}
\caption{Market-Implied Occupation Exposure and Task and Skill Measures}
\label{tab:occupation_task_alm_deming}
\begin{tabular}{lccccc}
\toprule
 & \multicolumn{2}{c}{Task content} & \multicolumn{3}{c}{Skill content} \\
\cmidrule(lr){2-3} \cmidrule(lr){4-6}
 & (1) & (2) & (3) & (4) & (5) \\
 & ALM & Acemoglu-Autor & Deming & $+$ interaction & detailed \\
\midrule
Routine & 0.004 &  &  &  &  \\
 & (0.09) &  &  &  &  \\
Nonroutine Analytic & -0.172*** & -0.230*** &  &  &  \\
 & (-4.02) & (-4.67) &  &  &  \\
Nonroutine Interactive & 0.163*** & 0.116** &  &  &  \\
 & (3.09) & (2.51) &  &  &  \\
Manual/Physical & 0.035 &  &  &  &  \\
 & (0.86) &  &  &  &  \\
Routine Cognitive &  & 0.052 &  &  &  \\
 &  & (1.08) &  &  &  \\
Routine Manual &  & -0.363*** &  &  &  \\
 &  & (-3.89) &  &  &  \\
Nonroutine Manual &  & 0.297*** &  &  &  \\
 &  & (3.81) &  &  &  \\
Social Skills &  &  & 0.111** & 0.106** & 0.205*** \\
 &  &  & (2.35) & (2.22) & (2.99) \\
Nonroutine Analytical / Math &  &  & -0.141*** & -0.140*** & 0.097 \\
 &  &  & (-4.02) & (-3.97) & (0.65) \\
Routine &  &  & 0.074 & 0.071 & 0.116* \\
 &  &  & (1.34) & (1.26) & (1.70) \\
Social Skills $\times$ Analytical / Math &  &  &  & -0.027 &  \\
 &  &  &  & (-0.70) &  \\
Service &  &  &  &  & -0.125*** \\
 &  &  &  &  & (-2.70) \\
Deductive / Inductive Reasoning &  &  &  &  & -0.340*** \\
 &  &  &  &  & (-3.72) \\
Number Facility &  &  &  &  & -0.043 \\
 &  &  &  &  & (-0.31) \\
Information Use &  &  &  &  & -0.267** \\
 &  &  &  &  & (-2.16) \\
Required Social Interaction &  &  &  &  & 0.059 \\
 &  &  &  &  & (1.07) \\
Coordination / Team Building &  &  &  &  & 0.119 \\
 &  &  &  &  & (1.50) \\
Interaction / Communication &  &  &  &  & 0.262*** \\
 &  &  &  &  & (3.06) \\
\midrule
Observations & 649 & 649 & 649 & 649 & 649 \\
$R^2$ & 0.025 & 0.063 & 0.020 & 0.020 & 0.108 \\
\bottomrule
\end{tabular}

\begin{minipage}{0.97\textwidth}
\footnotesize
\textit{Note:}
This table reports regressions of standardized market-implied occupation AI exposure on task and skill measures from the labor literature. Column (1) uses four Autor-Levy-Murnane (ALM) style task groups constructed from O*NET descriptors, following \citet{AutorLevyMurnane2003}: routine, nonroutine analytic, nonroutine interactive, and manual/physical. Column (2) uses an expanded Acemoglu-Autor style taxonomy that keeps the same nonroutine analytic and nonroutine interactive measures and splits the routine and manual margins into routine cognitive, routine manual, and nonroutine manual. Columns (1) and (2) use task-content measures. Columns (3)--(5) use the skill-based measures emphasized in \citet{deming2017growing}. Column (3) uses his three headline measures: social skills, nonroutine analytical and mathematical content, and routine tasks. Column (4) adds the interaction between standardized social skills and standardized nonroutine analytical/math, following Deming's social-cognitive complementarity. Column (5) uses the ten Deming-style O*NET task and skill measures. The first three have the same definitions as in columns (3) and (4). All right-hand-side variables are standardized. The interaction is the product of standardized components. $t$-statistics based on HC1 robust standard errors are in parentheses. *, **, and *** denote significance at the 10\%, 5\%, and 1\% levels, respectively. Underlying firm-level AI betas are estimated on the equity sample of 108 formation weeks from March 2024 through March 2026.
\end{minipage}
\end{table}

\begin{table}[!htbp]
\singlespacing
\centering
\footnotesize
\setlength{\tabcolsep}{4pt}
\renewcommand{\arraystretch}{0.85}
\caption{Market-Implied Occupation Exposure and AI-Exposure Measures}
\label{tab:occupation_ai_exposure_measures}
\begin{tabular}{lccc}
\toprule
 & (1) & (2) & (3) \\
 & Technical AI & ESZ (core \& suppl.) & Combined \\
\midrule
Felten AIOE & -0.150** &  & -0.133* \\
 & (-2.46) &  & (-1.81) \\
Eloundou GPT-4 & 0.131** &  & 0.068 \\
 & (2.33) &  & (0.82) \\
Webb AI & -0.067* &  & -0.085** \\
 & (-1.78) &  & (-2.06) \\
ESZ GenAI Exposure (Core Tasks) &  & -0.010 & 0.056 \\
 &  & (-0.28) & (0.77) \\
ESZ GenAI Exposure (Supplemental Tasks) &  & 0.024 & 0.022 \\
 &  & (0.96) & (0.66) \\
\midrule
Observations & 649 & 602 & 602 \\
$R^2$ & 0.012 & 0.001 & 0.013 \\
\bottomrule
\end{tabular}

\begin{minipage}{0.97\textwidth}
\footnotesize
\textit{Note:}
This table reports regressions of standardized market-implied occupation AI exposure on occupation-level AI-exposure measures. Column (1) uses technical AI exposure measures from \citet{felten2021occupational}, \citet{eloundou2023gpts}, and \citet{webb2020impact}. Column (2) uses the occupation-level generative-AI task exposure of \citet{eisfeldt2023generative} (ESZ), from their public data repository: exposure through core tasks and exposure through supplemental tasks, whose sum is the occupation's total exposed task share. In ESZ, exposure through core tasks proxies for labor substitution and exposure through supplemental tasks for complementarity. Column (3) combines the technical AI and ESZ measures. ESZ scores are mapped from 2010 SOC to 2018 SOC occupations using the Census occupation-code crosswalk, which reduces the sample in columns (2) and (3). All right-hand-side variables are standardized. $t$-statistics based on HC1 robust standard errors are in parentheses. *, **, and *** denote significance at the 10\%, 5\%, and 1\% levels, respectively. Underlying firm-level AI betas are estimated on the equity sample of 108 formation weeks from March 2024 through March 2026.
\end{minipage}
\end{table}

%% file: figure_oa_agentic_v16.tex
\begin{figure}[!htbp]
\centering
\includegraphics[width=0.95\textwidth]{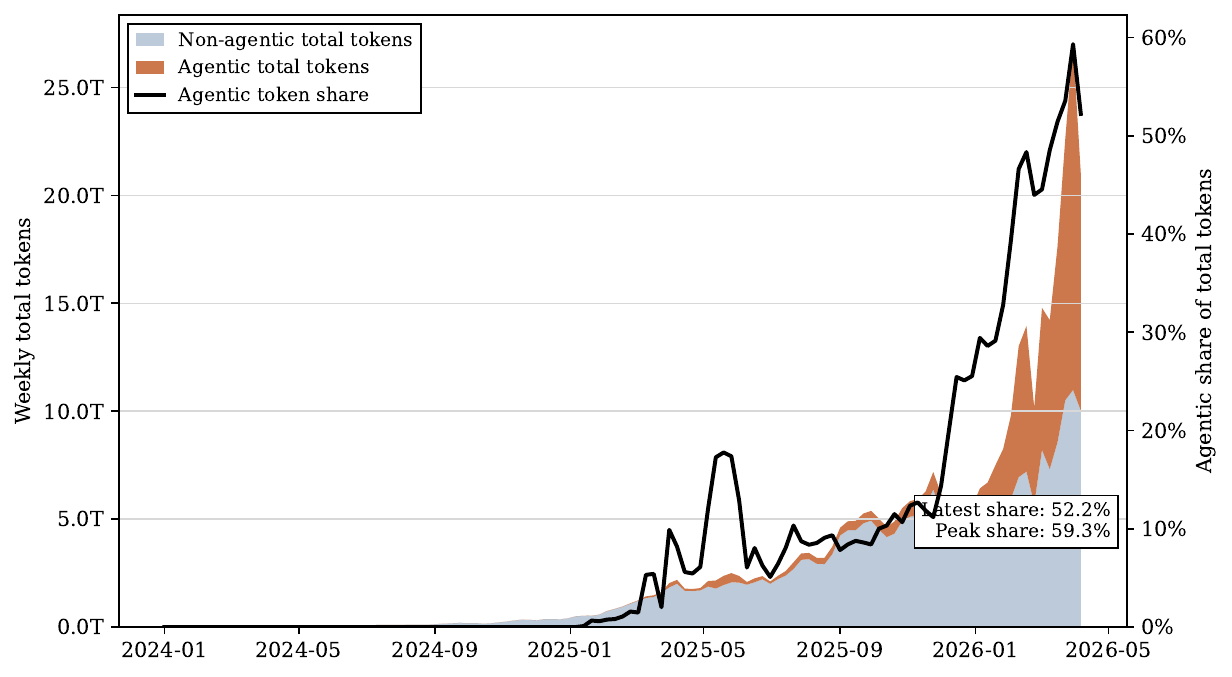}
\caption{Agentic and Non-Agentic Token Consumption}
\label{fig:agentic_total_token_usage}

\begin{minipage}{\textwidth}
\footnotesize
\textit{Note:} The figure plots weekly total token consumption in the OpenRouter dataset, decomposed into agentic and non-agentic total tokens. Total tokens are defined as prompt tokens plus completion tokens. Agentic total tokens are total tokens from requests whose normalized finish reason is \texttt{tool\_calls}; non-agentic total tokens are the remaining total tokens. The black line plots the agentic share of total tokens. Data is weekly from January 2024 through April 2026.
\end{minipage}
\end{figure}

%% file: figure_oa_agentic_paradox_v16.tex
\begin{figure}[!htbp]
\centering
\includegraphics[width=0.95\textwidth]{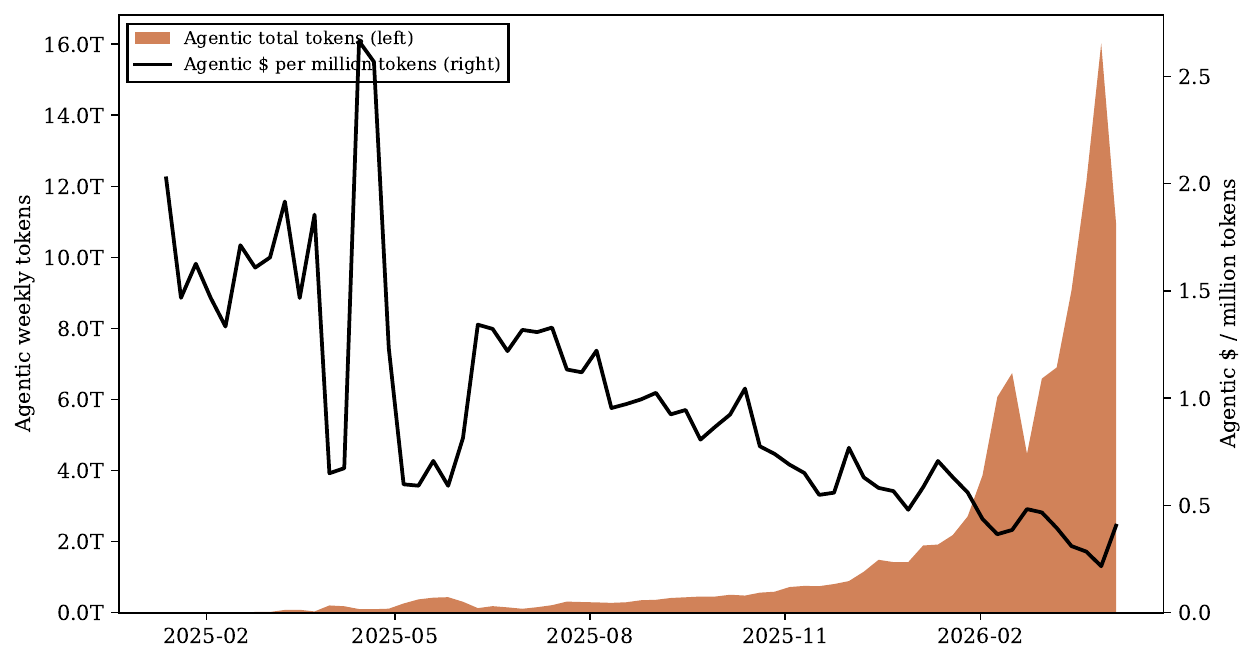}
\caption{Agentic Token Volume and Realized Price per Token}
\label{fig:agentic_paradox}

\begin{minipage}{\textwidth}
\footnotesize
\textit{Note:} The figure plots weekly agentic total tokens (left axis; prompt plus completion tokens from requests with normalized finish reason \texttt{tool\_calls}) against the realized dollars per million tokens on the same agentic requests (right axis; OpenRouter usage dollars divided by agentic total tokens, times one million). Agentic token volume rises by orders of magnitude while the realized price per token declines, consistent with prompt caching and routing toward cheaper models compressing revenue per token even as agentic usage expands.
\end{minipage}
\end{figure}

%% file: figure_oa_ai_data_v18.tex
\begin{figure}[!htbp]
\centering
\begin{subfigure}{0.50\textwidth}
\centering
\includegraphics[width=\textwidth]{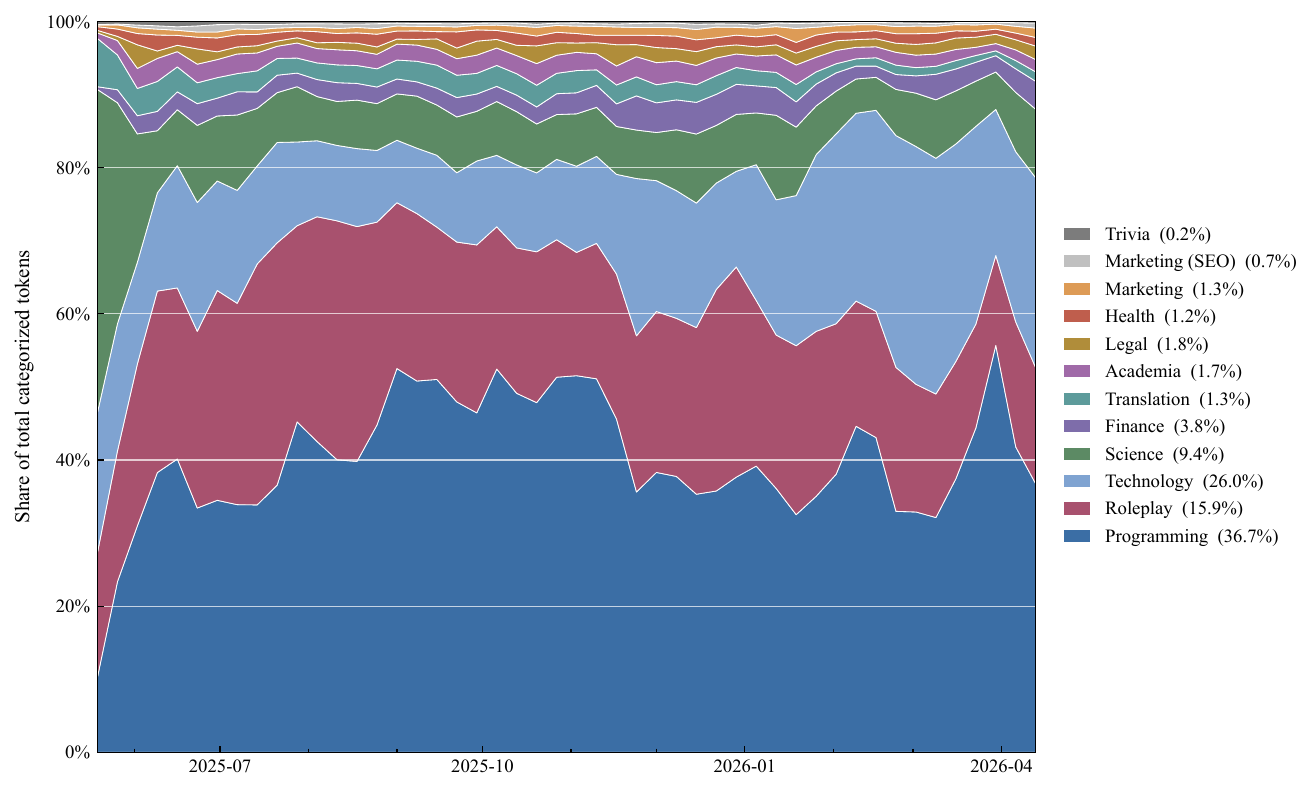}
\caption{All AI Models}
\label{fig:category_overall}
\end{subfigure}
\\[0.6em]
\begin{subfigure}{0.50\textwidth}
\centering
\includegraphics[width=\textwidth]{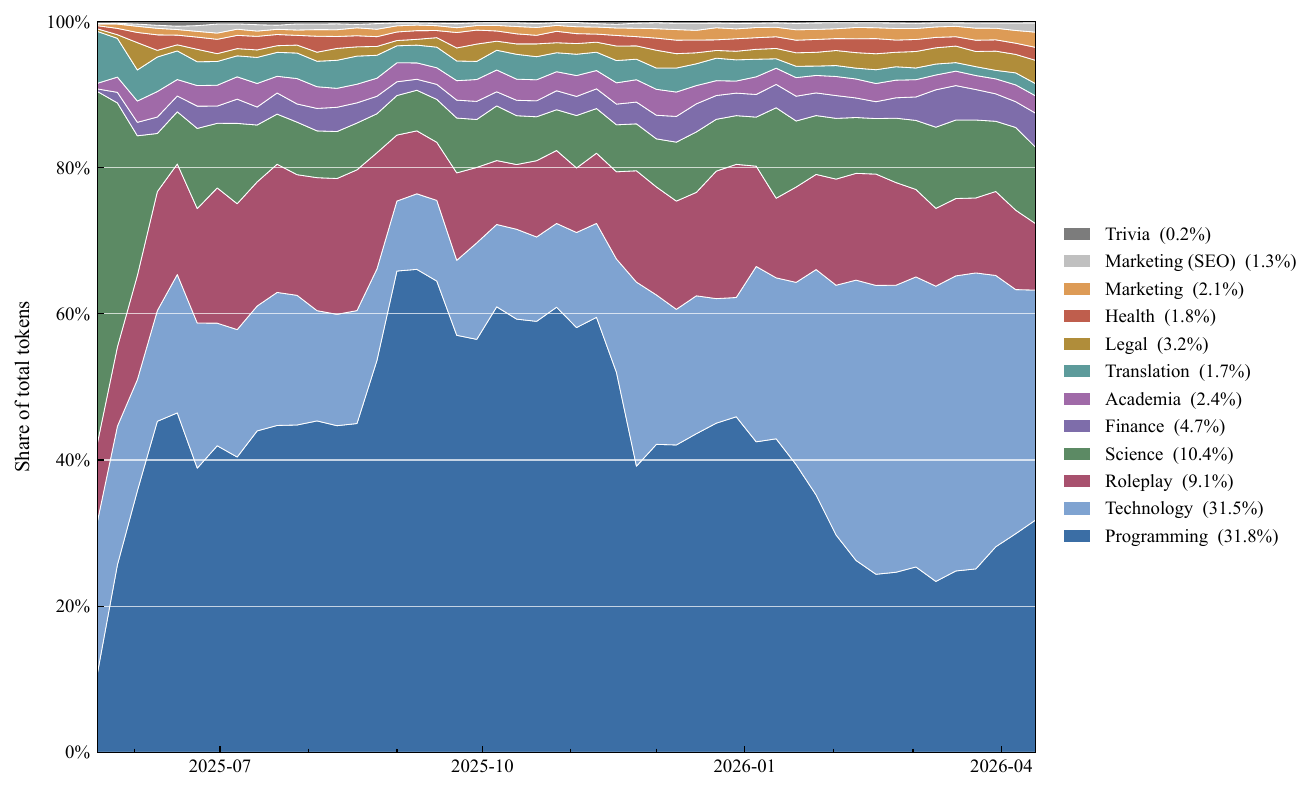}
\caption{Closed-Source Models Only}
\label{fig:category_closed}
\end{subfigure}
\\[0.6em]
\begin{subfigure}{0.50\textwidth}
\centering
\includegraphics[width=\textwidth]{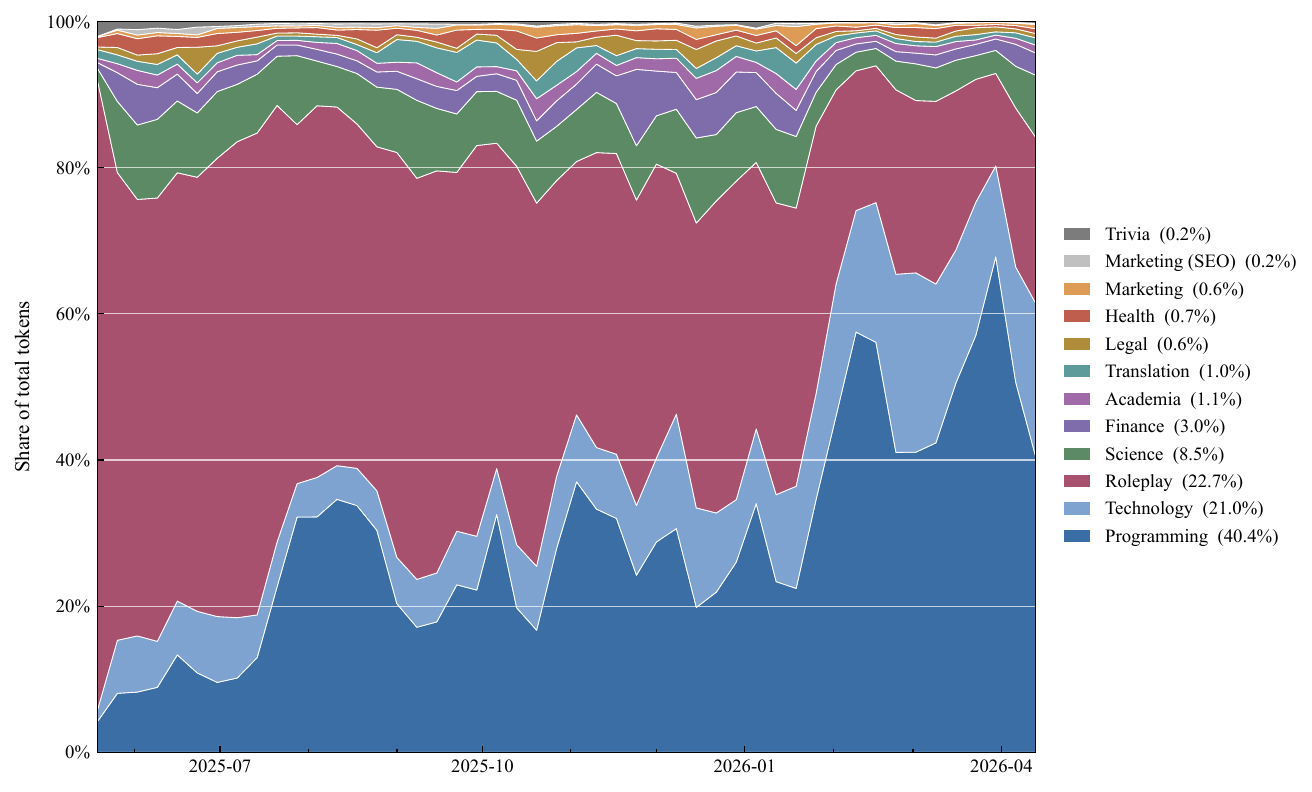}
\caption{Open-Source Models Only}
\label{fig:category_open}
\end{subfigure}
\caption{Evolution of AI Consumption by Prompt-Content Category}
\label{fig:category}
\label{fig:category_evolution}

\begin{minipage}{\textwidth}
\footnotesize
\textit{Note:} Each panel in the figure stacks the weekly share of total categorized tokens contributed by each of the 12 prompt-content categories tagged in the platform's category-token-share file. Panel~A pools all AI models; Panel~B restricts the universe to closed-weight models; Panel~C restricts the universe to open-weight models. Within each panel shares sum to 100\% by construction, and the right-hand legend reports the share at the final week (legend ordering matches the stacking order). Data is weekly from mid-May 2025 (when category coverage becomes available) through April 2026.
\end{minipage}
\end{figure}

%% file: tables_v22/table_user_model_granularity.tex
\begin{landscape}
\begin{table}[!p]
\centering
\caption{OpenRouter Account-Model Data Summary}
\label{tab:user_model_granularity}
\scriptsize
\setlength{\tabcolsep}{4pt}
\renewcommand{\arraystretch}{1.04}
\begin{minipage}{0.98\linewidth}
\begin{tabular*}{\linewidth}{@{\extracolsep{\fill}}lrrrrrrr}
\toprule
 & Mean & Std. & p10 & p25 & p50 & p75 & p90 \\
\midrule
\multicolumn{8}{l}{\textit{Panel A. Active account-day consumption}} \\
Requests per active account-day & 109.9 & 434.3 & 1.000 & 4.000 & 16.00 & 49.00 & 153.0 \\
Total tokens per active account-day (thous.) & 1,223 & 4,778 & 2.718 & 14.55 & 79.44 & 392.7 & 1,713 \\
User-days using at least two models (\%) & 31.66 &  &  &  &  &  &  \\
User-days using at least two providers (\%) & 44.43 &  &  &  &  &  &  \\
\addlinespace[3pt]
\multicolumn{8}{l}{\textit{Panel B. Account-level breadth and persistence}} \\
Active days per account & 16.73 & 35.73 & 1.000 & 1.000 & 3.000 & 13.00 & 47.00 \\
Active weeks per account & 5.300 & 8.131 & 1.000 & 1.000 & 2.000 & 5.000 & 15.00 \\
Observed account span (days) & 67.24 & 112.1 & 1.000 & 1.000 & 8.000 & 84.00 & 243.0 \\
Lifetime requests & 1,127 & 4,470 & 2.000 & 9.000 & 56.00 & 348.0 & 1,683 \\
Lifetime total tokens (thous.) & 15,916 & 64,301 & 2.258 & 22.36 & 288.7 & 2,723 & 22,607 \\
Top-model token share within account (\%) & 79.40 & 24.56 & 40.19 & 59.35 & 93.15 & 100.0 & 100.0 \\
Model-token HHI within account & 7,324 & 2,945 & 2,791 & 4,830 & 8,712 & 10,000 & 10,000 \\
Accounts using at least two models (\%) & 60.37 &  &  &  &  &  &  \\
Accounts using at least five models (\%) & 29.73 &  &  &  &  &  &  \\
Accounts using at least two providers (\%) & 70.00 &  &  &  &  &  &  \\
\addlinespace[3pt]
\multicolumn{8}{l}{\textit{Panel C. Account-model daily consumption and volatility}} \\
Active days per account-model pair & 5.066 & 10.07 & 1.000 & 1.000 & 1.000 & 4.000 & 12.00 \\
Observed pair span (days) & 20.79 & 44.47 & 1.000 & 1.000 & 1.000 & 16.00 & 65.00 \\
Active-day share within pair span (\%) & 74.38 & 35.73 & 12.50 & 41.18 & 100.0 & 100.0 & 100.0 \\
Requests per active account-model-day & 24.12 & 73.73 & 1.000 & 1.500 & 4.000 & 14.60 & 44.50 \\
Total tokens per active account-model-day (thous.) & 372.9 & 1,424 & 0.419 & 2.721 & 16.33 & 109.2 & 558.9 \\
SD of daily total tokens (thous.) & 971.0 & 3,385 & 3.401 & 13.28 & 67.03 & 311.4 & 1,614 \\
CV of daily total tokens & 1.132 & 0.561 & 0.561 & 0.749 & 1.005 & 1.379 & 1.865 \\
\bottomrule
\end{tabular*}

\begin{minipage}{\linewidth}
\footnotesize
\textit{Note:}
This table reports account-level and account-model-level summary statistics computed from the OpenRouter generations data. Non-share row variables are winsorized at the 1st and 99th percentiles before computing the reported statistics. An active account-day has positive recorded requests; requests are API calls; total tokens are prompt plus completion tokens. Active days and weeks count positive-use days and calendar weeks, and observed span is first-to-last positive-use days inclusive. Top-model share and model-token HHI use each account's lifetime token shares across models, with HHI scaled by 10,000. For account-model pairs, active-day share is active days divided by span; SD and CV are computed across active days, with CV equal to SD divided by the mean. Volatility rows require at least five active days. Share rows are percent means and leave distribution columns blank.
\end{minipage}
\end{minipage}
\end{table}
\end{landscape}

%% file: figure_oa_growth_v16.tex
\begin{figure}[!htbp]
\centering
\makebox[\linewidth][c]{\includegraphics[width=0.95\linewidth]{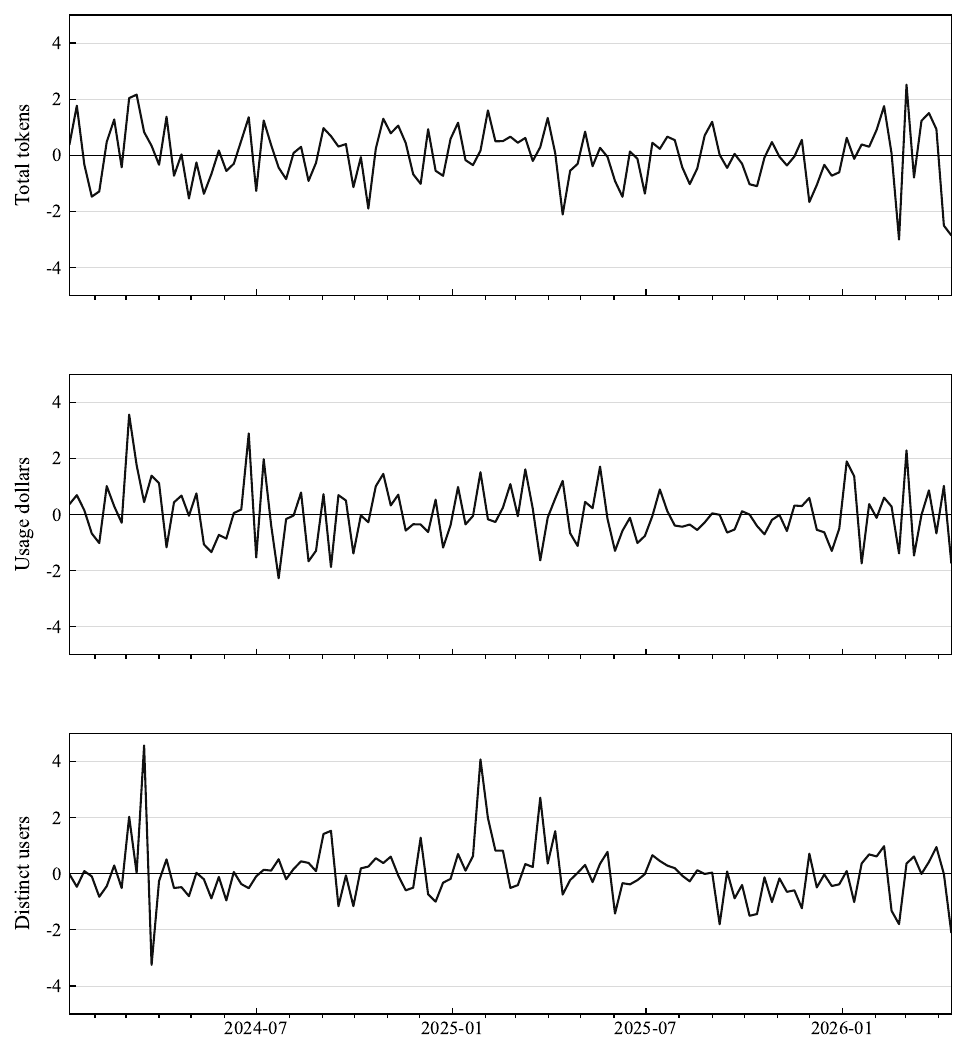}}
\caption{Standardized Weekly Natural-Log Growth of the Three AI-Factor Input Series}
\label{fig:pc3_growth}

\begin{minipage}{\textwidth}
\footnotesize
\textit{Note:} The figure plots the full-sample-standardized weekly natural-log growth of the three input components used to construct the headline AI factor: total tokens, usage dollars, and distinct users. The corresponding weekly levels are plotted in main-paper Figure~\ref{fig:pc3_levels}. Data is weekly from January 2024 through April 2026.
\end{minipage}
\end{figure}

%% file: tables_v22/table_ia_pca3_loadings.tex
\begin{table}[!htbp]
\centering
\small
\caption{PCA Loadings for Weekly Log Growth in AI Variables}
\label{tab:ia_pca3_loadings}
\begin{tabular}{lccc}
\toprule
 & Comp. 1 & Comp. 2 & Comp. 3 \\
\midrule

\\
$\Delta\ln Tok$    & 0.665 & $-$0.047 & $-$0.745 \\
$\Delta\ln Dol$     & 0.559 & $-$0.631 & 0.538 \\
$\Delta\ln User$  & 0.496 & 0.774 & 0.394 \\
\addlinespace[2pt]
Eigenvalue & 1.695 & 0.858 & 0.447 \\
\textit{Proportion} & \textit{0.565} & \textit{0.286} & \textit{0.149} \\
\textit{Cumulative} & \textit{0.565} & \textit{0.851} & \textit{1.000} \\
\\
\bottomrule
\end{tabular}

\begin{minipage}{\textwidth}
\footnotesize
\textit{Note:}
The table reports the loadings of each principal component on the three underlying input series $\Delta\ln Tok$, $\Delta\ln Dol$, and $\Delta\ln User$, where the inputs are the log growth in total tokens, usage dollars, and distinct active users, respectively. The table reports the weekly Monday-indexed decomposition used to construct the headline asset-pricing factor $AI_t$ (the first component of weekly inputs). The \textit{Proportion} and \textit{Cumulative} rows give the share of the input variance captured by each principal component. Loadings are in correlation form (eigenvectors of the input correlation matrix, with the sum of squared loadings within each component equal to 1). Data is weekly from January 2024 through April 2026.
\end{minipage}
\end{table}

%% file: figure_oa_decomposition_v18.tex
\begin{figure}[!t]
\centering
\begin{subfigure}{0.48\textwidth}
\centering
\includegraphics[width=\textwidth]{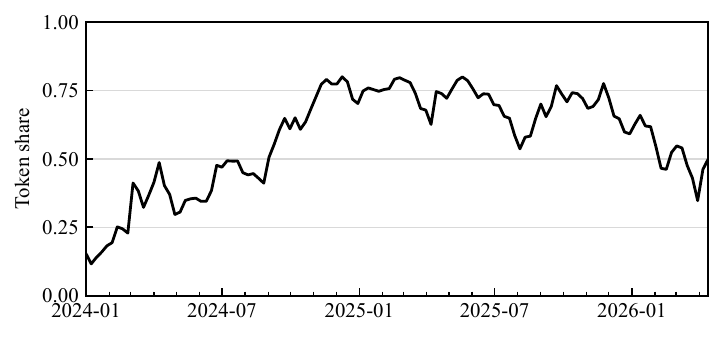}
\caption{Closed-Source Token Share}
\label{fig:sub_open_closed}
\end{subfigure}
\hfill
\begin{subfigure}{0.48\textwidth}
\centering
\includegraphics[width=\textwidth]{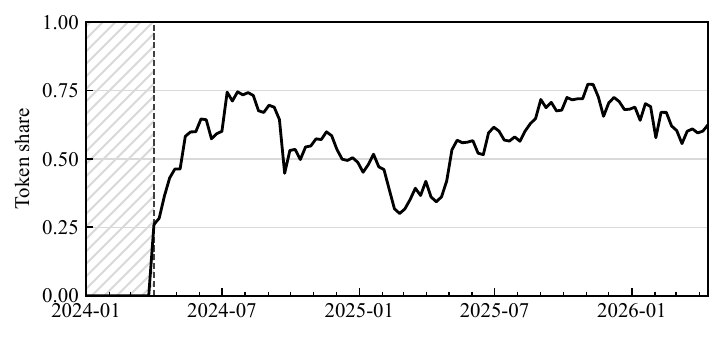}
\caption{Paid/Core Token Share}
\label{fig:sub_acct_paid}
\end{subfigure}
\\[0.6em]
\begin{subfigure}{0.48\textwidth}
\centering
\includegraphics[width=\textwidth]{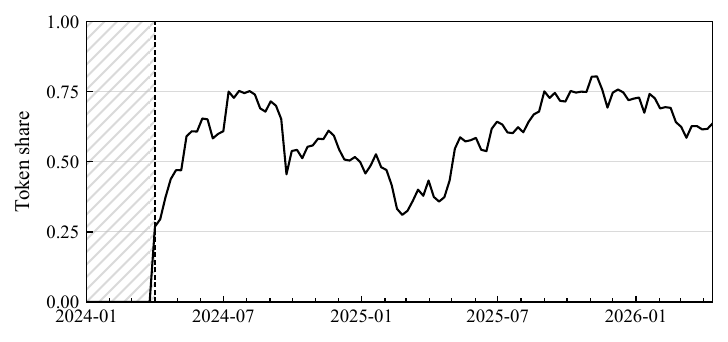}
\caption{Seasoned-User Token Share}
\label{fig:sub_seasoned}
\end{subfigure}
\hfill
\begin{subfigure}{0.48\textwidth}
\centering
\includegraphics[width=\textwidth]{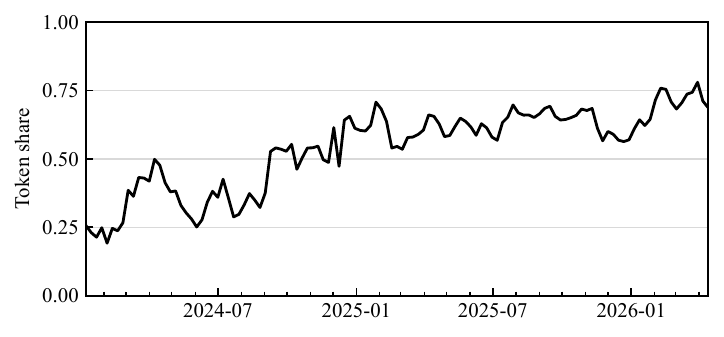}
\caption{Long-Prompt (Top 25\%) Token Share}
\label{fig:sub_prompt_length}
\end{subfigure}
\caption{Evolution of the AI Consumption Decompositions Used in Table~\ref{tab:subsamples_combined}}
\label{fig:decomposition}
\label{fig:subsample_evolution}

\begin{minipage}{\textwidth}
\footnotesize
\textit{Note:} Each panel plots, for one of the four splits used in Table~\ref{tab:subsamples_combined}, the weekly share of total tokens attributable to the focal leg of that split: Panel~A, closed-source AI models (Table~\ref{tab:subsamples_combined} Panel~A); Panel~B, paid/core accounts under the \texttt{acct\_paid} definition (Table~\ref{tab:subsamples_combined} Panel~B); Panel~C, seasoned users (Table~\ref{tab:subsamples_combined} Panel~C); Panel~D, the long-prompt segment defined as rows whose prompt-token count exceeds the prior-13-week rolling 75th percentile (Table~\ref{tab:subsamples_combined} Panel~D). Hatched regions and dashed vertical lines in Panels~B and C mark burn-in periods for the user classifications.
\end{minipage}
\end{figure}

%% file: tables_v22/table_us_components.tex
\begin{table}[!htbp]
\centering
\small
\caption{AI-Factor Input-Component Quintile Portfolios}
\label{tab:us_components}
\begin{tabular}{lcccccc}
\toprule
 & Low & 2 & 3 & 4 & High & H$-$L \\
\midrule
\multicolumn{7}{l}{\textit{Panel A. Sorted on $\beta^{Tok}$}} \\
Excess return & -0.111 & 0.207 & 0.245 & 0.081 & 0.383 & 0.494 \\
\textit{$t$} & (-0.46) & (1.00) & (1.27) & (0.40) & (1.35) & (2.31) \\
\addlinespace[2pt]
FF5 $\alpha$ & -0.243 & 0.005 & 0.039 & -0.098 & 0.143 & 0.386 \\
\textit{$t(\alpha)$} & (-1.94) & (0.07) & (0.57) & (-1.21) & (1.07) & (1.84) \\
\addlinespace[2pt]
FF5+Mom $\alpha$ & -0.207 & 0.021 & 0.030 & -0.103 & 0.128 & 0.335 \\
\textit{$t(\alpha)$} & (-1.73) & (0.28) & (0.44) & (-1.26) & (0.96) & (1.64) \\
\addlinespace[6pt]
\multicolumn{7}{l}{\textit{Panel B. Sorted on $\beta^{Dol}$}} \\
Excess return & 0.040 & 0.129 & 0.199 & 0.243 & 0.395 & 0.355 \\
\textit{$t$} & (0.17) & (0.61) & (1.06) & (1.14) & (1.40) & (1.46) \\
\addlinespace[2pt]
FF5 $\alpha$ & -0.092 & -0.028 & 0.008 & 0.031 & 0.172 & 0.264 \\
\textit{$t(\alpha)$} & (-0.69) & (-0.38) & (0.12) & (0.34) & (1.22) & (1.12) \\
\addlinespace[2pt]
FF5+Mom $\alpha$ & -0.104 & -0.018 & 0.016 & 0.027 & 0.163 & 0.267 \\
\textit{$t(\alpha)$} & (-0.77) & (-0.25) & (0.22) & (0.30) & (1.15) & (1.13) \\
\addlinespace[6pt]
\multicolumn{7}{l}{\textit{Panel C. Sorted on $\beta^{User}$}} \\
Excess return & -0.002 & 0.294 & 0.062 & 0.194 & 0.393 & 0.395 \\
\textit{$t$} & (-0.01) & (1.34) & (0.33) & (0.99) & (1.36) & (2.02) \\
\addlinespace[2pt]
FF5 $\alpha$ & -0.177 & 0.079 & -0.121 & 0.020 & 0.200 & 0.377 \\
\textit{$t(\alpha)$} & (-1.45) & (1.00) & (-1.75) & (0.23) & (1.43) & (1.88) \\
\addlinespace[2pt]
FF5+Mom $\alpha$ & -0.161 & 0.076 & -0.101 & 0.023 & 0.183 & 0.344 \\
\textit{$t(\alpha)$} & (-1.32) & (0.96) & (-1.53) & (0.26) & (1.31) & (1.73) \\
\bottomrule
\end{tabular}

\begin{minipage}{\textwidth}
\footnotesize
\textit{Note:}
The table decomposes the AI-factor result of Table~\ref{tab:us_portfolio_p5_p1} into its three input components. The AI factor is the first principal component of weekly log growth in total tokens, usage dollars, and distinct users. Here we replace the AI factor with each component individually: for $X \in \{Tok,Dol,User\}$, we estimate $r^{e}_{i,t} = a + b_{\mathrm{mkt}}\,r^{e}_{m,t} + \beta^{X}_{i,t}\,\Delta\ln X_{t}$ on a 13-week rolling window, then sort stocks into pooled quintiles by the estimated $\beta^{X}_{i,t}$. Returns and alphas are in percent per week, value-weighted by formation-week market capitalization, with all-stock breakpoints; FF5 and FF5+Mom denote alphas from \citet{famaFrench2015} and from that model augmented with \citet{carhart1997} momentum. AI consumption data is weekly from January 2024 through April 2026. The equity sample covers 108 formation weeks from March 2024 through March 2026.\end{minipage}
\end{table}

%% file: tables_v22/table_us_hml_ff_controls.tex
\begin{table}[!htbp]
\centering
\small
\caption{Fama-French Factor Regressions for the AI-Beta H$-$L Portfolio}
\label{tab:us_hml_ff_controls}
\begin{tabular}{lccc}
\toprule
 & FF3 & FF5 & FF5+Mom \\
\midrule
$\alpha$ & 0.540 & 0.545 & 0.526 \\
\textit{$t(\alpha)$} & (2.55) & (2.53) & (2.44) \\
\addlinespace[2pt]
Mkt-RF & 0.194 & 0.219 & 0.197 \\
\textit{$t$} & (1.82) & (1.95) & (1.72) \\
SMB & 0.070 & 0.128 & 0.165 \\
\textit{$t$} & (0.45) & (0.72) & (0.91) \\
HML & 0.219 & 0.205 & 0.194 \\
\textit{$t$} & (1.44) & (1.21) & (1.14) \\
\addlinespace[2pt]
RMW &  & 0.137 & 0.153 \\
\textit{$t$} &  & (0.64) & (0.71) \\
CMA &  & -0.085 & -0.047 \\
\textit{$t$} &  & (-0.41) & (-0.22) \\
\addlinespace[2pt]
Mom &  &  & 0.123 \\
\textit{$t$} &  &  & (0.99) \\
$R^2$ & 0.058 & 0.063 & 0.072 \\
\bottomrule
\end{tabular}

\begin{minipage}{\textwidth}
\footnotesize
\textit{Note:}
The dependent variable is the value-weighted all-stock-breakpoint H$-$L return from Table~\ref{tab:us_portfolio_p5_p1}. Returns and factors are in percent per week. FF3 is the Fama--French three-factor model (\citealp{famaFrench1993}). FF5 adds RMW and CMA (\citealp{famaFrench2015}). FF5+Mom additionally includes the momentum factor (\citealp{carhart1997}). The table reports OLS coefficient estimates, $t$-statistics in parentheses and $R^2$. AI consumption data is weekly from January 2024 through April 2026. The equity sample covers 108 formation weeks from March 2024 through March 2026.
\end{minipage}
\end{table}

%% file: tables_v22/table_saas_exposure_recent_v9.tex
\begin{table}[!htbp]
\centering
\small
\caption{SaaS Exposure to the AI Factor: Full Sample Versus Last Quarter}
\label{tab:saas_recent_exposure}
\begin{tabular}{lcc}
\toprule
 & Full sample & Last quarter \\
\midrule
\multicolumn{3}{l}{\textit{Value-weighted SaaS portfolio regression}} \\
$AI_t$ & 0.183 & 0.956 \\
\textit{$t$} & (0.83) & (3.10) \\
Mkt-RF & 1.152 & 0.867 \\
\textit{$t$} & (15.64) & (1.52) \\
$R^2$ & 0.565 & 0.154 \\
Avg. firms/week & 141.9 & 143.2 \\
\bottomrule
\end{tabular}

\begin{minipage}{\textwidth}
\footnotesize
\textit{Note:}
The table reports value-weighted portfolio regressions for public SaaS firms. The SaaS universe is the SaaSDB public-company list matched by ticker to the CRSP/firm-exposure sample. The dependent variable is the value-weighted SaaS portfolio excess log return, and the regressors are the baseline AI factor $AI_t$ and the weekly market excess return Mkt-RF. Standard errors use Newey-West five-week lags. The coefficient on $AI_t$ is multiplied by $100\times\sigma_w(AI)$, so the row reports the weekly return effect in percentage points for a one-standard-deviation increase in the baseline AI factor. AI consumption data is weekly from January 2024 through April 2026. Equity betas use 13-week rolling windows, so portfolio formation weeks begin in March 2024.
\end{minipage}
\end{table}

%% file: tables_v22/table_small_sample_robustness_v3.tex
\begin{table}[!htbp]
\centering
\small
\caption{Small-Sample Robustness of AI-Beta Portfolio}
\label{tab:ia_small_sample_robustness}
\begin{tabular}{@{}lrrrr@{}}
\toprule
\multicolumn{5}{@{}l}{\textit{Panel A. Baseline H$-$L: Block-Bootstrap Inference}} \\
 & Estimate & 2.5\% & 97.5\% & $t$ \\
Excess return & 0.604 & 0.241 & 0.966 & (2.85) \\
FF5 $\alpha$ & 0.545 & 0.158 & 0.923 & (2.53) \\
FF5+Mom $\alpha$ & 0.526 & 0.131 & 0.923 & (2.44) \\
\addlinespace[5pt]
\multicolumn{5}{@{}l}{\textit{Panel B. H$-$L Sorts on Shrunk AI Beta}} \\
 & VW All & VW NYSE & EW All & EW NYSE \\
Excess return & 0.462 & 0.420 & 0.256 & 0.246 \\
\textit{$t$} & (2.48) & (2.32) & (1.65) & (1.57) \\
\addlinespace[2pt]
FF5 $\alpha$ & 0.411 & 0.368 & 0.170 & 0.163 \\
\textit{$t(\alpha)$} & (2.18) & (2.02) & (1.13) & (1.07) \\
\addlinespace[2pt]
FF5+Mom $\alpha$ & 0.379 & 0.339 & 0.128 & 0.121 \\
\textit{$t(\alpha)$} & (2.03) & (1.87) & (0.89) & (0.82) \\
\bottomrule
\end{tabular}

\begin{minipage}{\textwidth}
\footnotesize
\textit{Note:}
The table examines the robustness of the baseline H$-$L portfolio in Table~\ref{tab:us_portfolio_p5_p1}. Panel A reports percentile confidence intervals from a circular moving-block bootstrap that resamples consecutive holding-week observations. Panel B repeats the quintile sort after replacing the raw rolling AI beta with an empirical-Bayes-shrunk beta. The shrinkage target is the formation-week cross-sectional mean beta, and the reliability weight uses the rolling-beta standard error and the formation-week cross-sectional signal variance. Returns and alphas are in percent per week; FF5 and FF5+Mom denote alphas from \citet{famaFrench2015} and from that model augmented with \citet{carhart1997} momentum. Breakpoint and weighting definitions match Table~\ref{tab:us_portfolio_p5_p1}. AI consumption data is weekly from January 2024 through April 2026. The equity sample covers 108 formation weeks from March 2024 through March 2026.
\end{minipage}
\end{table}

%% file: tables_v22/table_us_indadj.tex
\begin{table}[!htbp]
\centering
\small
\caption{Industry-Demeaned AI-Beta Quintile Portfolios}
\label{tab:us_indadj}
\begin{tabular}{lcccccc}
\toprule
 & Low & 2 & 3 & 4 & High & H$-$L \\
\midrule
\multicolumn{7}{l}{\textit{Panel A. Value-weighted, all-stock breakpoints}} \\
Excess return & -0.106 & 0.145 & 0.169 & 0.264 & 0.434 & 0.541 \\
\textit{$t$} & (-0.42) & (0.75) & (0.86) & (1.21) & (1.61) & (2.87) \\
\addlinespace[2pt]
FF5 $\alpha$ & -0.299 & -0.014 & -0.020 & 0.072 & 0.223 & 0.521 \\
\textit{$t(\alpha)$} & (-2.41) & (-0.20) & (-0.27) & (0.84) & (2.02) & (2.69) \\
\addlinespace[2pt]
FF5+Mom $\alpha$ & -0.280 & -0.011 & -0.012 & 0.064 & 0.214 & 0.495 \\
\textit{$t(\alpha)$} & (-2.27) & (-0.15) & (-0.16) & (0.74) & (1.94) & (2.56) \\
\addlinespace[4pt]
\multicolumn{7}{l}{\textit{Panel B. Value-weighted, NYSE breakpoints}} \\
Excess return & -0.098 & 0.146 & 0.182 & 0.280 & 0.436 & 0.534 \\
\textit{$t$} & (-0.41) & (0.76) & (0.90) & (1.35) & (1.65) & (3.08) \\
\addlinespace[2pt]
FF5 $\alpha$ & -0.269 & -0.016 & -0.018 & 0.095 & 0.225 & 0.494 \\
\textit{$t(\alpha)$} & (-2.52) & (-0.21) & (-0.23) & (1.16) & (2.12) & (2.79) \\
\addlinespace[2pt]
FF5+Mom $\alpha$ & -0.252 & -0.011 & -0.009 & 0.090 & 0.215 & 0.468 \\
\textit{$t(\alpha)$} & (-2.38) & (-0.14) & (-0.12) & (1.09) & (2.02) & (2.65) \\
\addlinespace[4pt]
\multicolumn{7}{l}{\textit{Panel C. Equal-weighted, all-stock breakpoints}} \\
Excess return & -0.082 & 0.127 & 0.153 & 0.164 & 0.134 & 0.215 \\
\textit{$t$} & (-0.29) & (0.59) & (0.75) & (0.72) & (0.48) & (1.61) \\
\addlinespace[2pt]
FF5 $\alpha$ & -0.215 & -0.016 & 0.002 & -0.020 & -0.047 & 0.168 \\
\textit{$t(\alpha)$} & (-2.33) & (-0.37) & (0.07) & (-0.44) & (-0.55) & (1.22) \\
\addlinespace[2pt]
FF5+Mom $\alpha$ & -0.190 & -0.004 & 0.008 & -0.016 & -0.052 & 0.138 \\
\textit{$t(\alpha)$} & (-2.14) & (-0.10) & (0.23) & (-0.35) & (-0.60) & (1.03) \\
\addlinespace[4pt]
\multicolumn{7}{l}{\textit{Panel D. Equal-weighted, NYSE breakpoints}} \\
Excess return & -0.068 & 0.156 & 0.170 & 0.135 & 0.155 & 0.223 \\
\textit{$t$} & (-0.25) & (0.74) & (0.82) & (0.60) & (0.57) & (1.83) \\
\addlinespace[2pt]
FF5 $\alpha$ & -0.202 & 0.015 & 0.013 & -0.044 & -0.024 & 0.178 \\
\textit{$t(\alpha)$} & (-2.43) & (0.35) & (0.36) & (-0.95) & (-0.31) & (1.43) \\
\addlinespace[2pt]
FF5+Mom $\alpha$ & -0.177 & 0.026 & 0.017 & -0.038 & -0.028 & 0.150 \\
\textit{$t(\alpha)$} & (-2.25) & (0.64) & (0.47) & (-0.82) & (-0.35) & (1.23) \\
\bottomrule
\end{tabular}

\begin{minipage}{\textwidth}
\footnotesize
\textit{Note:}
The table repeats the headline test of Table~\ref{tab:us_portfolio_p5_p1} after demeaning the sorting variable by FF30 industry. For each formation week $t$ and FF30 industry $j$ we compute the cross-sectional median AI beta $\widetilde{\beta}^{AI,(j)}_{t} = \mathrm{median}_{i \in j}(\beta^{AI}_{i,t})$ across all firms in industry $j$ that week, requiring at least five eligible firms per industry-week. The industry-demeaned beta is $\beta^{AI,Ind}_{i,t} = \beta^{AI}_{i,t} - \widetilde{\beta}^{AI,(j(i))}_{t}$, and stocks are sorted into quintiles on this within-industry deviation. Industry assignment follows the Kenneth French convention: the SIC code from the most recent Compustat fiscal year ending in calendar year $t-1$ (CRSP June-of-$t$ SICCD as fallback) is mapped to one of the 30 Fama-French industries via French's \texttt{Siccodes30}. Returns and alphas are in percent per week; FF5 and FF5+Mom denote alphas from \citet{famaFrench2015} and from that model augmented with \citet{carhart1997} momentum. AI consumption data is weekly from January 2024 through April 2026. The equity sample covers 108 formation weeks from March 2024 through March 2026.
\end{minipage}
\end{table}

%% file: tables_v22/table_ia_gtrend_controlled.tex
\begin{table}[!htbp]
\centering
\small
\caption{Google-Trends-Controlled AI-Beta Quintile Portfolios}
\label{tab:ia_gtrend_controlled}
\begin{tabular}{lcccccc}
\toprule
 & Low & 2 & 3 & 4 & High & H$-$L \\
\midrule
\multicolumn{7}{l}{\textit{Panel A. Value-weighted, all-stock breakpoints}} \\
Excess return & -0.118 & 0.104 & 0.164 & 0.393 & 0.573 & 0.691 \\
\textit{$t$} & (-0.47) & (0.49) & (0.82) & (1.84) & (2.09) & (3.34) \\
\addlinespace[2pt]
FF5 $\alpha$ & -0.324 & -0.085 & -0.050 & 0.172 & 0.309 & 0.633 \\
\textit{$t(\alpha)$} & (-2.57) & (-0.92) & (-0.69) & (1.93) & (2.38) & (3.01) \\
\addlinespace[2pt]
FF5+Mom $\alpha$ & -0.312 & -0.070 & -0.051 & 0.165 & 0.293 & 0.605 \\
\textit{$t(\alpha)$} & (-2.50) & (-0.79) & (-0.70) & (1.87) & (2.31) & (2.97) \\
\addlinespace[4pt]
\multicolumn{7}{l}{\textit{Panel B. Value-weighted, NYSE breakpoints}} \\
Excess return & -0.043 & 0.140 & 0.147 & 0.420 & 0.602 & 0.644 \\
\textit{$t$} & (-0.18) & (0.65) & (0.73) & (2.06) & (2.16) & (3.15) \\
\addlinespace[2pt]
FF5 $\alpha$ & -0.224 & -0.068 & -0.060 & 0.187 & 0.353 & 0.577 \\
\textit{$t(\alpha)$} & (-1.96) & (-0.76) & (-0.71) & (2.33) & (2.78) & (2.82) \\
\addlinespace[2pt]
FF5+Mom $\alpha$ & -0.206 & -0.063 & -0.061 & 0.182 & 0.338 & 0.544 \\
\textit{$t(\alpha)$} & (-1.89) & (-0.71) & (-0.72) & (2.27) & (2.72) & (2.79) \\
\addlinespace[4pt]
\multicolumn{7}{l}{\textit{Panel C. Equal-weighted, all-stock breakpoints}} \\
Excess return & -0.111 & 0.140 & 0.168 & 0.216 & 0.219 & 0.331 \\
\textit{$t$} & (-0.38) & (0.62) & (0.84) & (0.94) & (0.77) & (2.07) \\
\addlinespace[2pt]
FF5 $\alpha$ & -0.240 & -0.015 & 0.018 & 0.043 & 0.032 & 0.272 \\
\textit{$t(\alpha)$} & (-2.30) & (-0.30) & (0.49) & (0.77) & (0.34) & (1.72) \\
\addlinespace[2pt]
FF5+Mom $\alpha$ & -0.225 & -0.006 & 0.020 & 0.041 & 0.025 & 0.249 \\
\textit{$t(\alpha)$} & (-2.24) & (-0.12) & (0.55) & (0.73) & (0.26) & (1.64) \\
\addlinespace[4pt]
\multicolumn{7}{l}{\textit{Panel D. Equal-weighted, NYSE breakpoints}} \\
Excess return & -0.062 & 0.131 & 0.199 & 0.211 & 0.218 & 0.280 \\
\textit{$t$} & (-0.22) & (0.61) & (0.98) & (0.94) & (0.77) & (1.95) \\
\addlinespace[2pt]
FF5 $\alpha$ & -0.193 & -0.023 & 0.050 & 0.041 & 0.034 & 0.226 \\
\textit{$t(\alpha)$} & (-2.18) & (-0.50) & (1.28) & (0.75) & (0.38) & (1.59) \\
\addlinespace[2pt]
FF5+Mom $\alpha$ & -0.179 & -0.016 & 0.051 & 0.040 & 0.026 & 0.205 \\
\textit{$t(\alpha)$} & (-2.12) & (-0.35) & (1.30) & (0.74) & (0.30) & (1.50) \\
\bottomrule
\end{tabular}

\begin{minipage}{\textwidth}
\footnotesize
\textit{Note:}
The table extends the analysis of the properties of the AI-Beta quintile portfolios presented in Table~\ref{tab:us_portfolio_p5_p1} after expanding the first-pass rolling beta regression to include the weekly log growth in Google Trends search interest for the topic \textit{Artificial intelligence} as an additional control: $r^{e}_{i,t} = a + b_{\mathrm{mkt}}\,r^{e}_{m,t} + b_{\mathrm{gtrend}}\,\Delta\ln \mathrm{gtrend}^{\mathrm{AI}}_{t} + b_{\mathrm{AI}}\,AI_t$. Stocks are sorted into pooled quintiles by the Google-Trends-controlled $b_{\mathrm{AI}}$. Returns and alphas are in percent per week; FF5 and FF5+Mom denote alphas from \citet{famaFrench2015} and from that model augmented with \citet{carhart1997} momentum. The Google Trends weekly index covers 2023-10 to 2026-02 and is fetched via the Knowledge-Graph topic API. AI consumption data is weekly from January 2024 through April 2026. The equity sample covers 108 formation weeks from March 2024 through March 2026.
\end{minipage}
\end{table}

%% file: tables_v22/table_sec_attention_spanning_v9.tex
\begin{table}[!htbp]
\centering
\caption{Spanning Regressions with SEC-Based AI Mentions}
\label{tab:ia_sec_ai_mention_spanning}
\begingroup
\setlength{\tabcolsep}{4pt}
\renewcommand{\arraystretch}{0.94}
\scriptsize
\begin{tabular}{llccccc}
\toprule
 & Model & $\alpha$ & $t(\alpha)$ & $b_{\mathrm{Mention}}$ & $t(b_{\mathrm{Mention}})$ & $R^{2}$ \\
\midrule
\multicolumn{7}{l}{\textit{Panel A. Controlling for unscaled SEC-based AI mentions}} \\
(1) & Mean & 0.604 & (2.85) &  &  &  \\
(2) & SEC-Based AI Mention & 0.621 & (2.91) & -0.074 & (-0.73) & 0.005 \\
(3) & FF5 & 0.545 & (2.53) &  &  & 0.062 \\
(4) & FF5 + SEC-Based AI Mention & 0.552 & (2.52) & -0.037 & (-0.20) & 0.063 \\
(5) & FF5 + Mom & 0.526 & (2.44) &  &  & 0.071 \\
(6) & FF5 + Mom + SEC-Based AI Mention & 0.541 & (2.46) & -0.083 & (-0.43) & 0.073 \\
\addlinespace[4pt]
\multicolumn{7}{l}{\textit{Panel B. Controlling for word-scaled SEC-based AI mentions}} \\
(1) & Mean & 0.604 & (2.85) &  &  &  \\
(2) & SEC-Based AI Mention & 0.607 & (2.85) & -0.053 & (-0.46) & 0.002 \\
(3) & FF5 & 0.545 & (2.53) &  &  & 0.062 \\
(4) & FF5 + SEC-Based AI Mention & 0.544 & (2.52) & 0.039 & (0.17) & 0.063 \\
(5) & FF5 + Mom & 0.526 & (2.44) &  &  & 0.071 \\
(6) & FF5 + Mom + SEC-Based AI Mention & 0.525 & (2.42) & -0.031 & (-0.13) & 0.072 \\
\bottomrule
\end{tabular}
\endgroup

\begin{minipage}{\textwidth}
\footnotesize
\textit{Note:}
The table reports time-series spanning regressions of the baseline AI-beta H$-$L portfolio return on SEC-based AI-mention H$-$L portfolios. The dependent variable is the value-weighted H$-$L return from sorting stocks on the baseline rolling AI beta used in Table~\ref{tab:us_portfolio_p5_p1}, using all-stock breakpoints. The SEC-based AI-mention controls are constructed from the completed SEC filing text archive. We map firms to SEC CIKs using the latest non-missing CIK from Compustat quarterly data and use 10-K, 10-Q, and 8-K filing text. AI mentions are exact matches to \textit{artificial intelligence}, \textit{generative AI}, \textit{machine learning}, \textit{deep learning}, \textit{large language model}, \textit{large language models}, \textit{LLM}, \textit{OpenAI}, and \textit{AI model}. For firm $i$ in formation week $t$, let $H_{i,t}$ be the number of AI phrase mentions over weeks $t-13$ through $t-1$, let $F_{i,t}$ be the number of filings over the same lagged window, and let $W_{i,t}$ be the total number of words in those filings. Panel A sorts on the unscaled measure $\ln(1+H_{i,t})-\ln(1+F_{i,t})$. Panel B sorts on the word-scaled measure $10{,}000\times H_{i,t}/W_{i,t}$. In both panels, the SEC-based AI-mention H$-$L control buys the high-mention quintile and shorts the low-mention quintile among firms with at least one filing in the lagged window. The reported regression is $R^{AI,H-L}_{t}=\alpha+b_{\mathrm{Mention}}R^{SECMention,H-L}_{t}+\delta'X_t+\epsilon_t$, where $X_t$ is empty in row (2), contains the Fama-French five factors (\citealp{famaFrench2015}) in row (4), and contains the five factors plus \citet{carhart1997} momentum in row (6). Returns and factors are in percent per week. AI consumption data is weekly from January 2024 through April 2026. The equity sample covers 108 formation weeks from March 2024 through March 2026.
\end{minipage}
\end{table}

%% file: tables_v22/table_ia_avgstd_baseline.tex
\begin{table}[!htbp]
\centering
\small
\caption{Average-Standardized Usage-Growth Quintile Portfolios}
\label{tab:ia_avgstd_baseline}
\begin{tabular}{lcccccc}
\toprule
 & Low & 2 & 3 & 4 & High & H$-$L \\
\midrule
\multicolumn{7}{l}{\textit{Panel A. Value-weighted, all-stock breakpoints}} \\
Excess return & -0.141 & 0.163 & 0.166 & 0.311 & 0.444 & 0.585 \\
\textit{$t$} & (-0.58) & (0.79) & (0.84) & (1.48) & (1.64) & (2.70) \\
\addlinespace[2pt]
FF5 $\alpha$ & -0.293 & -0.001 & -0.028 & 0.109 & 0.229 & 0.523 \\
\textit{$t(\alpha)$} & (-2.22) & (-0.01) & (-0.40) & (1.38) & (1.74) & (2.38) \\
\addlinespace[2pt]
FF5+Mom $\alpha$ & -0.284 & 0.015 & -0.029 & 0.103 & 0.222 & 0.506 \\
\textit{$t(\alpha)$} & (-2.13) & (0.19) & (-0.40) & (1.29) & (1.68) & (2.29) \\
\addlinespace[4pt]
\multicolumn{7}{l}{\textit{Panel B. Value-weighted, NYSE breakpoints}} \\
Excess return & -0.032 & 0.089 & 0.228 & 0.267 & 0.466 & 0.498 \\
\textit{$t$} & (-0.14) & (0.43) & (1.16) & (1.26) & (1.78) & (2.49) \\
\addlinespace[2pt]
FF5 $\alpha$ & -0.188 & -0.079 & 0.042 & 0.053 & 0.252 & 0.440 \\
\textit{$t(\alpha)$} & (-1.56) & (-0.93) & (0.56) & (0.65) & (2.02) & (2.16) \\
\addlinespace[2pt]
FF5+Mom $\alpha$ & -0.171 & -0.072 & 0.043 & 0.045 & 0.248 & 0.419 \\
\textit{$t(\alpha)$} & (-1.42) & (-0.84) & (0.56) & (0.55) & (1.97) & (2.06) \\
\addlinespace[4pt]
\multicolumn{7}{l}{\textit{Panel C. Equal-weighted, all-stock breakpoints}} \\
Excess return & -0.108 & 0.095 & 0.171 & 0.162 & 0.179 & 0.286 \\
\textit{$t$} & (-0.37) & (0.44) & (0.88) & (0.72) & (0.64) & (1.84) \\
\addlinespace[2pt]
FF5 $\alpha$ & -0.227 & -0.049 & 0.024 & -0.021 & -0.008 & 0.219 \\
\textit{$t(\alpha)$} & (-2.22) & (-0.94) & (0.64) & (-0.41) & (-0.08) & (1.41) \\
\addlinespace[2pt]
FF5+Mom $\alpha$ & -0.205 & -0.029 & 0.027 & -0.020 & -0.017 & 0.187 \\
\textit{$t(\alpha)$} & (-2.05) & (-0.61) & (0.73) & (-0.39) & (-0.19) & (1.23) \\
\addlinespace[4pt]
\multicolumn{7}{l}{\textit{Panel D. Equal-weighted, NYSE breakpoints}} \\
Excess return & -0.057 & 0.096 & 0.166 & 0.173 & 0.179 & 0.236 \\
\textit{$t$} & (-0.20) & (0.46) & (0.84) & (0.78) & (0.65) & (1.65) \\
\addlinespace[2pt]
FF5 $\alpha$ & -0.180 & -0.047 & 0.016 & -0.004 & -0.008 & 0.172 \\
\textit{$t(\alpha)$} & (-1.95) & (-0.99) & (0.42) & (-0.07) & (-0.09) & (1.20) \\
\addlinespace[2pt]
FF5+Mom $\alpha$ & -0.156 & -0.031 & 0.018 & -0.002 & -0.016 & 0.140 \\
\textit{$t(\alpha)$} & (-1.76) & (-0.69) & (0.47) & (-0.04) & (-0.18) & (1.00) \\
\bottomrule
\end{tabular}

\begin{minipage}{\textwidth}
\footnotesize
\textit{Note:}
The table presents robustness evidence on the properties of the AI-Beta quintile portfolios of Table~\ref{tab:us_portfolio_p5_p1} after replacing the AI factor with a simple equal-weighted average of the three full-sample-standardized input series: $\Delta\widetilde{\ln}\,X_{t} = 1/3\bigl[z(\Delta\ln Tok_{t}) + z(\Delta\ln Dol_{t}) + z(\Delta\ln User_{t})\bigr]$, where $z(\cdot)$ denotes full-sample standardization (subtract sample mean, divide by sample standard deviation). This construction partly addresses the forward-looking concern that principal-component weights are estimated from the full sample, although it does not eliminate full-sample information because the input series are still standardized using their full-sample means and standard deviations. The first-pass rolling regression is $r^{e}_{i,t} = a + b_{\mathrm{mkt}}\,r^{e}_{m,t} + b_{\mathrm{avg}}\,\Delta\widetilde{\ln}\,X_{t}$ on a 13-week rolling window. Stocks are sorted into quintiles by $b_{\mathrm{avg}}$. The investment universe, weighting, and breakpoint definitions match Table~\ref{tab:us_portfolio_p5_p1}. Returns and alphas are in percent per week. AI consumption data is weekly from January 2024 through April 2026. The equity sample covers 108 formation weeks from March 2024 through March 2026.
\end{minipage}
\end{table}

%% file: tables_v22/table_ia_agentic_nonagentic_ttok.tex
\begin{table}[!htbp]
\centering
\small
\caption{Agentic and Non-Agentic Token-Growth Quintile Portfolios}
\label{tab:ia_agentic_nonagentic_ttok}
\begin{tabular}{lcccccc}
\toprule
 & Low & 2 & 3 & 4 & High & H$-$L \\
\midrule
\multicolumn{7}{l}{\textit{Value-weighted, all-stock breakpoints}} \\
\multicolumn{7}{l}{\quad\textit{$\beta^{Tok,Agentic}$}} \\
Excess return & 0.275 & 0.166 & 0.120 & 0.000 & 0.299 & 0.024 \\
\textit{$t$} & (0.81) & (0.62) & (0.41) & (0.00) & (0.62) & (0.06) \\
\addlinespace[2pt]
FF5 $\alpha$ & -0.056 & -0.065 & -0.102 & -0.269 & 0.006 & 0.061 \\
\textit{$t(\alpha)$} & (-0.31) & (-0.58) & (-0.93) & (-2.57) & (0.03) & (0.22) \\
\addlinespace[2pt]
FF5+Mom $\alpha$ & 0.003 & -0.078 & -0.097 & -0.291 & -0.021 & -0.024 \\
\textit{$t(\alpha)$} & (0.01) & (-0.68) & (-0.87) & (-2.75) & (-0.12) & (-0.09) \\
\addlinespace[6pt]
\multicolumn{7}{l}{\quad\textit{$\beta^{Tok,Nonagentic}$}} \\
Excess return & -0.118 & 0.272 & 0.154 & 0.163 & 0.343 & 0.461 \\
\textit{$t$} & (-0.48) & (1.30) & (0.80) & (0.79) & (1.25) & (2.22) \\
\addlinespace[2pt]
FF5 $\alpha$ & -0.251 & 0.081 & -0.061 & -0.027 & 0.110 & 0.361 \\
\textit{$t(\alpha)$} & (-2.05) & (0.99) & (-0.88) & (-0.32) & (0.88) & (1.77) \\
\addlinespace[2pt]
FF5+Mom $\alpha$ & -0.221 & 0.102 & -0.065 & -0.031 & 0.093 & 0.313 \\
\textit{$t(\alpha)$} & (-1.86) & (1.29) & (-0.93) & (-0.37) & (0.75) & (1.57) \\
\bottomrule
\end{tabular}

\begin{minipage}{\textwidth}
\footnotesize
\textit{Note:}
The table repeats the value-weighted, all-stock-breakpoint portfolio test of Table~\ref{tab:us_portfolio_p5_p1} after replacing the baseline AI factor with weekly log growth in agentic or non-agentic total tokens. Total tokens are prompt tokens plus completion tokens. Agentic total tokens are total tokens from requests whose normalized finish reason is \texttt{tool\_calls}; non-agentic total tokens are the remaining total tokens. Firm betas are estimated from a 13-week rolling regression of weekly log-excess returns on the log-excess market return and the token-growth factor. Stocks are sorted into quintiles by this beta. Returns and alphas are in percent per week; FF5 and FF5+Mom denote alphas from \citet{famaFrench2015} and from that model augmented with \citet{carhart1997} momentum. Data is weekly from January 2024 through April 2026.
\end{minipage}
\end{table}

%% file: tables_v22/table_ia_agentic_pricing_v15.tex
\begin{table}[H]
\centering
\footnotesize
\caption{Cross-Sectional Pricing of Agentic Factors with Fama--French \citep{famaFrench1993,famaFrench2015} and Momentum \citep{carhart1997} Adjustments}
\label{tab:agentic_pricing}
\begin{tabular}{@{}l cc cc cc cc@{}}
\toprule
 & \multicolumn{2}{c}{Raw} & \multicolumn{2}{c}{FF3} & \multicolumn{2}{c}{FF5} & \multicolumn{2}{c}{FF5+Mom} \\
\cmidrule(lr){2-3}\cmidrule(lr){4-5}\cmidrule(lr){6-7}\cmidrule(lr){8-9}
Factor & $\alpha$ & $t$ & $\alpha$ & $t$ & $\alpha$ & $t$ & $\alpha$ & $t$ \\
\midrule
Agentic tokens & 0.024 & (0.06) & 0.072 & (0.23) & 0.061 & (0.22) & -0.024 & (-0.09) \\
Cache-read tokens & 0.283 & (0.75) & 0.155 & (0.39) & 0.340 & (0.80) & 0.079 & (0.23) \\
Reasoning tokens & 0.301 & (1.35) & 0.275 & (1.21) & 0.270 & (1.15) & 0.251 & (1.11) \\
Agentic dollars & 0.095 & (0.27) & 0.113 & (0.43) & 0.078 & (0.30) & 0.100 & (0.37) \\
\bottomrule
\end{tabular}

\begin{minipage}{\textwidth}
\footnotesize
\textit{Note:}
The table reports value-weighted high-minus-low quintile spreads, in percent per week, for salient components of agentic AI token consumption factors; for details see Table~\ref{tab:agentic_defs} in the Internet Appendix. Firm betas are estimated from a 13-week rolling regression of weekly log-excess returns on the log-excess market return and the factor; stocks are sorted into quintiles by this beta. Columns report the raw spread and alphas from the Fama--French three-factor model (FF3), the Fama--French five-factor model (FF5), and the five-factor model augmented with momentum (FF5+Mom), all value-weighted under all-stock breakpoints. $t$-statistics in parentheses.
\end{minipage}
\end{table}

%% file: tables_v22/table_ia_baseline_decile.tex
\begin{landscape}
\begin{table}[!htbp]
\centering
\small
\caption{AI-Beta Decile Portfolios}
\label{tab:ia_baseline_decile}
\begin{tabular}{lccccccccccc}
\toprule
 & Low & 2 & 3 & 4 & 5 & 6 & 7 & 8 & 9 & High & H$-$L \\
\midrule
\multicolumn{12}{l}{\textit{Panel A. Value-weighted, all-stock breakpoints}} \\
Excess return & -0.028 & -0.172 & 0.295 & 0.020 & 0.223 & 0.128 & 0.229 & 0.342 & 0.271 & 0.666 & 0.694 \\
\textit{$t$} & (-0.09) & (-0.74) & (1.43) & (0.09) & (1.10) & (0.59) & (1.03) & (1.54) & (1.07) & (1.97) & (2.37) \\
\addlinespace[2pt]
FF5 $\alpha$ & -0.183 & -0.335 & 0.130 & -0.132 & 0.060 & -0.099 & 0.034 & 0.119 & 0.057 & 0.422 & 0.605 \\
\textit{$t(\alpha)$} & (-0.88) & (-2.32) & (1.18) & (-1.14) & (0.58) & (-1.03) & (0.32) & (1.13) & (0.42) & (2.17) & (2.03) \\
\addlinespace[2pt]
FF5+Mom $\alpha$ & -0.157 & -0.323 & 0.161 & -0.133 & 0.066 & -0.102 & 0.026 & 0.113 & 0.069 & 0.381 & 0.538 \\
\textit{$t(\alpha)$} & (-0.75) & (-2.23) & (1.54) & (-1.14) & (0.63) & (-1.06) & (0.24) & (1.06) & (0.50) & (2.00) & (1.85) \\
\addlinespace[4pt]
\multicolumn{12}{l}{\textit{Panel B. Value-weighted, NYSE breakpoints}} \\
Excess return & -0.107 & 0.059 & 0.068 & 0.129 & 0.305 & 0.163 & 0.235 & 0.319 & 0.260 & 0.576 & 0.682 \\
\textit{$t$} & (-0.40) & (0.26) & (0.31) & (0.59) & (1.57) & (0.75) & (1.12) & (1.33) & (1.09) & (1.75) & (2.58) \\
\addlinespace[2pt]
FF5 $\alpha$ & -0.239 & -0.109 & -0.096 & -0.019 & 0.125 & -0.057 & 0.016 & 0.109 & 0.074 & 0.332 & 0.571 \\
\textit{$t(\alpha)$} & (-1.47) & (-0.79) & (-0.85) & (-0.16) & (1.23) & (-0.58) & (0.17) & (0.87) & (0.56) & (1.79) & (2.16) \\
\addlinespace[2pt]
FF5+Mom $\alpha$ & -0.222 & -0.078 & -0.088 & -0.010 & 0.127 & -0.059 & 0.013 & 0.091 & 0.088 & 0.302 & 0.524 \\
\textit{$t(\alpha)$} & (-1.36) & (-0.58) & (-0.78) & (-0.09) & (1.24) & (-0.59) & (0.13) & (0.73) & (0.66) & (1.64) & (2.00) \\
\addlinespace[4pt]
\multicolumn{12}{l}{\textit{Panel C. Equal-weighted, all-stock breakpoints}} \\
Excess return & -0.210 & -0.017 & 0.072 & 0.123 & 0.151 & 0.195 & 0.154 & 0.171 & 0.167 & 0.192 & 0.402 \\
\textit{$t$} & (-0.62) & (-0.07) & (0.32) & (0.58) & (0.76) & (1.00) & (0.73) & (0.72) & (0.64) & (0.62) & (1.93) \\
\addlinespace[2pt]
FF5 $\alpha$ & -0.309 & -0.154 & -0.072 & -0.029 & 0.016 & 0.047 & -0.022 & -0.020 & -0.021 & 0.000 & 0.309 \\
\textit{$t(\alpha)$} & (-2.10) & (-1.99) & (-1.12) & (-0.54) & (0.32) & (1.09) & (-0.37) & (-0.34) & (-0.29) & (0.00) & (1.49) \\
\addlinespace[2pt]
FF5+Mom $\alpha$ & -0.286 & -0.129 & -0.049 & -0.015 & 0.027 & 0.048 & -0.025 & -0.020 & -0.021 & -0.017 & 0.269 \\
\textit{$t(\alpha)$} & (-1.96) & (-1.78) & (-0.84) & (-0.29) & (0.55) & (1.09) & (-0.44) & (-0.33) & (-0.29) & (-0.13) & (1.32) \\
\addlinespace[4pt]
\multicolumn{12}{l}{\textit{Panel D. Equal-weighted, NYSE breakpoints}} \\
Excess return & -0.158 & 0.048 & 0.122 & 0.125 & 0.156 & 0.191 & 0.178 & 0.162 & 0.166 & 0.183 & 0.341 \\
\textit{$t$} & (-0.51) & (0.20) & (0.55) & (0.62) & (0.78) & (0.95) & (0.85) & (0.68) & (0.66) & (0.61) & (1.93) \\
\addlinespace[2pt]
FF5 $\alpha$ & -0.272 & -0.088 & -0.024 & -0.023 & 0.013 & 0.043 & 0.016 & -0.026 & -0.031 & -0.000 & 0.272 \\
\textit{$t(\alpha)$} & (-2.32) & (-1.16) & (-0.42) & (-0.48) & (0.29) & (0.87) & (0.27) & (-0.43) & (-0.46) & (-0.00) & (1.52) \\
\addlinespace[2pt]
FF5+Mom $\alpha$ & -0.251 & -0.061 & -0.007 & -0.008 & 0.019 & 0.041 & 0.013 & -0.022 & -0.032 & -0.014 & 0.236 \\
\textit{$t(\alpha)$} & (-2.17) & (-0.87) & (-0.12) & (-0.17) & (0.41) & (0.83) & (0.22) & (-0.36) & (-0.48) & (-0.12) & (1.34) \\
\bottomrule
\end{tabular}

\begin{minipage}{\linewidth}
\footnotesize
\textit{Note:}
The table repeats Table~\ref{tab:us_portfolio_p5_p1} with decile, rather than quintile, sorts on 13-week rolling AI beta. Universe, weighting, and breakpoints match Table~\ref{tab:us_portfolio_p5_p1}. Returns and alphas are in percent per week; FF5 and FF5+Mom denote alphas from \citet{famaFrench2015} and that model augmented with \citet{carhart1997} momentum. H$-$L buys High and shorts Low. AI consumption data is weekly from January 2024 through April 2026. The equity sample covers 108 formation weeks from March 2024 through March 2026.
\end{minipage}
\end{table}
\end{landscape}

%% file: tables_v22/table_ia_aietf_spanning.tex
\begin{table}[!htbp]
\centering
\caption{Spanning Regressions for the AI-Beta H$-$L Portfolio with High-Tech Returns}
\label{tab:ia_aietf_spanning}
\begingroup
\setlength{\tabcolsep}{4pt}
\footnotesize
\begin{tabular}{llccccc}
\toprule
 &  & $\alpha$ & $t(\alpha)$ & $b_{\mathrm{Add}}$ & $t(b_{\mathrm{Add}})$ & $\bar{R}^{2}$ \\
\midrule
\multicolumn{7}{l}{\textit{Panel A. FF10 HiTec return as additional control}} \\
\multicolumn{7}{l}{\quad\textit{Value-weighted, all-stock breakpoints}} \\
(1) & HiTec & 0.595 & (2.79) & 0.038 & (0.56) & 0.003  \\
(2) & Mkt + HiTec & 0.558 & (2.70) & -0.454 & (-2.48) & 0.076  \\
(3) & FF5 + HiTec & 0.554 & (2.59) & -0.450 & (-1.43) & 0.081  \\
(4) & FF5 + Mom + HiTec & 0.532 & (2.48) & -0.499 & (-1.57) & 0.094  \\
\addlinespace[3pt]
\multicolumn{7}{l}{\quad\textit{Value-weighted, NYSE breakpoints}} \\
(1) & HiTec & 0.454 & (2.23) & 0.031 & (0.47) & 0.002  \\
(2) & Mkt + HiTec & 0.414 & (2.12) & -0.485 & (-2.79) & 0.090  \\
(3) & FF5 + HiTec & 0.418 & (2.06) & -0.573 & (-1.91) & 0.094  \\
(4) & FF5 + Mom + HiTec & 0.391 & (1.93) & -0.634 & (-2.11) & 0.115  \\
\addlinespace[6pt]
\multicolumn{7}{l}{\textit{Panel B. NAICS semiconductor return as additional control}} \\
\multicolumn{7}{l}{\quad\textit{Value-weighted, all-stock breakpoints}} \\
(1) & Semi & 0.600 & (2.78) & 0.004 & (0.09) & 0.000  \\
(2) & Mkt + Semi & 0.603 & (2.86) & -0.132 & (-1.96) & 0.057  \\
(3) & FF5 + Semi & 0.574 & (2.63) & -0.085 & (-1.05) & 0.073  \\
(4) & FF5 + Mom + Semi & 0.560 & (2.58) & -0.133 & (-1.53) & 0.093  \\
\addlinespace[3pt]
\multicolumn{7}{l}{\quad\textit{Value-weighted, NYSE breakpoints}} \\
(1) & Semi & 0.454 & (2.20) & 0.004 & (0.11) & 0.000  \\
(2) & Mkt + Semi & 0.457 & (2.27) & -0.124 & (-1.92) & 0.056  \\
(3) & FF5 + Semi & 0.433 & (2.08) & -0.080 & (-1.03) & 0.070  \\
(4) & FF5 + Mom + Semi & 0.416 & (2.02) & -0.134 & (-1.63) & 0.099  \\
\addlinespace[6pt]
\multicolumn{7}{l}{\textit{Panel C. AI/robotics ETF basket as additional control}} \\
\multicolumn{7}{l}{\quad\textit{Value-weighted, all-stock breakpoints}} \\
(1) & AI ETF & 0.589 & (2.77) & 0.057 & (0.91) & 0.008  \\
(2) & Mkt + AI ETF & 0.570 & (2.70) & -0.181 & (-1.20) & 0.035  \\
(3) & FF5 + AI ETF & 0.546 & (2.53) & -0.042 & (-0.23) & 0.063  \\
(4) & FF5 + Mom + AI ETF & 0.526 & (2.43) & -0.087 & (-0.45) & 0.073  \\
\addlinespace[3pt]
\multicolumn{7}{l}{\quad\textit{Value-weighted, NYSE breakpoints}} \\
(1) & AI ETF & 0.441 & (2.18) & 0.076 & (1.28) & 0.015  \\
(2) & Mkt + AI ETF & 0.431 & (2.13) & -0.049 & (-0.34) & 0.024  \\
(3) & FF5 + AI ETF & 0.404 & (1.96) & 0.070 & (0.39) & 0.062  \\
(4) & FF5 + Mom + AI ETF & 0.383 & (1.85) & 0.022 & (0.12) & 0.075  \\
\bottomrule
\end{tabular}
\endgroup

\begin{minipage}{\textwidth}
\footnotesize
\textit{Note:}
The dependent variable is the H$-$L value-weighted AI-beta quintile spread. Panel~A adds the Fama-French 10-industry HiTec return, Panel~B adds the value-weighted weekly return on the NAICS-defined semiconductor portfolio, and Panel~C adds the simple weekly excess return on an equal-weighted basket of seven AI ETFs (BOTZ, AIQ, IRBO, ROBO, ARTY, WTAI, CHAT). The ETF basket return is the simple-return aggregation of the daily basket minus the Fama-French weekly risk-free rate. $b_{\mathrm{Add}}$ is the coefficient on the panel-specific additional control. \textit{Mkt} is the Fama-French weekly market excess return. FF5 is the Fama-French five-factor model (\citealp{famaFrench2015}), and FF5+Mom adds the momentum factor (\citealp{carhart1997}). Each row is one OLS spanning regression, and no row combines multiple additional controls. AI consumption data is weekly from January 2024 through April 2026. The equity sample covers 108 formation weeks from March 2024 through March 2026.
\end{minipage}
\end{table}

%% file: tables_v22/table_ia_china_hml.tex
\begin{table}[!htbp]
\centering
\small
\caption{China A-Share Quintile Spreads Sorted on AI-Consumption Betas}
\label{tab:ia_china_hml}
\begin{tabular}{lcccc}
\toprule
 & \multicolumn{2}{c}{Raw} & \multicolumn{2}{c}{China-market $\alpha$} \\
\cmidrule(lr){2-3} \cmidrule(lr){4-5}
Weight & $\alpha$ & $t$ & $\alpha$ & $t$ \\
\midrule
\multicolumn{5}{l}{\textit{Panel A. China A-shares sorted on $\beta^{AI}$}} \\
Value-weighted & -0.159 & (-0.61) & -0.167 & (-0.63) \\
Equal-weighted & -0.072 & (-0.37) & -0.073 & (-0.36) \\
\addlinespace[4pt]
\multicolumn{5}{l}{\textit{Panel B. China A-shares sorted on $\beta^{AI,China}$}} \\
Value-weighted & -0.241 & (-1.03) & -0.113 & (-0.45) \\
Equal-weighted & -0.300 & (-1.59) & -0.219 & (-1.07) \\
\bottomrule
\end{tabular}

\begin{minipage}{\textwidth}
\footnotesize
\textit{Note:}
The table reports value-weighted and equal-weighted High$-$Low spreads for China A-share stocks sorted into pooled quintiles by their 13-week rolling log-excess AI beta. Panel~A uses the overall AI factor. Panel~B uses a China-specific AI factor constructed from the subset of AI consumption originating from users in mainland China. The sample is the Datastream China A-share common-stock panel. \textit{Raw} is the time-series mean of the H$-$L spread. \textit{China-market $\alpha$} is the intercept from regressing the H$-$L spread on the value-weighted Datastream A-share market log-excess return. Returns and alphas are in percent per week. AI consumption data is weekly from January 2024 through April 2026. Equity betas use 13-week rolling windows, so portfolio formation weeks begin in March 2024.
\end{minipage}
\end{table}

%% file: figure_oa_ai_vix_v22.tex
\begin{figure}[!htbp]
\centering
\includegraphics[width=0.95\textwidth]{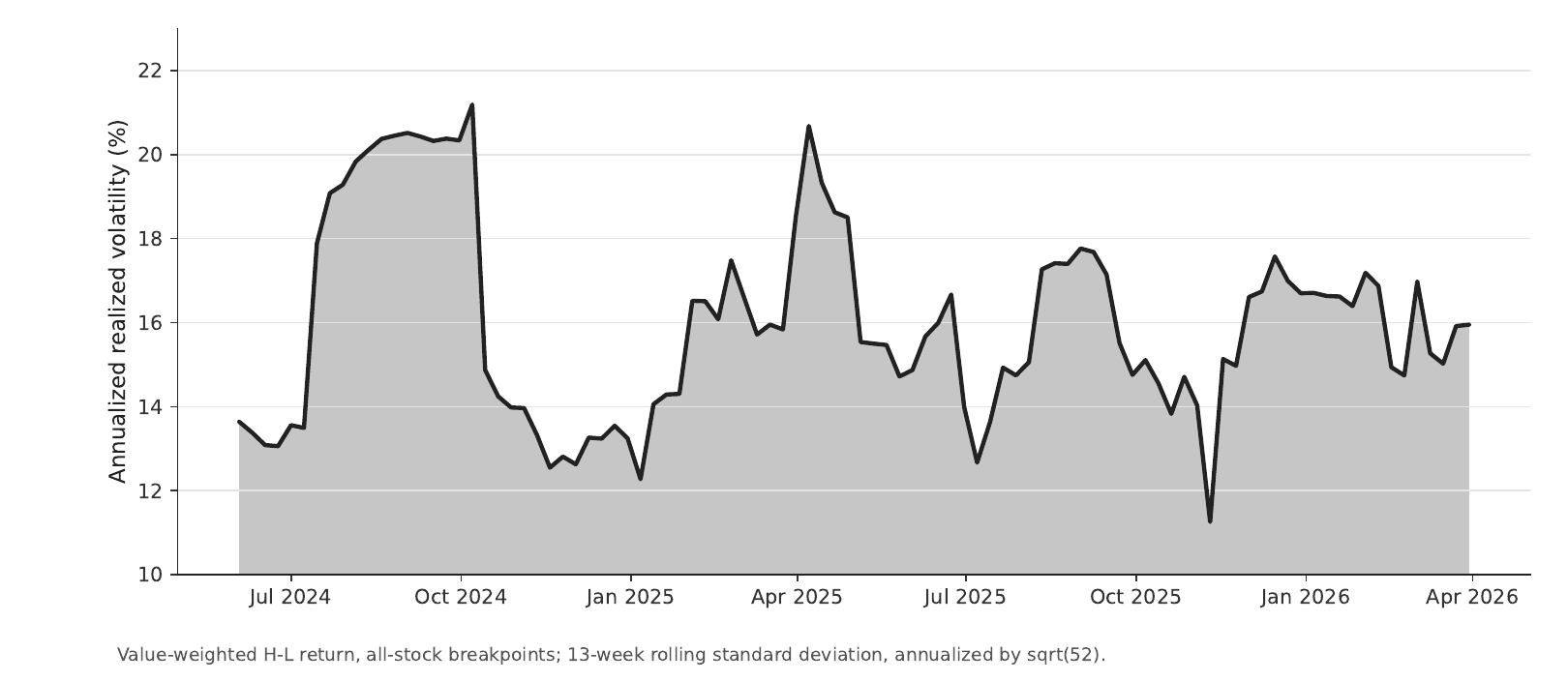}
\caption{AI VIX: Rolling Volatility of the AI-Premium Portfolio}
\label{fig:ai_vix_hml_realized_volatility}

\begin{minipage}{\textwidth}
\footnotesize
\textit{Note:} The figure plots the AI VIX index, constructed as the 13-week rolling standard deviation of the value-weighted high-minus-low AI-beta portfolio return, annualized by $\sqrt{52}$. The high-minus-low portfolio uses all-stock breakpoints, the baseline AI beta, and the same value-weighted specification as the baseline portfolio-sort tests. AI consumption data is weekly from January 2024 through April 2026. The equity sample covers 108 formation weeks from March 2024 through March 2026.
\end{minipage}
\end{figure}

%% file: tables_v22/table_event_study_all.tex
\begin{table}[!htbp]
\centering
\begingroup
\small
\renewcommand{\arraystretch}{1.08}
\setlength{\tabcolsep}{12pt}
\caption{AI-Beta Long-Short Portfolio Returns Around AI Model Releases}
\label{tab:event_study_all}
\begin{tabular*}{0.74\textwidth}{@{\extracolsep{\fill}}lccc@{}}
\toprule
\multicolumn{4}{l}{\textit{Panel A. Frontier model releases (29 events)}} \\
 & $(-1,+1)$ & $(0,+1)$ & $(0,+5)$ \\
\midrule
Raw & 1.127 & 0.704 & 1.731 \\
\textit{t} & (2.59) & (1.95) & (3.31) \\
\addlinespace[2pt]
Market model & 1.089 & 0.675 & 1.225 \\
\textit{t} & (2.21) & (1.78) & (2.12) \\
\addlinespace[2pt]
FF3 & 0.787 & 0.476 & 0.726 \\
\textit{t} & (1.79) & (1.40) & (1.37) \\
\addlinespace[2pt]
FF5 & 0.799 & 0.529 & 0.748 \\
\textit{t} & (1.86) & (1.64) & (1.39) \\
\addlinespace[2pt]
FF5+Mom & 0.609 & 0.452 & 0.932 \\
\textit{t} & (1.62) & (1.53) & (1.80) \\
\addlinespace[6pt]
\midrule
\multicolumn{4}{l}{\textit{Panel B. Non-frontier model releases (36 events)}} \\
 & $(-1,+1)$ & $(0,+1)$ & $(0,+5)$ \\
\midrule
Raw & 0.464 & 0.384 & 0.982 \\
\textit{t} & (1.49) & (1.58) & (2.60) \\
\addlinespace[2pt]
Market model & 0.285 & 0.194 & 0.705 \\
\textit{t} & (0.86) & (0.78) & (1.75) \\
\addlinespace[2pt]
FF3 & 0.244 & 0.135 & 0.425 \\
\textit{t} & (0.99) & (0.75) & (1.37) \\
\addlinespace[2pt]
FF5 & 0.072 & 0.103 & 0.237 \\
\textit{t} & (0.28) & (0.54) & (0.71) \\
\addlinespace[2pt]
FF5+Mom & 0.281 & 0.148 & 0.251 \\
\textit{t} & (1.19) & (0.86) & (0.72) \\
\bottomrule
\end{tabular*}

\begin{minipage}{\textwidth}
\footnotesize
\textit{Note:}
Panel~A uses 29 frontier AI model release dates from Anthropic, OpenAI, Google, Meta, and DeepSeek. Panel~B uses 36 verified non-frontier releases from non-top-5 providers. For each event, firms in the eligible U.S.\ universe are sorted by pre-event 13-week rolling AI beta and formed into value-weighted High-minus-Low portfolios. Windows are trading days from the event anchor. \textit{Raw} is the cumulative simple return. The Market model, Fama--French three-factor model (FF3; \citealp{famaFrench1993}), Fama--French five-factor model (FF5; \citealp{famaFrench2015}), and five-factor model augmented with momentum (FF5+Mom; \citealp{carhart1997}) rows subtract expected returns from the corresponding model estimated over the prior $(-250,-30)$ trading-day window. Reported numbers are cross-event mean H$-$L returns in percent, with $t$-statistics in parentheses. AI consumption data is weekly from January 2024 through April 2026. The equity sample covers 108 formation weeks from March 2024 through March 2026.
\end{minipage}
\endgroup
\end{table}

%% file: tables_v22/table_ia_release_calendar.tex
\begin{table}[p]
\centering
\setstretch{1}
\fontsize{9.5}{10.5}\selectfont
\setlength{\tabcolsep}{3pt}
\renewcommand{\arraystretch}{1.03}
\caption{AI Model Release Calendar and Primary Sources}
\label{tab:ia_release_calendar}
\begin{tabular}{@{}p{47pt}p{120pt}p{\dimexpr\linewidth-179pt\relax}@{}}
\toprule
\multicolumn{3}{l}{\textit{Panel A. Frontier model release calendar (29 events)}} \\
Date & Provider and model & Note \\
\midrule
2024-04-18 & \raggedright Meta: \href{https://ai.meta.com/blog/meta-llama-3/}{Llama 3} & \raggedright Open model release; strong open-weight benchmark step. \tabularnewline
2024-05-13 & \raggedright OpenAI: \href{https://openai.com/index/gpt-4o-and-more-tools-to-chatgpt-free/}{GPT-4o} & \raggedright OpenAI flagship omni model; cheaper and faster API. \tabularnewline
2024-05-14 & \raggedright Google: \href{https://blog.google/technology/ai/google-gemini-update-flash-ai-assistant-io-2024/}{Gemini 1.5 Flash} & \raggedright Fast/efficient Gemini family update at Google I/O. \tabularnewline
2024-06-21 & \raggedright Anthropic: \href{https://www.anthropic.com/news/claude-3-5-sonnet}{Claude 3.5 Sonnet} & \raggedright Anthropic frontier Sonnet model. \tabularnewline
2024-07-23 & \raggedright Meta: \href{https://ai.meta.com/blog/meta-llama-3-1/}{Llama 3.1 405B} & \raggedright First frontier-level open-source AI model per Meta. \tabularnewline
2024-09-12 & \raggedright OpenAI: \href{https://openai.com/index/introducing-openai-o1-preview/}{o1-preview} & \raggedright OpenAI first public reasoning-model series. \tabularnewline
2024-10-22 & \raggedright Anthropic: \href{https://www.anthropic.com/news/3-5-models-and-computer-use}{Claude 3.5 Sonnet new} & \raggedright Upgraded Sonnet plus public beta computer use. \tabularnewline
2024-12-06 & \raggedright Meta: \href{https://ai.meta.com/blog/future-of-ai-built-with-llama/}{Llama 3.3 70B} & \raggedright Efficiency/cost-improved Llama model; source is Meta year-end review. \tabularnewline
2024-12-11 & \raggedright Google: \href{https://blog.google/innovation-and-ai/models-and-research/google-deepmind/google-gemini-ai-update-december-2024/}{Gemini 2.0} & \raggedright Google agentic-era model release. \tabularnewline
2024-12-27 & \raggedright DeepSeek: \href{https://arxiv.org/abs/2412.19437}{DeepSeek-V3} & \raggedright Open-weight frontier MoE model; arXiv date used for public paper/checkpoint availability. \tabularnewline
2025-01-20 & \raggedright DeepSeek: \href{https://docs.aws.amazon.com/bedrock/latest/userguide/model-card-deepseek-deepseek-r1.html}{DeepSeek-R1 release} & \raggedright Official model launch date per AWS Bedrock model card. \tabularnewline
2025-01-27 & \raggedright DeepSeek: \href{https://www.axios.com/2025/01/27/deepseek-ai-model-china-openai-rival}{DeepSeek-R1 market recognition} & \raggedright US market-recognition shock one week after the Jan. 20 R1 release. \tabularnewline
2025-02-24 & \raggedright Anthropic: \href{https://www.anthropic.com/news/claude-3-7-sonnet}{Claude 3.7 Sonnet} & \raggedright Hybrid reasoning model. \tabularnewline
2025-02-27 & \raggedright OpenAI: \href{https://openai.com/index/introducing-gpt-4-5/}{GPT-4.5} & \raggedright OpenAI largest GPT model research preview. \tabularnewline
2025-03-25 & \raggedright Google: \href{https://blog.google/innovation-and-ai/models-and-research/google-deepmind/gemini-model-thinking-updates-march-2025/}{Gemini 2.5 Pro} & \raggedright Google thinking model. \tabularnewline
2025-04-05 & \raggedright Meta: \href{https://ai.meta.com/blog/llama-4-multimodal-intelligence/}{Llama 4 Scout/Maverick} & \raggedright Open-weight natively multimodal Llama 4 models. \tabularnewline
2025-04-14 & \raggedright OpenAI: \href{https://openai.com/index/gpt-4-1/}{GPT-4.1} & \raggedright API model with coding, instruction-following, and long-context gains. \tabularnewline
2025-04-16 & \raggedright OpenAI: \href{https://openai.com/index/introducing-o3-and-o4-mini/}{o3 and o4-mini} & \raggedright OpenAI o-series reasoning models with tool access. \tabularnewline
2025-05-22 & \raggedright Anthropic: \href{https://www.anthropic.com/news/claude-4}{Claude 4} & \raggedright Claude Opus 4 and Sonnet 4. \tabularnewline
2025-08-07 & \raggedright OpenAI: \href{https://openai.com/index/introducing-gpt-5/}{GPT-5} & \raggedright OpenAI unified GPT-5 system. \tabularnewline
2025-08-21 & \raggedright DeepSeek: \href{https://api-docs.deepseek.com/news/news250821/}{DeepSeek V3.1} & \raggedright Hybrid thinking model; DeepSeek's first step toward the agent era. \tabularnewline
2025-09-29 & \raggedright Anthropic: \href{https://www.anthropic.com/news/claude-sonnet-4-5}{Claude Sonnet 4.5} & \raggedright Anthropic frontier Sonnet model for complex agents and computer use. \tabularnewline
2025-11-12 & \raggedright OpenAI: \href{https://openai.com/index/gpt-5-1/}{GPT-5.1} & \raggedright GPT-5 series update; agentic-focused API models followed the next day. \tabularnewline
2025-11-18 & \raggedright Google: \href{https://blog.google/products/gemini/gemini-3/}{Gemini 3 Pro} & \raggedright Gemini 3 flagship; Antigravity agentic platform launched the same day. \tabularnewline
2025-11-24 & \raggedright Anthropic: \href{https://www.anthropic.com/news/claude-opus-4-5}{Claude Opus 4.5} & \raggedright Anthropic flagship Opus model for coding, agents, and computer use. \tabularnewline
2025-12-01 & \raggedright DeepSeek: \href{https://api-docs.deepseek.com/news/news251201/}{DeepSeek V3.2} & \raggedright Stable V3.2 and V3.2-Speciale; reasoning-first open-weight models. \tabularnewline
2025-12-11 & \raggedright OpenAI: \href{https://openai.com/index/introducing-gpt-5-2/}{GPT-5.2} & \raggedright GPT-5 series update for professional work and long-running agents. \tabularnewline
2026-02-05 & \raggedright Anthropic: \href{https://www.anthropic.com/news/claude-opus-4-6}{Claude Opus 4.6} & \raggedright Agent teams and 1M-token context; same-day general availability. \tabularnewline
2026-02-19 & \raggedright Google: \href{https://blog.google/innovation-and-ai/models-and-research/gemini-models/gemini-3-1-pro/}{Gemini 3.1 Pro} & \raggedright Reasoning-focused Gemini flagship update; preview in the API and AI Studio. \tabularnewline
2026-03-05 & \raggedright OpenAI: \href{https://openai.com/index/introducing-gpt-5-4/}{GPT-5.4} & \raggedright GPT-5.4, GPT-5.4 Thinking, and GPT-5.4 Pro; mini and nano followed on March 17. \tabularnewline
\bottomrule
\end{tabular}
\end{table}

\begin{table}[p]
\centering
\setstretch{1}
\fontsize{9.5}{10.5}\selectfont
\setlength{\tabcolsep}{3pt}
\renewcommand{\arraystretch}{1.03}
\begin{tabular}{@{}p{47pt}p{120pt}p{\dimexpr\linewidth-179pt\relax}@{}}
\toprule
\multicolumn{3}{l}{\textit{Panel B. Non-frontier model release calendar (36 events)}} \\
Date & Provider and model & Note \\
\midrule
2024-03-11 & \raggedright Cohere: \href{https://cohere.com/blog/command-r}{Command R} & \raggedright RAG / tool-use enterprise LLM (128K context). \tabularnewline
2024-03-17 & \raggedright xAI: \href{https://x.ai/news/grok-os}{Grok-1 (open weights)} & \raggedright 314B-parameter MoE base model, Apache 2.0. \tabularnewline
2024-03-27 & \raggedright Databricks: \href{https://www.databricks.com/blog/introducing-dbrx-new-state-art-open-llm}{DBRX} & \raggedright 132B-parameter MoE open LLM (36B active). \tabularnewline
2024-03-28 & \raggedright AI21: \href{https://www.ai21.com/blog/announcing-jamba/}{Jamba} & \raggedright First production-grade hybrid SSM-Transformer LLM. \tabularnewline
2024-04-04 & \raggedright Cohere: \href{https://cohere.com/blog/command-r-plus-microsoft-azure}{Command R+} & \raggedright Scaled-up RAG/tool-use enterprise LLM (104B). \tabularnewline
2024-04-15 & \raggedright Reka: \href{https://reka.ai/news/reka-core-our-frontier-class-multimodal-language-model}{Reka Core} & \raggedright Frontier-class multimodal model (text/image/video/audio). \tabularnewline
2024-04-23 & \raggedright Microsoft: \href{https://azure.microsoft.com/en-us/blog/introducing-phi-3-redefining-whats-possible-with-slms/}{Phi-3 (mini)} & \raggedright 3.8B small language model launched on Azure. \tabularnewline
2024-05-13 & \raggedright 01.AI: \href{https://x.com/01AI_Yi/status/1789894091620458667}{Yi-Large} & \raggedright 01.AI's first proprietary closed-source frontier dense LLM. \tabularnewline
2024-06-05 & \raggedright Zhipu: \href{https://github.com/THUDM/GLM-4}{GLM-4-9B (open weights)} & \raggedright Open-weight 9B variant of the GLM-4 family. \tabularnewline
2024-06-07 & \raggedright Alibaba: \href{https://qwenlm.github.io/blog/qwen2/}{Qwen 2} & \raggedright Qwen 2 family (0.5B-72B) open-source release. \tabularnewline
2024-07-24 & \raggedright Mistral: \href{https://mistral.ai/news/mistral-large-2407}{Mistral Large 2} & \raggedright 123B-parameter flagship multilingual / code-strong dense LLM. \tabularnewline
2024-08-01 & \raggedright Black Forest Labs: \href{https://bfl.ai/announcing-black-forest-labs/}{FLUX.1} & \raggedright 12B rectified-flow text-to-image model family. \tabularnewline
2024-08-13 & \raggedright xAI: \href{https://x.ai/news/grok-2}{Grok-2} & \raggedright Grok-2 / Grok-2 mini early-preview release on X. \tabularnewline
2024-08-22 & \raggedright AI21: \href{https://www.ai21.com/blog/announcing-jamba-model-family/}{Jamba 1.5 family} & \raggedright Hybrid SSM-Transformer family with 256K context. \tabularnewline
2024-09-19 & \raggedright Alibaba: \href{https://qwenlm.github.io/blog/qwen2.5/}{Qwen 2.5} & \raggedright Open-source family announced at Apsara Conference. \tabularnewline
2024-10-21 & \raggedright IBM: \href{https://newsroom.ibm.com/2024-10-21-ibm-introduces-granite-3-0-high-performing-ai-models-built-for-business}{Granite 3.0} & \raggedright Apache-2.0 enterprise LLM family launched at TechXchange. \tabularnewline
2024-11-05 & \raggedright Tencent: \href{https://huggingface.co/tencent/Tencent-Hunyuan-Large}{Hunyuan-Large} & \raggedright 389B-parameter MoE (52B active) open-sourced. \tabularnewline
2024-11-18 & \raggedright Mistral: \href{https://mistral.ai/news/pixtral-large}{Pixtral Large} & \raggedright 124B open-weights multimodal model (123B + 1B vision). \tabularnewline
2024-11-28 & \raggedright Alibaba: \href{https://qwenlm.github.io/blog/qwq-32b-preview/}{QwQ-32B-Preview} & \raggedright Open-weight reasoning model preview from Qwen. \tabularnewline
2024-12-03 & \raggedright Amazon: \href{https://press.aboutamazon.com/2024/12/introducing-amazon-nova-a-new-generation-of-foundation-models}{Nova family} & \raggedright Foundation-model family launched on Amazon Bedrock at re:Invent. \tabularnewline
2024-12-12 & \raggedright Microsoft: \href{https://huggingface.co/microsoft/phi-4}{Phi-4} & \raggedright 14B reasoning-focused SLM (Azure preview, fully open Jan 2025). \tabularnewline
2025-01-22 & \raggedright ByteDance: \href{https://seed.bytedance.com/en/special/doubao_1_5_pro}{Doubao 1.5 Pro} & \raggedright Large-scale sparse MoE flagship on Volcano Engine. \tabularnewline
2025-01-28 & \raggedright Alibaba: \href{https://qwenlm.github.io/blog/qwen2.5-max/}{Qwen 2.5-Max} & \raggedright Closed-source MoE flagship API model. \tabularnewline
2025-02-17 & \raggedright xAI: \href{https://x.ai/news/grok-3}{Grok-3} & \raggedright Flagship reasoning model trained on Colossus. \tabularnewline
2025-03-13 & \raggedright Cohere: \href{https://cohere.com/blog/command-a}{Command A} & \raggedright 111B enterprise model (256K context, 2 GPU inference). \tabularnewline
2025-04-29 & \raggedright Alibaba: \href{https://qwenlm.github.io/blog/qwen3/}{Qwen 3} & \raggedright Hybrid thinking/non-thinking dense + MoE family. \tabularnewline
2025-06-10 & \raggedright Mistral: \href{https://mistral.ai/news/magistral}{Magistral (Small + Medium)} & \raggedright Mistral's first reasoning-focused models. \tabularnewline
2025-07-09 & \raggedright xAI: \href{https://x.ai/news/grok-4}{Grok 4} & \raggedright Grok 4 / Grok 4 Heavy livestream release. \tabularnewline
2025-07-11 & \raggedright Moonshot: \href{https://moonshotai.github.io/Kimi-K2/}{Kimi K2} & \raggedright 1T-parameter open-weight MoE flagship built for agentic tasks. \tabularnewline
2025-07-28 & \raggedright Zhipu: \href{https://z.ai/blog/glm-4.5}{GLM-4.5} & \raggedright 355B open-weight flagship with native agent abilities (GLM-4.5 and GLM-4.5-Air). \tabularnewline
2025-10-02 & \raggedright IBM: \href{https://www.ibm.com/new/announcements/ibm-granite-4-0-hyper-efficient-high-performance-hybrid-models}{Granite 4.0} & \raggedright Hybrid Mamba/transformer enterprise model family, general availability. \tabularnewline
2025-10-27 & \raggedright MiniMax: \href{https://www.minimax.io/news/minimax-m2}{MiniMax M2} & \raggedright Open-weight model built for agents and code. \tabularnewline
2025-12-02 & \raggedright Amazon: \href{https://www.aboutamazon.com/news/aws/aws-re-invent-2025-ai-news-updates}{Nova 2 family} & \raggedright Nova 2 Lite, Pro, Sonic, and Omni announced at re:Invent. \tabularnewline
2025-12-02 & \raggedright Mistral: \href{https://mistral.ai/news/mistral-3}{Mistral Large 3} & \raggedright 675B sparse MoE flagship (41B active), released with the Ministral 3 series. \tabularnewline
2026-01-27 & \raggedright Moonshot: \href{https://www.kimi.ai/blog/kimi-k2-5}{Kimi K2.5} & \raggedright Visual agentic model; agent swarms of up to 100 sub-agents. \tabularnewline
2026-02-12 & \raggedright MiniMax: \href{https://www.minimax.io/blog/minimax-m25}{MiniMax M2.5} & \raggedright Coding, agentic tool use, and search. \tabularnewline
2026-02-12 & \raggedright Zhipu: \href{https://z.ai/blog/glm-5}{GLM-5} & \raggedright Flagship for complex systems engineering and long-horizon agentic tasks. \tabularnewline
2026-02-14 & \raggedright ByteDance: \href{https://www.caixinglobal.com/2026-02-15/bytedance-unveils-doubao-20-ai-model-to-tackle-complex-tasks-102414865.html}{Doubao-Seed-2.0} & \raggedright Doubao-Seed-2.0 series (Pro, Lite, Mini, Code); press source. \tabularnewline
2026-02-16 & \raggedright Alibaba: \href{https://qwen.ai/blog?id=qwen3.5}{Qwen3.5} & \raggedright Qwen3.5 family: Plus API model and 397B-A17B open weights. \tabularnewline
\bottomrule
\end{tabular}

\vspace{5pt}
\begin{minipage}{\linewidth}
\fontsize{9}{10}\selectfont
\textit{Note:} Model names link to the cited sources. Dates are announced public release dates and sources are primary citations. Panel~A is the frontier-event sample. Panel~A lists both the DeepSeek-R1 release date and the U.S. market-recognition date one week later. The event studies anchor DeepSeek-R1 on the January 20, 2025 release date, so the two DeepSeek-R1 rows document a single event and Panel~A contains 29 events. Releases that share an event day (Panel~B: the rows dated December 2, 2025; the rows dated February 12, 2026; the rows dated February 14, 2026 and February 16, 2026 (both anchored on February 17, 2026)) are listed separately but enter the event studies once, so Panel~B contains 36 events. Panel~B excludes Anthropic, OpenAI, Google, Meta, and DeepSeek and keeps flagship or series-defining releases. AI consumption data is weekly from January 2024 through April 2026. The equity sample covers 108 formation weeks from March 2024 through March 2026.
\end{minipage}
\end{table}

%% file: figure_oa_industry_v22.tex
\begin{figure}[!htbp]
\centering
\caption{Industry Exposure to AI Factor}
\label{fig:ff10_beta_heatmap}
\includegraphics[width=0.80\textwidth]{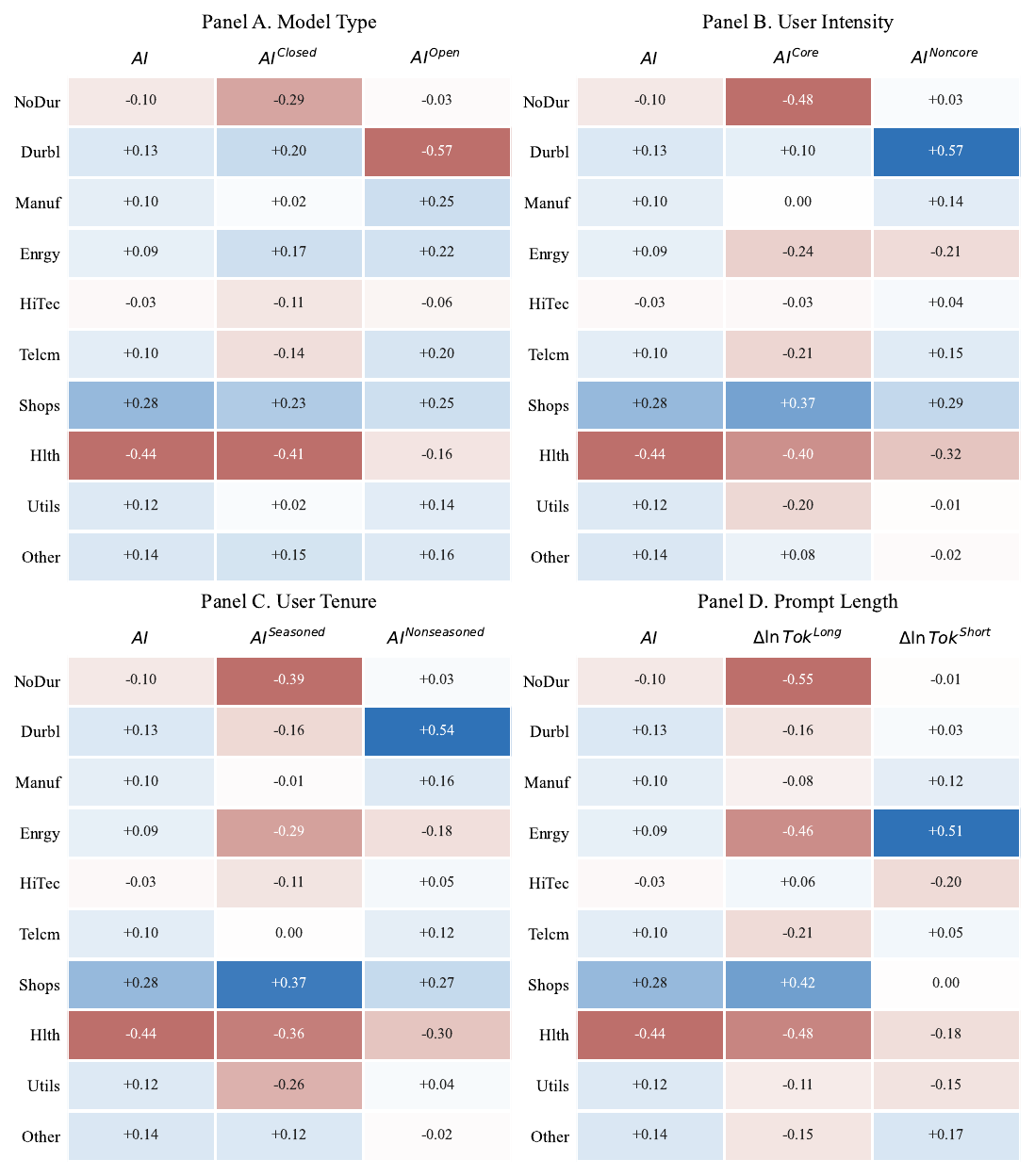}

\begin{minipage}{\textwidth}
\footnotesize
\textit{Note:} The figure reports the implied weekly industry return in percent associated with a one-standard-deviation increase in each AI variable for the Fama-French 10 industries. For each firm-week, the relevant beta is estimated from the same 13-week rolling regression of weekly log-excess returns on the log-excess market return and the AI variable used in the portfolio sorts. Within each FF10 industry, we take the value-weighted mean of firm-level betas using formation-week market equity weights; each cell reports $100\times\overline{\beta}^{X}_{j,\mathrm{vw}}\times\sigma_w(X)$ for the column variable $X$. The baseline $AI$ factor is the first principal component of weekly log growth in total tokens, usage dollars, and distinct users. Panel A compares $AI$, $AI^{Closed}$, and $AI^{Open}$. Panel B compares $AI$, $AI^{Core}$, and $AI^{Noncore}$. Panel C compares $AI$, $AI^{Seasoned}$, and $AI^{Nonseasoned}$. Panel D compares $AI$, $\Delta\ln Tok^{Long}$, and $\Delta\ln Tok^{Short}$. Samples vary by split according to non-missing rolling-beta availability. Cell shading is scaled within each column; blue denotes positive responses and red denotes negative responses. AI consumption data is weekly from January 2024 through April 2026. The equity sample covers 108 formation weeks from March 2024 through March 2026.
\end{minipage}
\end{figure}

%% file: figure_oa_industry_full_sample.tex
\begin{figure}[!htbp]
\centering
\caption{Industry AI Exposure: Average Firm Exposure and Full-Sample Regressions}
\label{fig:industry_full_sample}
\makebox[\linewidth][c]{\includegraphics[width=\linewidth]{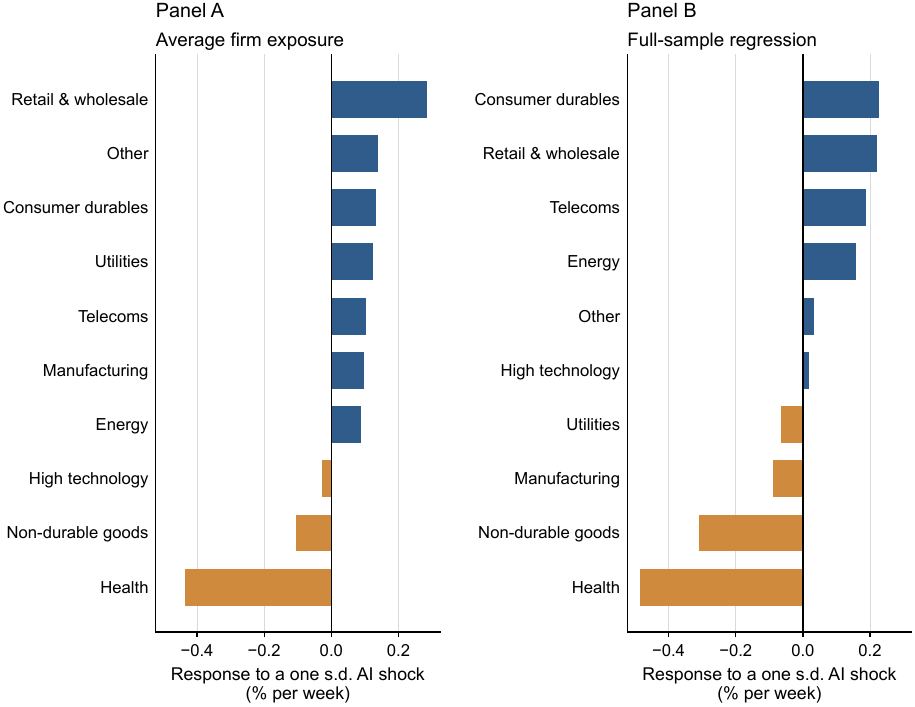}}

\begin{minipage}{\textwidth}
\footnotesize
\textit{Note:} The figure reports AI exposure for the Fama-French 10 industries, as the return response, in percentage points per week, to a one-standard-deviation AI-factor shock. Panel~A repeats Panel~A of Figure~\ref{fig:exposure_industries_firms}, the value-weighted average of firm-level exposures, where a firm's exposure is the coefficient on the AI factor in 13-week rolling regressions of its weekly log-excess return on the AI factor and the log-excess market return. Panel~B reports full-sample exposure. For each industry, the value-weighted weekly log-excess return is regressed once on the AI factor, the Fama-French five factors, and momentum over the 108 holding weeks from March 2024 through March 2026. Industry returns use the same firms and formation-week market equity weights as Panel~A. The two panels share the horizontal axis.
\end{minipage}
\end{figure}

%% file: figure_us_baseline_v22.tex
\begin{landscape}
\begin{figure}[p]
\centering
\textit{Panel A. Top 50 S\&P 500 Firms}\\[2pt]
\makebox[\linewidth][c]{\includegraphics[width=0.95\linewidth]{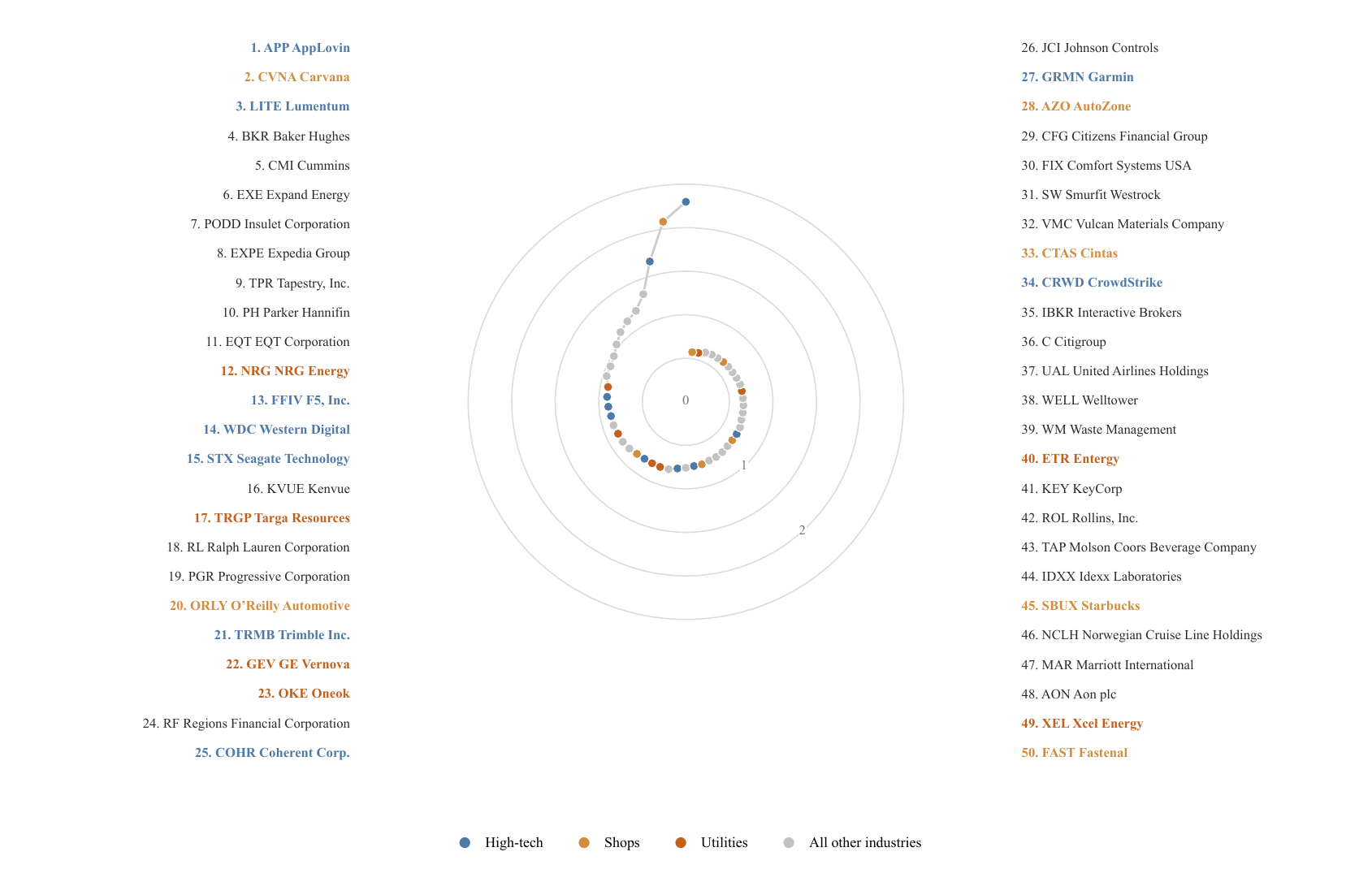}}
\caption{S\&P 500 Firm-Level AI Exposure}
\label{fig:sp500_top_bottom_company_ai_exposure}
\end{figure}

\clearpage
\begin{figure}[p]
\ContinuedFloat
\centering
\textit{Panel B. Bottom 50 S\&P 500 Firms}\\[2pt]
\makebox[\linewidth][c]{\includegraphics[width=0.95\linewidth]{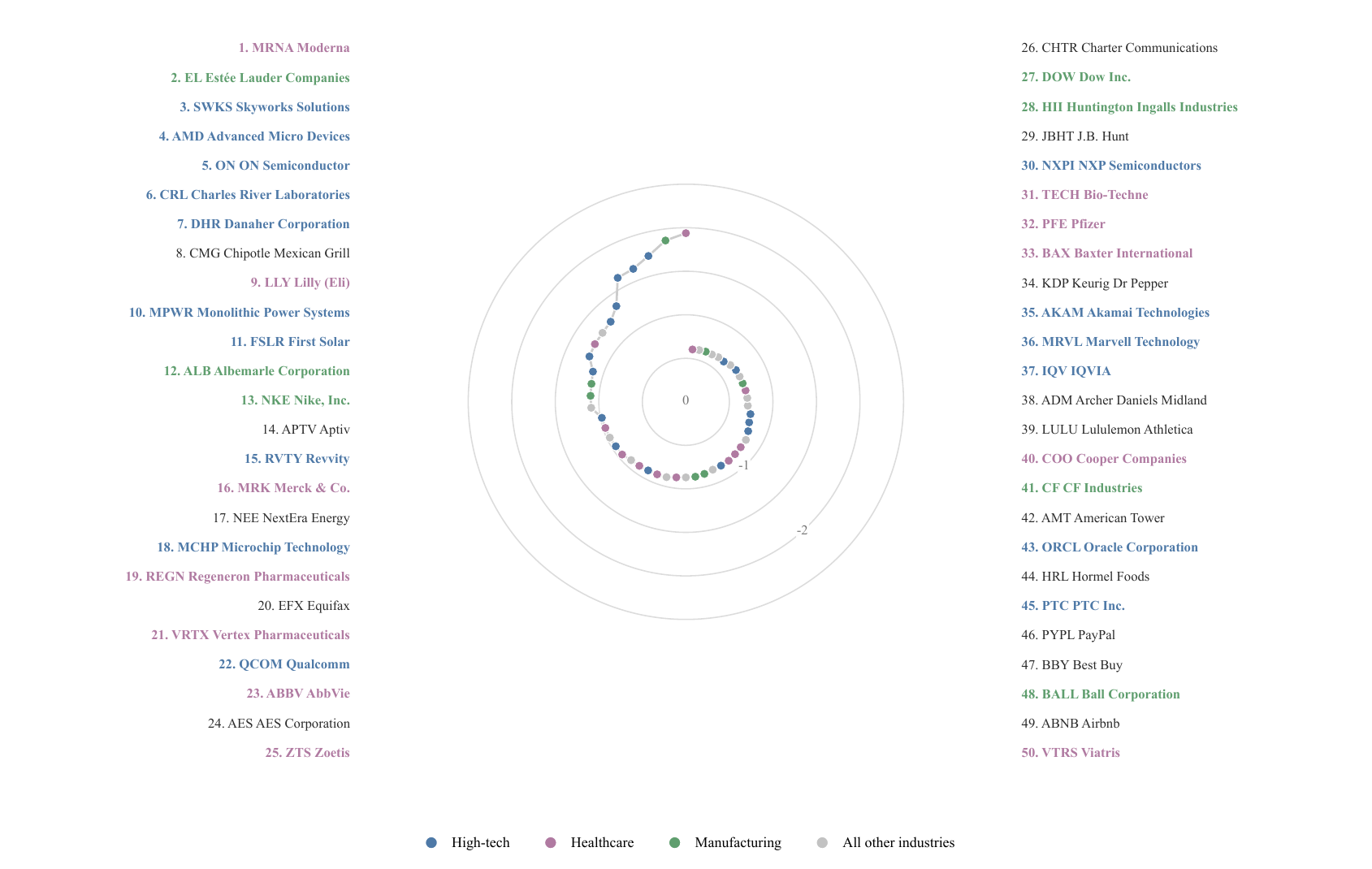}}
\caption[]{S\&P 500 Firm-Level AI Exposure (continued)}

\begin{minipage}{\textwidth}
\footnotesize
\textit{Note:} The figure plots the 50 current S\&P 500 firms with the largest positive (Panel~A) and most negative (Panel~B) average rolling exposures to the baseline AI factor. Firm-level $\beta^{AI}$ is estimated from 13-week rolling regressions of weekly log-excess returns on the market log-excess return and the AI factor, then averaged across valid weeks. Radial distance reports $100\times\overline{\beta}^{AI}_{i}\times\sigma_w(AI)$, the weekly return in percentage points associated with a one-standard-deviation AI-factor shock. Label colors denote Fama-French 10 industries, as shown in the legend. Both panels use the same radial scale.
\end{minipage}
\end{figure}
\end{landscape}

%% file: tables_v22/table_ia_agentic_definitions_v15.tex
\begin{table}[!htbp]
\centering
\footnotesize
\caption{Agentic-Economy Concepts and Factor Definitions}
\label{tab:agentic_defs}
\begin{tabular}{@{}l l p{8.0cm}@{}}
\toprule
Factor & Concept & Construction \\
\midrule
Agentic tokens & Tool use & Log growth in total tokens on requests whose normalized finish reason is \texttt{tool\_calls}. \\
\addlinespace[2pt]
Cache-read tokens & Context reuse & Log growth in cache-read tokens---prompt-prefix tokens served from cache rather than recomputed. \\
\addlinespace[2pt]
Reasoning tokens & Test-time compute & Log growth in reasoning tokens---the model's internal reasoning (``work plan'') tokens. \\
\addlinespace[2pt]
Agentic dollars & Token economics & Log growth in usage dollars on requests whose normalized finish reason is \texttt{tool\_calls}. \\
\addlinespace[2pt]
\bottomrule
\end{tabular}

\begin{minipage}{\textwidth}
\footnotesize
\textit{Note:}
Each row is one weekly factor built from the OpenRouter inference logs; the corresponding token and dollar shares are plotted in Figure~\ref{fig:agentic_combined}. Total tokens are prompt plus completion tokens. Each factor is a weekly log growth, $\ln L_t-\ln L_{t-1}$, of a positive weekly level. The agentic, cache-read, and reasoning slices are populated only after their instrumentation onset.
\end{minipage}
\end{table}

%% file: tables_v22/table_nber_fact_rewrites_v9.tex
\begin{landscape}
\begin{table}[!p]
\centering
\caption{Granular OpenRouter Facts for Comparison with \citet{demirer2025emerging}}
\label{tab:nber_fact_rewrites}
\scriptsize
\setlength{\tabcolsep}{4pt}
\renewcommand{\arraystretch}{1.02}
\begin{minipage}{0.98\linewidth}
\raggedright

\noindent\textit{Panel A. Open-Source Token and Dollar Shares}\\[-0.25em]

\begin{tabular*}{\linewidth}{@{\extracolsep{\fill}}lllccc}
\toprule
Period & Week & Class & Request share & Token share & Dollar share \\
\midrule
First week & 2024-01-01 & Closed & 9.4 & 15.6 & 51.9 \\
First week & 2024-01-01 & Open & 90.6 & 84.4 & 48.1 \\
Final week & 2026-04-13 & Closed & 54.8 & 51.6 & 84.7 \\
Final week & 2026-04-13 & Open & 45.2 & 48.4 & 15.3 \\
\bottomrule
\end{tabular*}

\vspace{0.35em}
\noindent\textit{Panel B. Market Dynamism in Tokens and Dollars}\\[-0.25em]

\begin{tabular*}{\linewidth}{@{\extracolsep{\fill}}lcccccc}
\toprule
Metric & Model HHI first & Model HHI latest & Change & Author HHI first & Author HHI latest & Change \\
\midrule
Requests & 1321.7 & 267.5 & -79.8 & 1764.7 & 1565.1 & -11.3 \\
Tokens & 1057.9 & 333.4 & -68.5 & 1508.6 & 1117.0 & -26.0 \\
Usage dollars & 975.6 & 959.3 & -1.7 & 1996.2 & 3435.8 & 72.1 \\
\bottomrule
\end{tabular*}

\vspace{0.15em}
\begin{tabular*}{\linewidth}{@{\extracolsep{\fill}}lccc}
\toprule
Metric & Horizon & Model top-10 retention & Author top-10 retention \\
\midrule
Requests & 4 weeks & 87.0 & 91.0 \\
Requests & 8 weeks & 85.0 & 90.0 \\
Requests & 12 weeks & 80.0 & 90.0 \\
Tokens & 4 weeks & 58.0 & 78.0 \\
Tokens & 8 weeks & 36.7 & 73.3 \\
Tokens & 12 weeks & 35.0 & 85.0 \\
Usage dollars & 4 weeks & 70.0 & 92.0 \\
Usage dollars & 8 weeks & 50.0 & 86.7 \\
Usage dollars & 12 weeks & 35.0 & 80.0 \\
\bottomrule
\end{tabular*}

\vspace{0.35em}
\noindent\textit{Panel C. Prompt-Price Elasticity Comparison}\\[-0.25em]

\begin{tabular*}{\linewidth}{@{\extracolsep{\fill}}lcccc}
\toprule
Specification & Paper coef. & Our coef. & Our $t$-stat. & Obs. \\
\midrule
Date FE & -0.55 & -0.450 & (-4.67) & 52,728 \\
Date + model FE & -1.08 & -1.170 & (-5.66) & 52,728 \\
Model-date + model-provider FE & -1.11 & 0.280 & (0.79) & 53,800 \\
\bottomrule
\end{tabular*}

\vspace{0.35em}
\noindent\textit{Panel D. Multihoming and Concentration Across OpenRouter Entities}\\[-0.25em]

\begin{tabular*}{\linewidth}{@{\extracolsep{\fill}}lcccc}
\toprule
Entity & First & Latest & $\geq$2 first & $\geq$2 latest \\
\midrule
API key & 2025-01-01 & 2026-04-01 & 48.7 & 51.8 \\
Application & 2025-01-01 & 2026-04-01 & 38.3 & 35.8 \\
\bottomrule
\end{tabular*}

\vspace{0.35em}
\begin{minipage}{\linewidth}
\scriptsize
\textit{Note:}
Panel A reports shares among the open-plus-closed model universe using realized requests, prompt-plus-completion tokens, and usage dollars. Panel B reports Herfindahl-Hirschman indices across models and authors and average top-10 retention in the latest 90 days of the sample. Panel C compares the prompt-price elasticity estimates in \citet{demirer2025emerging}, Table~2, with the corresponding granular OpenRouter reconstruction. The sample is open-source model-provider-days, the dependent variable is log token quantity, and the price regressor is log posted prompt price. Panel D reports multihoming among OpenRouter applications and API keys. Usage dollars are the realized OpenRouter \texttt{usage} field, not a recomputation from posted prices. Shares and changes are reported in percent. Data is weekly from January 2024 through April 2026.
\end{minipage}
\end{minipage}
\end{table}
\end{landscape}

%% file: figure_oa_retention_providers.tex
\begin{figure}[!htbp]
\centering
\includegraphics[width=0.85\textwidth]{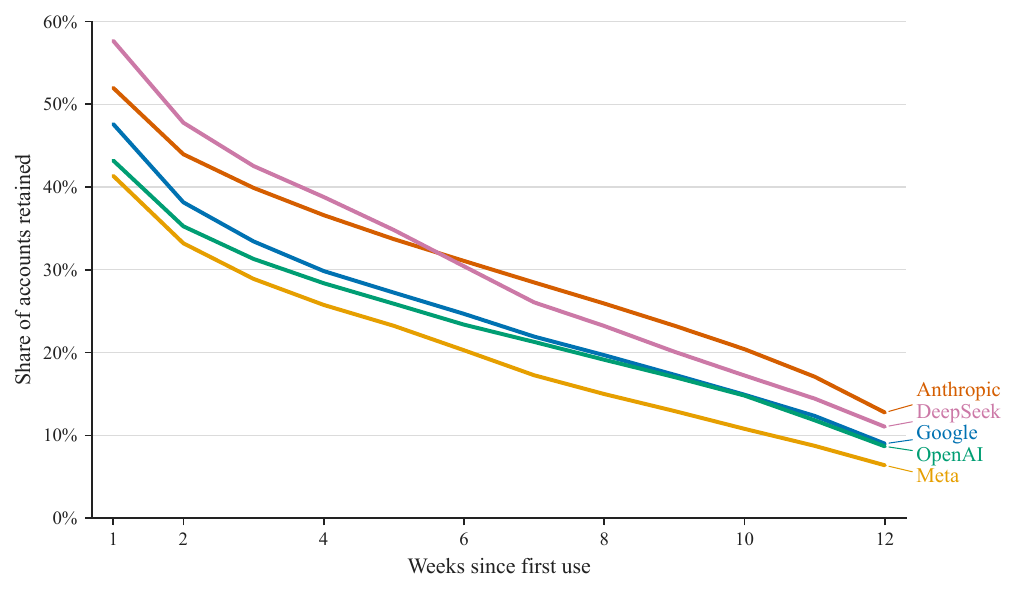}
\caption{User Retention by Model Provider}
\label{fig:retention_providers}

\begin{minipage}{\textwidth}
\footnotesize
\textit{Note:} The figure plots twelve-week user retention for the five frontier providers in the event study of Section~\ref{sec:event_study}: Anthropic, DeepSeek, Google, Meta, and OpenAI. For each provider, a cohort consists of the accounts that use any of the provider's models for the first time in a given week. Retention at week $k$ is the share of the cohort that is active on any of the provider's models in week $k$ or in a later week within twelve weeks of first use. Cohorts are pooled over the weeks from January 29, 2024 through January 12, 2026, the last week that can be followed for twelve weeks before the data end in April 2026. Cohorts from the first four weeks of 2024 are left out because accounts first seen in those weeks may have used the provider before the sample starts. The pooled cohorts contain between 0.5 million (Anthropic) and 2.1 million (DeepSeek) accounts.
\end{minipage}
\end{figure}

%% file: figure_oa_retention_margins.tex
\begin{figure}[!htbp]
\centering
\textit{Panel A. Closed-Source versus Open-Weight Models}\\[2pt]
\makebox[\linewidth][c]{\includegraphics[width=0.80\linewidth]{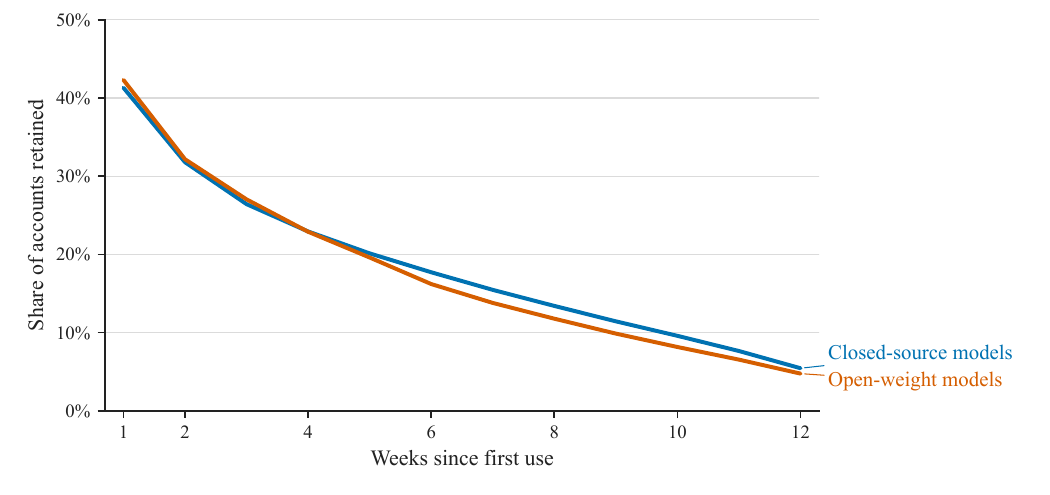}}

\vspace{0.3em}

\textit{Panel B. Seasoned versus Non-Seasoned Users}\\[2pt]
\makebox[\linewidth][c]{\includegraphics[width=0.80\linewidth]{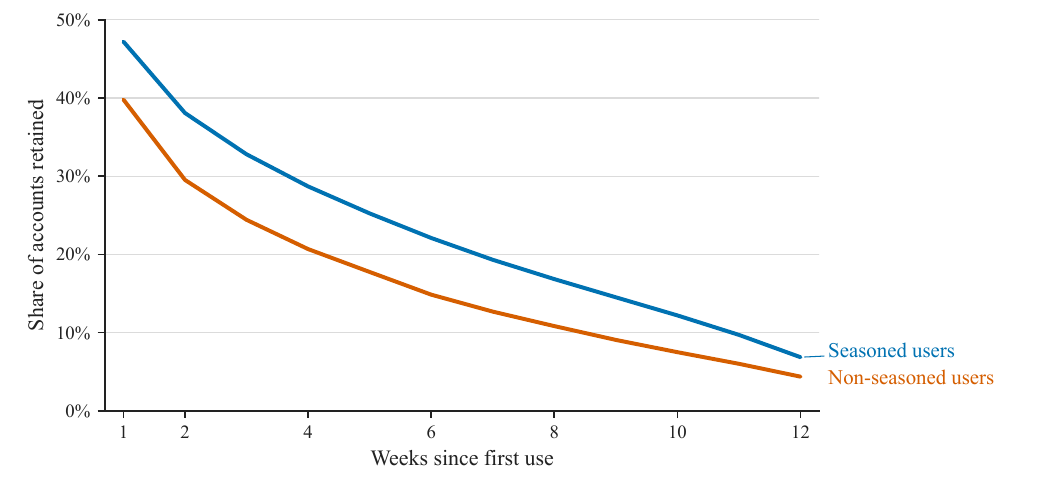}}
\caption{User Retention by Model Type and User Experience}
\label{fig:retention_margins}

\begin{minipage}{\textwidth}
\footnotesize
\textit{Note:} The two panels plot twelve-week retention of model adopters along two of the margins in Section~\ref{sec:ai_subcomponents}. A cohort consists of the accounts that use a given model for the first time in a given week, and each curve pools the cohorts of all models in the group. Retention at week $k$ is the share of the cohort that is active on the same model in week $k$ or in a later week within twelve weeks of first use. Panel~A splits models into closed-source and open-weight models following OpenRouter's model taxonomy. Panel~B splits adopters by whether the account is seasoned in the week of first use, that is, observed for at least 13 weeks, active in at least eight prior weeks, and with at least 20 prior requests. Cohorts are pooled over the weeks from January 29, 2024 through January 12, 2026. Panel~A pools 6.9 million first uses of closed-source models and 11.5 million first uses of open-weight models (models without a closed or open classification are left out), and Panel~B pools 3.9 million first uses by seasoned accounts and 15.0 million by non-seasoned accounts.
\end{minipage}
\end{figure}

%% file: figure_oa_skill_exposure_v22.tex
\begin{figure}[p]
\centering
\textit{Panel A. Ranks 1--55}\\[2pt]
\makebox[\linewidth][c]{\includegraphics[width=\linewidth]{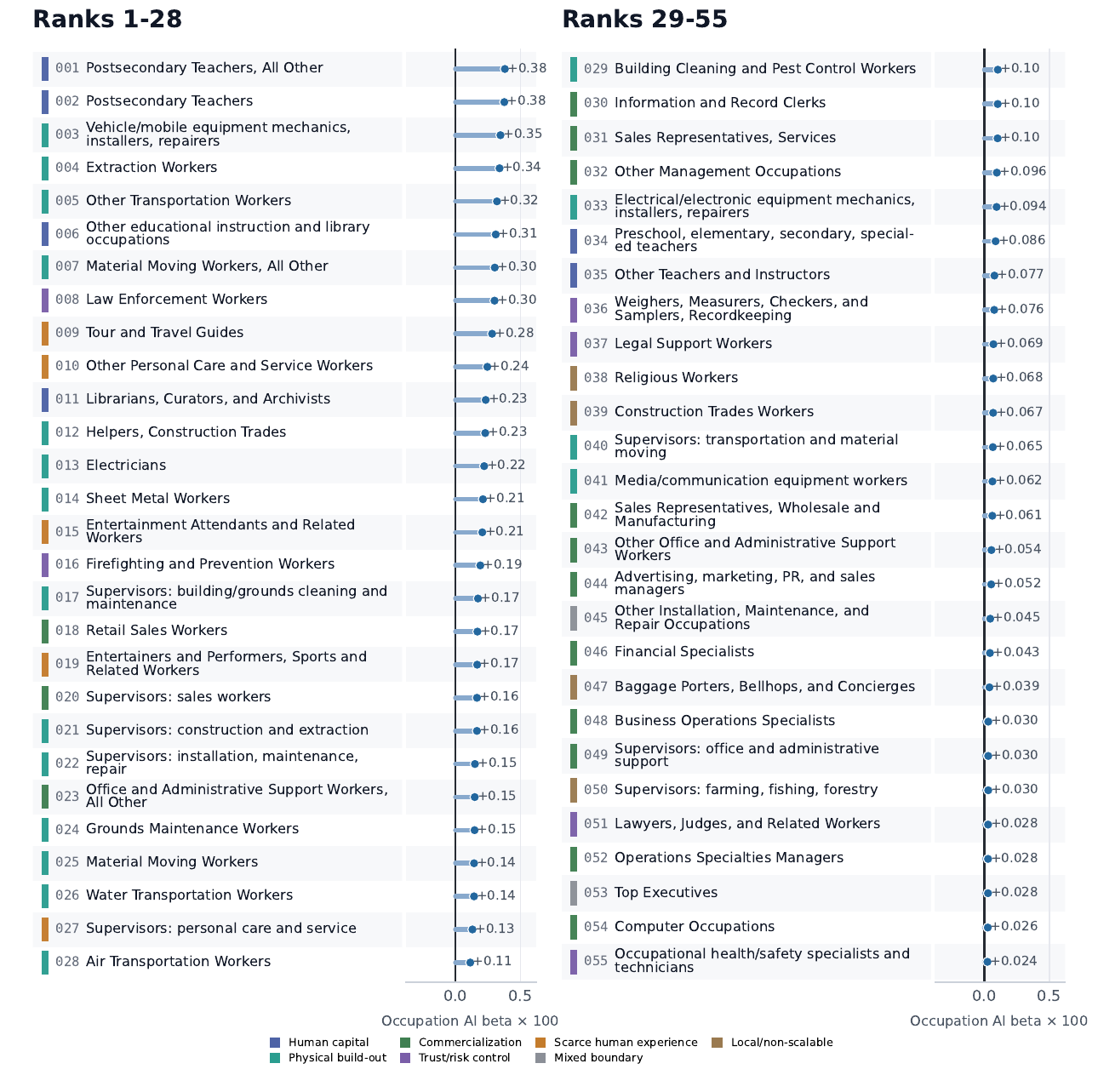}}
\end{figure}

\begin{figure}[p]
\centering
\textit{Panel B. Ranks 56--109}\\[2pt]
\makebox[\linewidth][c]{\includegraphics[width=\linewidth]{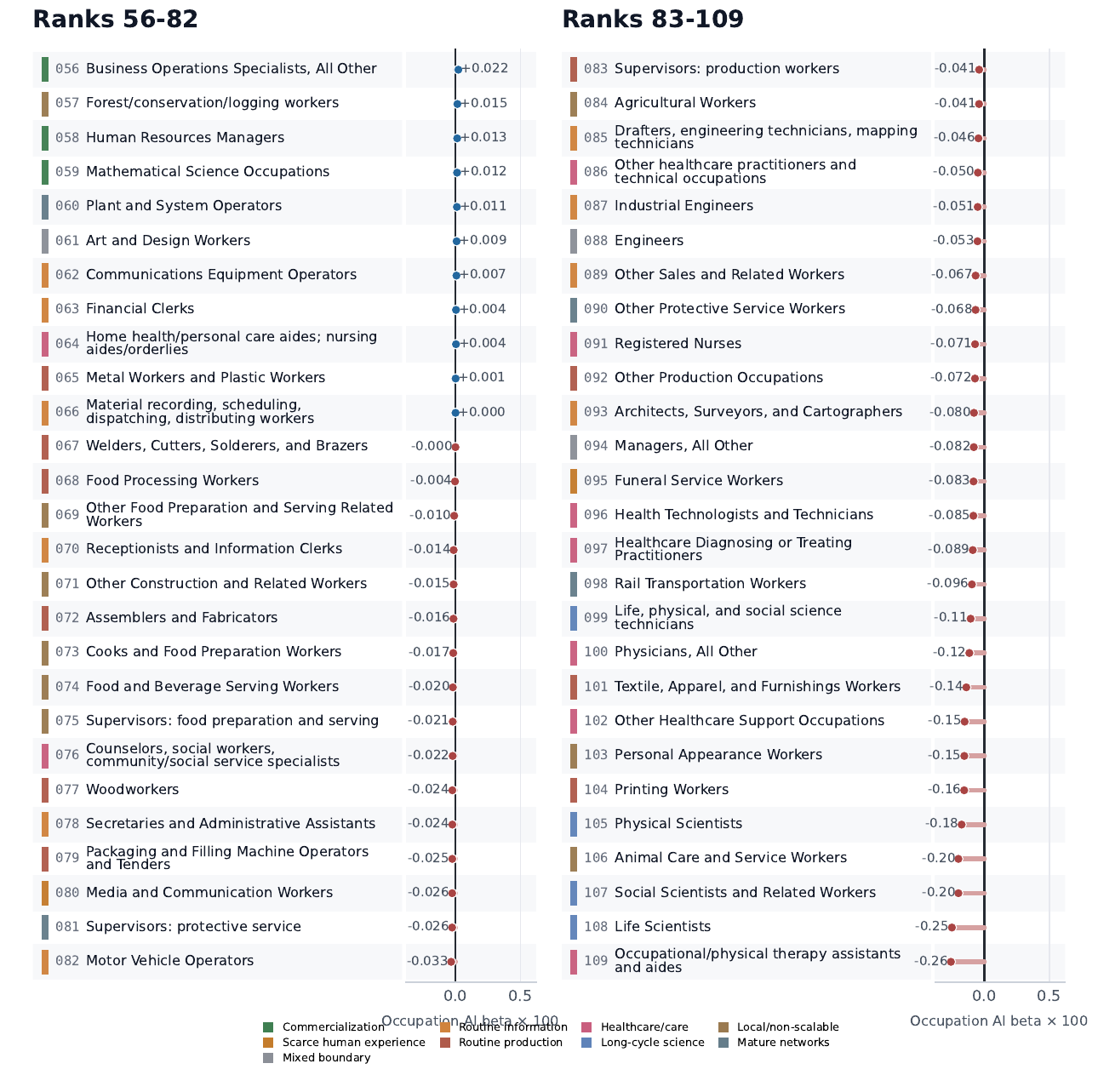}}
\caption{Occupation Exposure to the AI Factor}
\label{fig:occupation_ai_exposure}

\begin{minipage}{\textwidth}
\footnotesize
\textit{Note:} The figure plots the full-firm occupation ranking based on SOC minor occupation groups. NAICS-level exposure is the market-equity-weighted mean of firm-level AI betas within NAICS-week cells, requiring at least three firms in a NAICS-week, averaged over time and mapped to occupations using BLS OEWS occupation-by-industry employment weights. Detailed SOC occupations are aggregated to SOC minor groups using employment weights. The plotted value is $100\times\beta^{AI}_{o}$, where $\beta^{AI}_{o}$ is the employment-weighted occupation exposure to the baseline AI factor. The colored rail marks the coherent economic theme assigned to each occupation group. Panel A reports ranks 1--55, and Panel B reports ranks 56--109. Underlying firm-level AI betas are estimated on the equity sample of 108 formation weeks from March 2024 through March 2026.
\end{minipage}
\end{figure}

%% file: tables_v22/table_beta_determinants.tex
\begin{landscape}
\begin{table}[!p]
\centering
\caption{Determinants of Firm-Level AI Betas}
\label{tab:beta_determinants}
\begingroup
\setlength{\tabcolsep}{3pt}
\renewcommand{\arraystretch}{0.92}
\scriptsize
\begin{minipage}[t]{0.48\linewidth}
\vspace{0pt}
\centering
\begin{tabular}{lccccc}
\toprule
 & (1) & (2) & (3) & (4) & (5) \\
\midrule
\multicolumn{6}{l}{\textit{Panel A. Week fixed effects}} \\
$\ln ME$ & 0.056 & 0.059 & 0.064 & 0.010 & 0.010 \\
 & (6.95) & (5.61) & (5.89) & (1.08) & (1.05) \\
\addlinespace[1pt]
$\ln BM$ &  & 0.079 & 0.135 & 0.093 & 0.055 \\
 &  & (3.01) & (4.93) & (3.96) & (2.27) \\
\addlinespace[1pt]
$Mom$ &  & 0.456 & 0.488 & 0.474 & 0.464 \\
 &  & (6.24) & (6.67) & (6.71) & (6.55) \\
\addlinespace[1pt]
$Profit$ &  &  & 0.507 & 0.492 & 0.402 \\
 &  &  & (5.80) & (5.66) & (4.66) \\
\addlinespace[1pt]
$AG$ &  &  & -0.045 & 0.068 & 0.040 \\
 &  &  & (-0.77) & (1.24) & (0.73) \\
\addlinespace[1pt]
$ROE$ &  &  &  & 0.011 & -0.053 \\
 &  &  &  & (0.27) & (-1.24) \\
\addlinespace[1pt]
$Lev$ &  &  &  & 0.481 & 0.358 \\
 &  &  &  & (6.93) & (5.20) \\
\addlinespace[1pt]
$Ivol$ &  &  &  & -12.264 & -10.814 \\
 &  &  &  & (-6.90) & (-6.11) \\
\addlinespace[1pt]
$Turn$ &  &  &  & 0.001 & 0.000 \\
 &  &  &  & (0.36) & (0.07) \\
\addlinespace[1pt]
$Illiq$ &  &  &  & 0.007 & 0.004 \\
 &  &  &  & (0.40) & (0.23) \\
\addlinespace[1pt]
$Capx$ &  &  &  &  & -0.480 \\
 &  &  &  &  & (-0.78) \\
\addlinespace[1pt]
$R\&D$ &  &  &  &  & -1.568 \\
 &  &  &  &  & (-5.00) \\
\addlinespace[1pt]
Obs & 367,059 & 263,850 & 257,421 & 255,939 & 255,084 \\
$R^{2}$ & 0.054 & 0.078 & 0.083 & 0.099 & 0.101 \\
\bottomrule
\end{tabular}
\end{minipage}
\hfill
\begin{minipage}[t]{0.48\linewidth}
\vspace{0pt}
\centering
\begin{tabular}{lccccc}
\toprule
 & (1) & (2) & (3) & (4) & (5) \\
\midrule
\multicolumn{6}{l}{\textit{Panel B. Week and industry fixed effects}} \\
$\ln ME$ & 0.054 & 0.040 & 0.046 & 0.017 & 0.017 \\
 & (6.71) & (4.05) & (4.50) & (1.82) & (1.79) \\
\addlinespace[1pt]
$\ln BM$ &  & -0.004 & 0.051 & 0.052 & 0.031 \\
 &  & (-0.17) & (2.05) & (2.18) & (1.32) \\
\addlinespace[1pt]
$Mom$ &  & 0.409 & 0.420 & 0.442 & 0.438 \\
 &  & (5.67) & (5.90) & (6.23) & (6.15) \\
\addlinespace[1pt]
$Profit$ &  &  & 0.631 & 0.576 & 0.520 \\
 &  &  & (6.41) & (6.17) & (5.68) \\
\addlinespace[1pt]
$AG$ &  &  & 0.002 & 0.055 & 0.038 \\
 &  &  & (0.04) & (0.99) & (0.67) \\
\addlinespace[1pt]
$ROE$ &  &  &  & -0.050 & -0.083 \\
 &  &  &  & (-1.24) & (-1.97) \\
\addlinespace[1pt]
$Lev$ &  &  &  & 0.205 & 0.148 \\
 &  &  &  & (2.88) & (2.10) \\
\addlinespace[1pt]
$Ivol$ &  &  &  & -9.710 & -9.113 \\
 &  &  &  & (-5.41) & (-5.12) \\
\addlinespace[1pt]
$Turn$ &  &  &  & -0.000 & -0.000 \\
 &  &  &  & (-0.01) & (-0.08) \\
\addlinespace[1pt]
$Illiq$ &  &  &  & 0.003 & 0.001 \\
 &  &  &  & (0.20) & (0.08) \\
\addlinespace[1pt]
$Capx$ &  &  &  &  & -0.906 \\
 &  &  &  &  & (-1.44) \\
\addlinespace[1pt]
$R\&D$ &  &  &  &  & -1.067 \\
 &  &  &  &  & (-3.37) \\
\addlinespace[1pt]
Obs & 355,322 & 261,095 & 254,798 & 253,329 & 252,474 \\
$R^{2}$ & 0.069 & 0.093 & 0.099 & 0.106 & 0.106 \\
\bottomrule
\end{tabular}
\end{minipage}
\endgroup

\begin{minipage}{0.98\linewidth}
\footnotesize\textit{Note:}
\footnotesize This table reports pooled panel regressions of firm-level AI betas on firm characteristics. The dependent variable is the rolling 13-week AI beta, winsorized at the $1$st/$99$th percentiles within each week. Coefficients are multiplied by $100$. $\ln ME$ is log market equity, $\ln BM$ is log book-to-market, $Mom$ is prior $52$-to-$5$-week return, $Profit$ is gross profitability, $AG$ is asset growth, $ROE$ is return on equity, $Lev$ is total debt over total assets, $Ivol$ is idiosyncratic volatility, $Turn$ is average daily share turnover, $Illiq$ is Amihud illiquidity, $Capx$ is capital expenditures over total assets, and $R\&D$ is research and development expense over total assets and is set to zero when missing. All characteristics are winsorized at the $1$st/$99$th percentiles within each week. Panel A includes week fixed effects. Panel B includes week and Fama-French 30 industry fixed effects. $t$-statistics are reported in parentheses and are based on standard errors clustered by week and firm. Intercepts are absorbed by the fixed effects.
\end{minipage}
\end{table}
\end{landscape}

%% file: tables_v22/table_gtrend_residual_sort.tex
\begin{table}[!htbp]
\centering
\small
\caption{Attention-Orthogonalized AI Factor: Spanning Regression and Portfolio Sorts}
\label{tab:gtrend_residual_sort}
\begingroup
\footnotesize
\textit{Panel A. Spanning regression of the AI factor on Google Trends attention}
\par\medskip
\begin{tabular}{lcccc}
\toprule
 & $b_{\mathrm{gtrend}}$ & $t$ & $R^{2}$ & $N$ \\
\midrule
$AI_t$ on $\Delta\ln \mathrm{gtrend}^{\mathrm{AI}}_{t}$ & 2.419 & (1.72) & 0.027 & 112 \\
\bottomrule
\end{tabular}
\par\bigskip
\textit{Panel B. Value-weighted quintile portfolios sorted on the residual-factor beta}
\par\medskip
\setlength{\tabcolsep}{6pt}
\begin{tabular}{lcccccc}
\toprule
 & Low & 2 & 3 & 4 & High & H$-$L \\
\midrule
Excess return & -0.148 & 0.216 & 0.149 & 0.459 & 0.435 & 0.582 \\
\textit{$t$} & (-0.57) & (1.03) & (0.75) & (2.11) & (1.59) & (2.82) \\
\addlinespace[2pt]
FF5 $\alpha$ & -0.350 & 0.020 & -0.061 & 0.233 & 0.183 & 0.533 \\
\textit{$t(\alpha)$} & (-2.69) & (0.22) & (-0.88) & (2.59) & (1.42) & (2.52) \\
\addlinespace[2pt]
FF5+Mom $\alpha$ & -0.334 & 0.028 & -0.062 & 0.224 & 0.177 & 0.511 \\
\textit{$t(\alpha)$} & (-2.63) & (0.31) & (-0.90) & (2.52) & (1.37) & (2.45) \\
\bottomrule
\end{tabular}
\endgroup

\begin{minipage}{\textwidth}
\footnotesize\textit{Note:}
\footnotesize This table reports the two-step attention-orthogonalization exercise. Panel~A regresses the weekly AI factor on the weekly log growth in Google Trends search interest for the topic \textit{Artificial intelligence}: $AI_{t}=a+b_{\mathrm{gtrend}}\,\Delta\ln \mathrm{gtrend}^{\mathrm{AI}}_{t}+e_{t}$. The $t$-statistic is based on Newey-West standard errors with five lags and is reported in parentheses. The residual $e_{t}$ is the attention-orthogonalized AI factor. Panel~B repeats the baseline portfolio exercise of Table~\ref{tab:us_portfolio_p5_p1} with the residual replacing the raw AI factor in the first-pass rolling beta regression: $r^{e}_{i,t}=a+b_{\mathrm{mkt}}\,r^{e}_{m,t}+b_{\mathrm{AI},\perp}\,e_{t}$, estimated over rolling $13$-week windows. Stocks are sorted into pooled quintiles by $b_{\mathrm{AI},\perp}$ using all-stock breakpoints, and portfolios are value-weighted. Returns and alphas are in percent per week. FF5 is the Fama-French five-factor model and FF5+Mom adds the momentum factor. $t$-statistics are reported in parentheses.
\end{minipage}
\end{table}

%% file: tables_v22/table_ia_imc_ist.tex
\begin{table}[!htbp]
\centering
\small
\caption{The AI Factor and Investment-Specific Technology Risk}
\label{tab:ia_imc_ist}
\begin{tabular}{lccc}
\toprule
 & GKY 1987 & BEA 2012 & BEA 2017 \\
\midrule
\multicolumn{4}{l}{\textit{Panel A. Factor-level correlation with IMC}} \\
corr($AI_t$, $R^{IMC}_t$), net of the market & $-0.016$ & $-0.134$ & $-0.138$ \\
 & $(-0.17)$ & $(-1.44)$ & $(-1.49)$ \\
\addlinespace[1pt]
\multicolumn{4}{l}{\textit{Panel B. Cross-section of exposures}} \\
corr($\beta^{AI}_i$, $\beta^{IMC}_i$) & $-0.008$ & $-0.123$ & $-0.089$ \\
 & $(-0.30)$ & $(-4.04)$ & $(-2.80)$ \\
\addlinespace[1pt]
\multicolumn{4}{l}{\textit{Panel C. Spanning of the AI premium}} \\
H$-$L $\alpha$ (FF5$+$Mom$+$IMC) & 0.521 & 0.508 & 0.494 \\
 & (2.60) & (2.50) & (2.56) \\
$b_{IMC}$ & 0.045 & 0.094 & 0.280 \\
 & (0.33) & (0.66) & (2.22) \\
\bottomrule
\end{tabular}

\begin{minipage}{\textwidth}
\footnotesize
\textit{Note:}
The table asks whether the AI factor is the investment-specific technology shock of \citet{papanikolaou2011investment} and \citet{koganPapanikolaou2014growth} by comparing it with the investment-minus-consumption (IMC) portfolio that mimics that shock. Columns differ in the industry classification defining the legs: the published \citet{gomesKoganYogo2009} classification based on the 1987 benchmark input-output accounts, and the same algorithm applied to the 2012 and 2017 benchmark accounts, in which investment is private fixed investment excluding inventories and net exports are exports less imports. IMC is the value-weighted return of investment-good producers minus consumption-good producers, weighting the following week's return by formation-week market equity. The legs hold 658 and 1,491 firms (1987), 553 and 2,968 (2012), and 765 and 2,822 (2017), and IMC correlates $-0.24$ to $-0.43$ with HML, as an investment portfolio should. Panel~A reports the weekly correlation of the AI factor with the IMC return after residualizing the AI factor on the market. Panel~B reports the mean weekly cross-sectional correlation between firm-level AI betas and IMC betas, both estimated from 13-week rolling regressions with at least 9 non-missing weeks. Panel~C reports the intercept and IMC loading from regressing the value-weighted, all-stock-breakpoint H$-$L return on IMC, the \citet{famaFrench2015} factors, and momentum \citep{carhart1997}, with Newey-West standard errors with five lags. Returns and alphas are in percent per week. $t$-statistics in parentheses. AI consumption data is weekly from January 2024 through April 2026. The equity sample covers 108 formation weeks from March 2024 through March 2026.
\end{minipage}
\end{table}

%% file: tables_v22/variable_definitions_v9.tex
\begingroup
\begin{singlespace}
\footnotesize
\setlength{\LTpre}{0pt}
\setlength{\LTpost}{0pt}
\setlength{\tabcolsep}{3pt}
\renewcommand{\arraystretch}{0.90}

\begin{longtable}{p{0.22\textwidth}p{0.49\textwidth}p{0.23\textwidth}}
\caption{Variable Definitions}\label{tab:definitions_appendix}\\
\toprule
Name & Definition & Dataset used \\
\midrule
\endfirsthead
\toprule
Name & Definition & Dataset used \\
\midrule
\endhead
\bottomrule
\endfoot
\bottomrule
\endlastfoot

\multicolumn{3}{l}{\textit{AI consumption and attention variables}} \\
\addlinespace[1pt]
$Tok_{t}$ & Total OpenRouter prompt plus completion tokens in week $t$. & OpenRouter \\
$Dol_{t}$ & Total dollar-denominated OpenRouter usage in week $t$. & OpenRouter \\
$User_{t}$ & Distinct active OpenRouter users in week $t$. & OpenRouter \\
$\Delta\ln Tok_{t}$ & Weekly log growth in total OpenRouter tokens. & OpenRouter \\
$\Delta\ln Dol_{t}$ & Weekly log growth in OpenRouter usage dollars. & OpenRouter \\
$\Delta\ln User_{t}$ & Weekly log growth in distinct active OpenRouter users. & OpenRouter \\
$AI_t$ & First principal component of $\Delta\ln Tok_{t}$, $\Delta\ln Dol_{t}$, and $\Delta\ln User_{t}$. & OpenRouter \\
$AI^{Closed}_{t}$, $AI^{Open}_{t}$ & AI factors constructed separately from closed-weight and open-weight model consumption. & OpenRouter; model catalog \\
$AI^{Core}_{t}$, $AI^{New}_{t}$ & AI factors constructed separately from paid/core-user and new-user consumption. & OpenRouter account data \\
$AI^{Noncore}_{t}$ & AI factor constructed from non-paid/core account consumption, the complement of paid/core account consumption. & OpenRouter account data \\
$AI^{Seasoned}_{t}$, $AI^{Nonseasoned}_{t}$ & AI factors constructed separately from seasoned-user and non-seasoned-user consumption. & OpenRouter account data \\
$\Delta\ln Tok^{Long}_{t}$, $\Delta\ln Tok^{Short}_{t}$ & Weekly token-growth series for long prompts and short prompts, split by the rolling prompt-length cutoff. & OpenRouter row-level usage \\
$Share^{cat}_{c,t}$ & Share of categorized tokens assigned to prompt category $c$ in week $t$. & OpenRouter category-token file \\
$Share^{Closed}_{t}$, $Share^{Core}_{t}$, $Share^{Seasoned}_{t}$, $Share^{Long}_{t}$ & Weekly usage shares for closed-weight models, paid/core users, seasoned users, and long prompts. & OpenRouter; model catalog; account data \\
$\Delta\ln \mathrm{gtrend}^{AI}_{t}$ & Weekly log growth in Google Trends interest for the topic ``Artificial intelligence.'' & Google Trends \\

\addlinespace[3pt]
\multicolumn{3}{l}{\textit{Return, exposure, and control variables}} \\
\addlinespace[1pt]
$r^e_{i,t}$ & Weekly log excess return of stock $i$ over the risk-free rate. & CRSP; Fama-French \\
$r^e_{m,t}$ or $rm_t$ & Weekly log excess market return. & Fama-French \\
$\beta^{rm}_{i,t}$ & Rolling market beta of stock $i$ from the first-pass return regression. & CRSP; Fama-French \\
$\beta^{AI}_{i,t}$ & Rolling AI exposure of stock $i$: coefficient on the AI factor in a 13-week regression of $r^e_{i,t}$ on $r^e_{m,t}$ and $AI_t$. & CRSP; OpenRouter; Fama-French \\
$\beta^{Tok}_{i,t}$ & Rolling exposure of stock $i$ to $\Delta\ln Tok_{t}$. & CRSP; OpenRouter; Fama-French \\
$\beta^{Dol}_{i,t}$ & Rolling exposure of stock $i$ to $\Delta\ln Dol_{t}$. & CRSP; OpenRouter; Fama-French \\
$\beta^{User}_{i,t}$ & Rolling exposure of stock $i$ to $\Delta\ln User_{t}$. & CRSP; OpenRouter; Fama-French \\
$\beta^{AI,Ind}_{i,t}$ & AI beta minus the same-week median AI beta of stock $i$'s Fama-French 30 industry. & CRSP; Compustat; OpenRouter; Fama-French \\
$r^e_{\mathrm{HiTec},t}$ & Fama-French HiTec industry excess return used as a first-pass control. & Fama-French \\
$r_{\mathrm{semi},t}$ & Weekly log return of the U.S. semiconductor industry used as a first-pass control. & CRSP; Compustat \\
$\beta^{AI,Closed}_{i,t}$, $\beta^{AI,Open}_{i,t}$ & Rolling exposures of stock $i$ to $AI^{Closed}_{t}$ and $AI^{Open}_{t}$. & CRSP; OpenRouter; model catalog \\
$\beta^{AI,Core}_{i,t}$, $\beta^{AI,New}_{i,t}$ & Rolling exposures of stock $i$ to $AI^{Core}_{t}$ and $AI^{New}_{t}$. & CRSP; OpenRouter account data \\
$\beta^{AI,Seasoned}_{i,t}$, $\beta^{AI,Nonseasoned}_{i,t}$ & Rolling exposures of stock $i$ to $AI^{Seasoned}_{t}$ and $AI^{Nonseasoned}_{t}$. & CRSP; OpenRouter account data \\
$\beta^{Tok,Long}_{i,t}$, $\beta^{Tok,Short}_{i,t}$ & Rolling exposures of stock $i$ to $\Delta\ln Tok^{Long}_{t}$ and $\Delta\ln Tok^{Short}_{t}$. & CRSP; OpenRouter row-level usage \\
$\ln ME_{i,t}$ & Natural log of stock $i$'s formation-week market equity. & CRSP; Compustat \\
$\ln BM_{i,t}$ & Natural log of stock $i$'s book-to-market equity. & Compustat; CRSP \\
$Profit_{i,t}$ & Gross profitability, revenue minus cost of goods sold scaled by assets. & Compustat \\
$Inv_{i,t}$ & Asset growth, measured as the year-over-year change in total assets scaled by lagged assets. & Compustat \\
$Mom_{i,t}$ & Prior stock return from week $t-52$ through week $t-5$. & CRSP \\
$Rev_{i,t}$ & Prior one-month stock return used as a short-term reversal control. & CRSP \\
$Lev_{i,t}$ & Long-term debt scaled by assets. & Compustat \\
$Accr_{i,t}$ & Accruals. & Compustat \\

\addlinespace[3pt]
\multicolumn{3}{l}{\textit{International and event-study variables}} \\
\addlinespace[1pt]
$r^{local}_{i,t}$ & Weekly local-currency return of non-U.S. stock $i$. & Compustat Global \\
$r^{local}_{m,t}$ & Weekly regional local-market return. & Compustat Global; market returns \\
$AI^{China}_{t}$ & AI factor constructed from OpenRouter usage originating from mainland China. & OpenRouter country-coded usage \\
$r^{China}_{m,t}$ & Weekly Datastream China A-share market log excess return. & Datastream; Fama-French \\
$r^{AIETF}_{t}$ & Weekly excess return on the AI and robotics ETF basket used as a first-pass and spanning control. & ETF return data; Fama-French \\
$CAR_{e,[a,b]}$ & Cumulative abnormal H-minus-Low return for event $e$ over trading-day window $[a,b]$. & CRSP; Fama-French; release calendar \\
$CAAR_{[a,b]}$ & Cross-event average of $CAR_{e,[a,b]}$ over the specified event window. & Constructed \\

\end{longtable}
\begin{minipage}{\textwidth}
\footnotesize
\textit{Note:} The table defines the variables used in the analysis and identifies the dataset used for each.
\end{minipage}
\end{singlespace}
\endgroup